\documentclass[final,1p,times,authoryear]{elsarticle}

\usepackage{amssymb}
\usepackage{amsmath}
\usepackage{appendix}

\usepackage{siunitx}
\usepackage{url}
\journal{Icarus}

\begin{document}

\begin{frontmatter}




\title{Inferring Meteoroid Properties with Dynamic Nested Sampling: A Case Study of Orionid and Capricornid Shower Meteors}


\author[UWO,WS]{Maximilian Vovk\corref{cor1}}\ead{mvovk@uwo.ca}

\author[UWO,WS]{Peter G. Brown}\ead{pbrown@uwo.ca}

\author[UWO,WS]{Denis Vida}\ead{dvida@uwo.ca}

\author[NASA]{Daeyoung Lee}\ead{daeyoung.lee@nasa.gov}

\author[UWO]{Emma G. Harmos}\ead{eharmos2@uwo.ca}

\cortext[cor1]{Corresponding author}

\affiliation[UWO]{organization={Department of Physics and Astronomy, University of Western Ontario},
            addressline={1151 Richmond Street}, 
            city={London},
            postcode={N6A 3K7}, 
            state={Ontario},
            country={Canada}}

\affiliation[WS]{organization={Institute for Earth and Space Exploration, University of Western Ontario},
            addressline={Perth Drive}, 
            city={London},
            postcode={N6A 5B8}, 
            state={Ontario},
            country={Canada}}
            
\affiliation[NASA]{organization={Space Environments Team and Meteoroid Environment Office, NASA},
            addressline={Marshall Space Flight Center}, 
            city={Huntsville},
            postcode={35812}, 
            state={Alabama},
            country={USA}}
            
\begin{abstract}

\textbf{Importance:} Accurate estimation of meteoroid bulk density is crucial for assessing spacecraft impact hazards of sub-millimeter to millimeter-sized meteoroids, which represent the bulk of the hazard.

\textbf{Research Gap:} Previous studies utilized manual or optimization methods for fitting numerical meteoroid ablation and fragmentation models to optical meteor observations. However, these methods struggled with reliably estimating meteoroid physical properties and the associated uncertainties due to the subjectivity of the modeling approach.

\textbf{Objective:} We aim to develop a global and statistically robust optimization method for inferring the physical properties of individual meteors, focusing on bulk density and fragmentation behavior, using multi-instrument optical data.

\textbf{Methodology:} We apply Dynamic Nested Sampling to fit an erosion-fragmentation model to measurements of meteor light curve and deceleration. The method was applied to 15 shower meteors observed by the Canadian Automated Meteor Observatory’s (CAMO) mirror tracking and Electron-Multiplied Charge Coupled Device (EMCCD) systems. The method yields posterior distributions and Bayesian evidences for single and double fragmentations.

\textbf{Key Findings:} Validation against four synthetic test cases demonstrated accurate recovery of known inputs, with best-guess solutions matching true parameters. We applied this method to 9 Orionids (ORI) and 6 $\alpha$ Capricornids (CAP) ranging in mass from $10^{-6}$ to $10^{-5}$~kg. The median bulk density was measured as $159^{+558}_{-57}$~kg/m$^3$ for Orionid meteors and $333^{+1089}_{-114}$~kg/m$^3$ for Capricornid meteors. These results are consistent with earlier studies: Orionids exhibit characteristics expected for meteoroids of cometary origin, whereas $\alpha$~Capricornids show systematically higher bulk densities.
The CAP results show a second cluster around $1300$~kg/m$^3$, more inline with higher density asteroidal material, but our method achieves this using a more consistent and statistically robust estimation of uncertainties. 

\textbf{Implications:} The developed framework enables automated, statistically rigorous characterization of meteoroid physical properties. The method will be applied to more shower and sporadic meteors to characterize material properties of objects across orbital classes.

\end{abstract}
\begin{graphicalabstract}
\includegraphics[width = \textwidth]{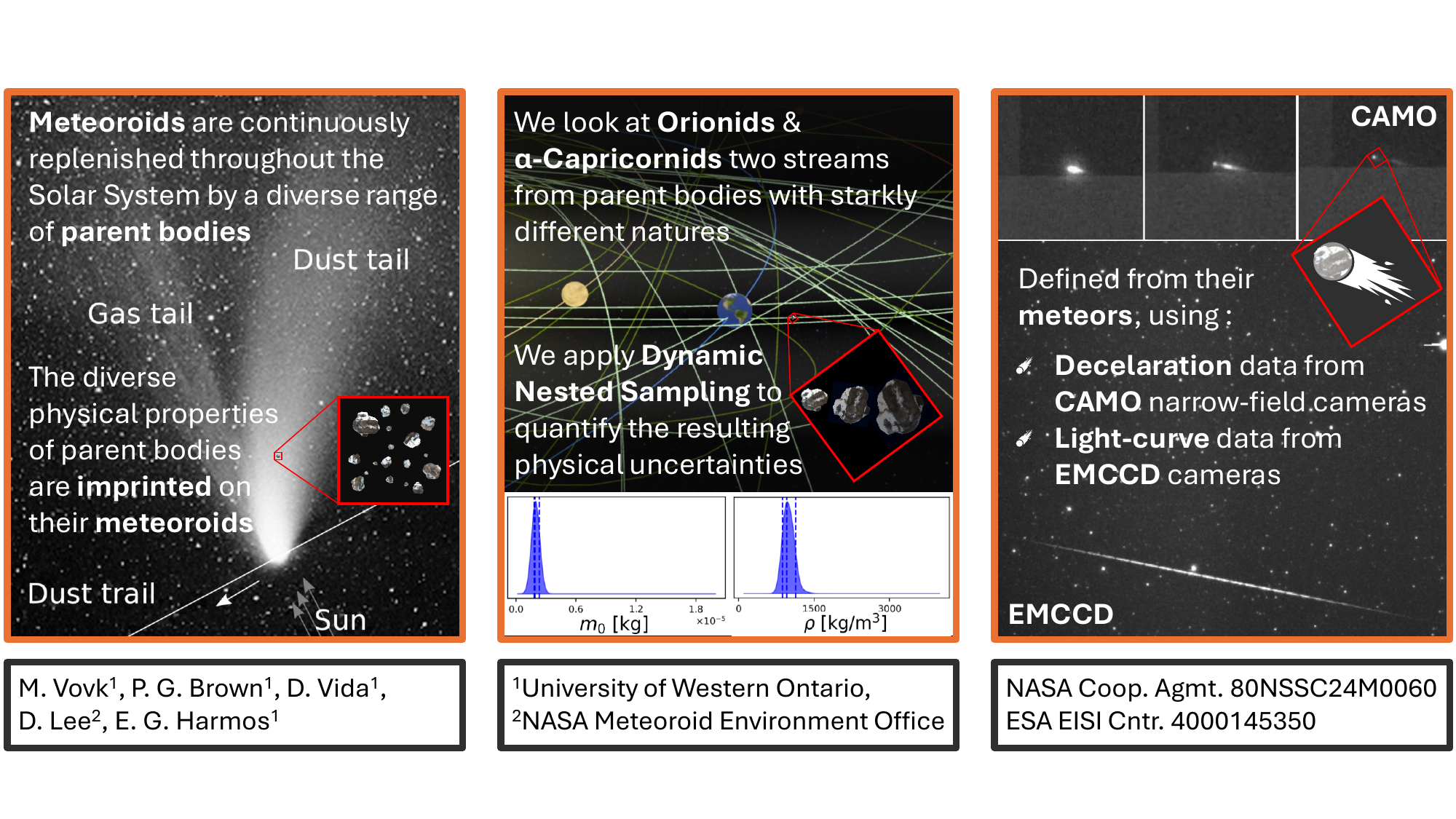}
\end{graphicalabstract}

\begin{highlights}
\item Fully automated recovery of meteoroid properties using Dynamic Nested Sampling.
\item Simultaneous fitting of meteor luminosity and dynamics using EMCCD and CAMO observations.
\item Method validated on four synthetic cases with accurate recovery of known parameters.
\item Posterior distributions of physical parameters fitted for 9 Orionids and 6 Capricornids match previously published values.
\end{highlights}


\end{frontmatter}



\section{Introduction}\label{sec:introduction}




Meteoroid impacts pose a significant hazard to spacecraft, with larger meteoroids ($>0.2$~mm in size) striking operational satellites every few years \citep{moorhead2019meteoroid}. These impacts can cause severe damage by severing wires, compromising spacecraft components, and even puncturing spacesuits \citep{moorhead2019meteoroid}. The extent of damage is governed by the physical properties of the meteoroid---mass, bulk density, and material composition---in addition to  meteoroid velocity \citep{Christiansen2001, moorhead2020nasa}. Sub-mm to mm-sized meteoroids, which account for a substantial fraction of the meteoroid population, are too small to be directly observed in interplanetary space. As a result, their physical characteristics must be inferred indirectly through the meteor phenomenon when entering the Earth's atmosphere \citep{Ceplecha1998e}. 


Meteoroid ablation at cm sizes and larger is dominated by fragmentation. High-resolution imagery of faint meteors reveals that over 90\% exhibit clear evidence of fragmentation \citep{Subasinghe2016}, with the remainder likely fragmenting below the detection threshold. Classical models, such as the dust-ball fragmentation model of \citet{Hawkes1975}, along with more recent variants including the thermal erosion model of \citet{Campbell-Brown_Koschny_2004} and the erosion fragmentation model of \citet{borovivcka2007atmospheric}, incorporate fragmentation explicitly. However, these models rely on forward modeling to fit ablation parameters, making rigorous uncertainty quantification difficult and leaving the uniqueness of solutions unresolved \citep{vida2024first, buccongello2024physical}. Despite past efforts to apply numerical fitting to meteor data \citep[see][for a summary]{popova2019modelling}, very few approaches have attempted to automated this process \citep{Kikwaya_2011, Henych_Borovička_Spurný_2023, vovk_PER_2025}. Automation can improve uncertainty quantification and solution uniqueness. 

In this paper, we introduce the dynamic nested sampling optimization method within the framework of the erosion fragmentation model of \cite{borovivcka2007atmospheric}. Dynamic nested sampling, implemented via the \texttt{dynesty} software package \citep{speagle2020dynesty}, efficiently explores complex, high-dimensional parameter spaces by adaptively reallocating live points and employing dynamic stopping criteria \citep{higson2019dynamic}. 
Dynamic nested sampling has demonstrated superior efficiency in finding model fits and uncertainties in high-dimensional astrophysical problems, including galaxy modeling, cosmological parameter inference, and gravitational wave analysis \citep{abuter2024dynamical, vilardi2025discriminating, morras2025orbital}.

The purpose of the current study is to measure physical properties of two very different meteor showers with different entry velocities and bulk densities, namely the Orionids (ORI) and the $\alpha$ Capricornids (CAP). Physical parameters are inverted via the erosion ablation model for individual meteoroids using observational data and address two longstanding challenges in meteoroid modeling: uncertainty quantification and the inherent non-uniqueness of inverse solutions. Rather than seeking a single best-fit parameter set, we use Bayesian inference to estimate posterior distributions, which reveal parameter constraints and degeneracies. 

We use double-station observations from Electron Multiplying Charge-Coupled Device cameras \citep[EMCCD;][]{vida2020new, gural2022development} and the mirror tracking system of the Canadian Automated Meteor Observatory \citep[CAMO;][]{vida2021high, subasinghe2018luminous, subasinghe2017luminous}. We have fused data from these two distinct systems for common events, as luminosity data from EMCCD cameras provides video observations of meteors down to a limiting magnitude of +7, which is much fainter than the detection limit for the CAMO mirror-system. These high-sensitivity measurements allow detection of meteors at higher altitudes, extending the observed luminous flight path. The process is repeated for measurements at both CAMO sites, yielding meteor trajectories and orbits. Complementing the EMCCD sensitivity is the high precision deceleration data provided by CAMO. The CAMO mirror data provide ultra-precise (spatial resolutions of order 3-4 m) and high-cadence (10 ms exposures) measurements for the last half of meteor trajectories. 

During the model inversion process, we select broad priors allowing for a wide parameter space to be explored. The likelihood function as used within the \texttt{dynesty} package combines two observational constraints: the observed luminosity from EMCCD recordings and the lag from CAMO. Lag is a measure of deceleration and represents the cumulative distance the observed meteor falls behind a hypothetical non-decelerating body. 


We verified our method against synthetic data designed to mimic the temporal cadence, observational noise, and resolution of the combined EMCCD and CAMO systems. These runs demonstrated that the precision of parameter recovery is primarily governed by the number of data points available per meteor and the uncertainty in both lag and luminosity measurements. The results confirm that our approach manages to invert the correct physical parameters and delivers robust physical estimates, even under realistic noise.


We find that our results are consistent with previous work using the same fragmentation model and luminous efficiency parameters \citep[e.g.,][]{buccongello2024physical,vida2024first}. We confirm that our method works in a completely automated manner and is able to broadly reproduce previous forward-modeling solutions, in addition to providing a rigorous statistical uncertainty for each model parameter. Moreover, the results for the two showers produce similar estimates for common parameters between members of these two meteor showers.

This paper is organized as follows: in Section~\ref{sec:methods}, we describe the methodology, including the integration of the erosion-fragmentation model with dynamic nested sampling and the validation on synthetic test cases. Section~\ref{sec:results} presents the physical results obtained for a suite of observed Orionid and $\alpha$ Capricornid meteors, including the posterior distributions and the correlation between physical parameters. In Section~\ref{sec:discussion}, we discuss the physical interpretation of these findings, the agreement and differences with previous work, and the methodological implications of our approach. Finally, in Section~\ref{sec:conclusions}, we summarize the main conclusions and outline future directions, including extending this framework to broader meteoroid populations.


\section{Methods} \label{sec:methods}





In this section, we describe the instruments used for data collection, briefly review the used meteoroid fragmentation and ablation model, and summarize the dynamic nested sampling method as implemented for our purpose. We base our meteoroid ablation model on the \cite{borovivcka2007atmospheric} approach which assumes meteoroids fragment quasi-continously into constituent refractory grains. Our implementation of the model allows for multiple fragmentation events along the meteoroid trajectory, as they are sometimes necessary to explain the observed behaviour. We then fit this model to observations of \emph{luminosity} and \emph{lag} data simultaneously. It also naturally incorporates observational uncertainties into parameter estimation.

\subsection{Equipment, Data Collection and Reduction} \label{subsec:obs}

\begin{figure}[h!]
\centering
\includegraphics[width=0.34\linewidth]{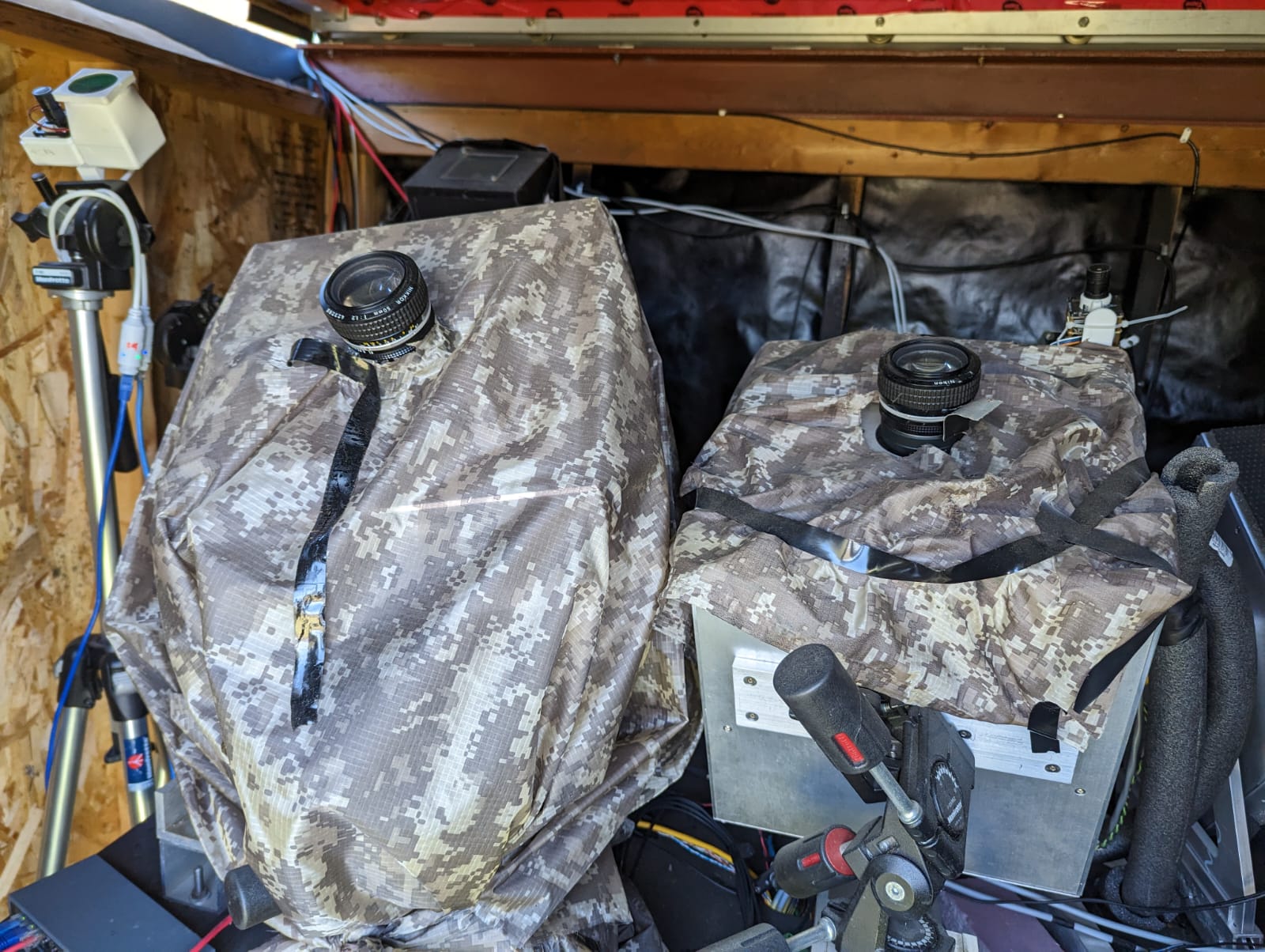}
\includegraphics[width=0.34\linewidth]{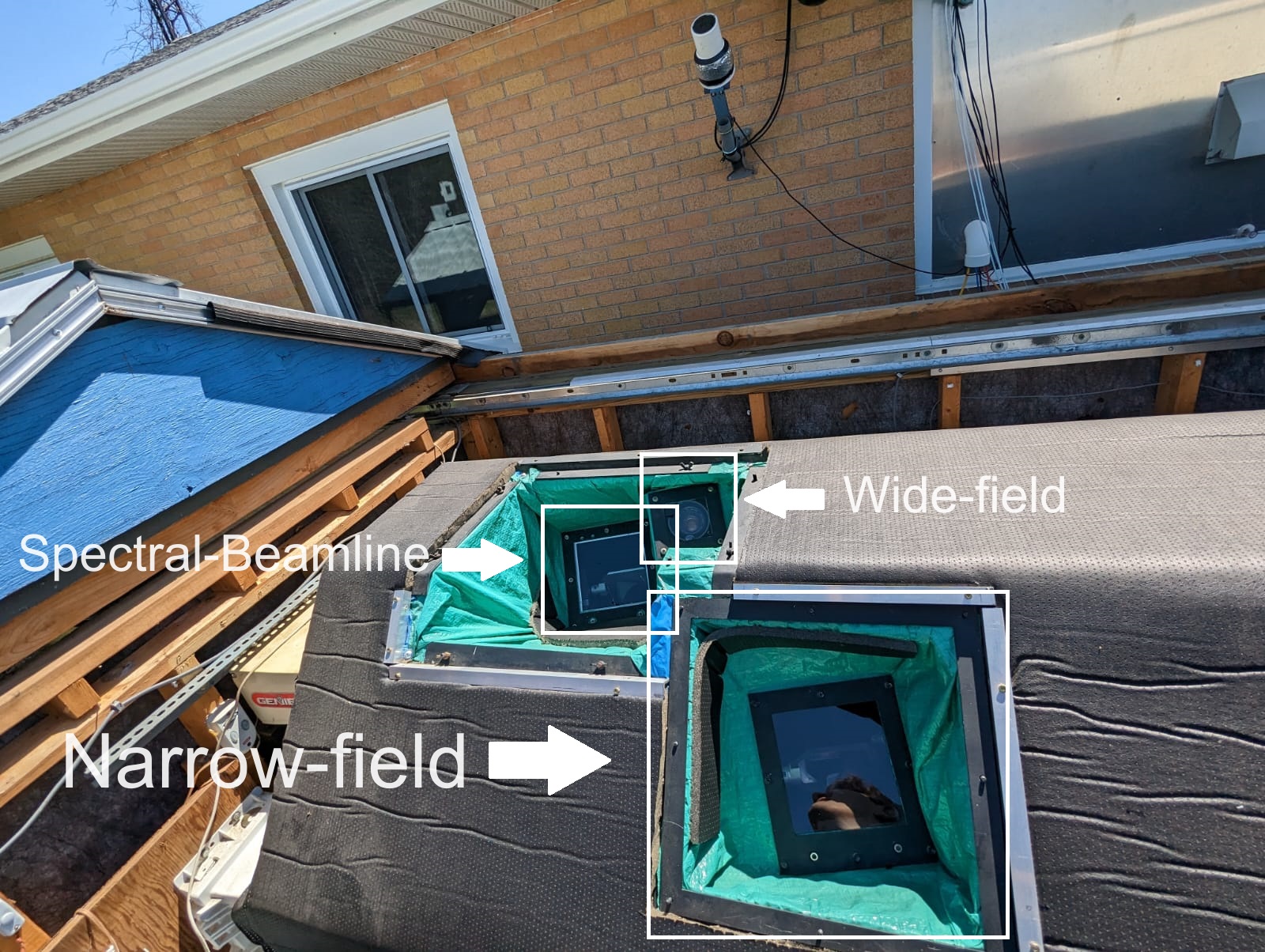}
\includegraphics[width=0.29\linewidth]{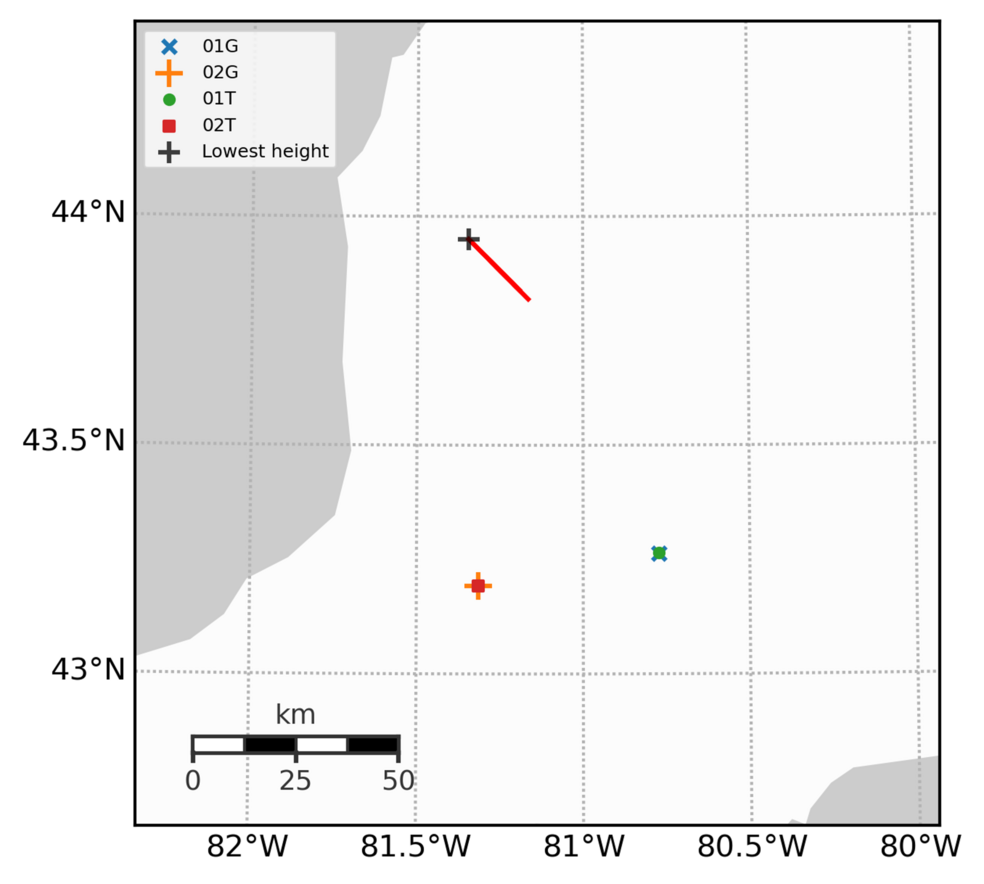}
\caption{On the left is shown two EMCCD cameras (02G, 02F) at Elginfield observatory positioned at 70$^{\circ}$ and 40$^{\circ}$ elevations. On the middle, the window below which the narrow-field CAMO camera (02T) at Elginfield observatory is shown. The image on the right shows the map where the cameras are located and the meteor trail of a detected meteor.}
\label{img:EMCCDcamera}
\label{img:CAMOcamera}
\end{figure}

For this study, we make use of two pairs of Electron Multiplying Charge-Coupled Device (EMCCD) cameras which are part of the dual-station Canadian Automated Meteor Observatory facility. The Western Meteor Physics Group (WMPG) operates four N\"uv\"u HN\"u1024 EMCCD cameras\footnote{https://www.nuvucameras.com/products/hnu-1024/} at two sites (Elginfield 43.19279$^{\circ}$ N, 81.31565$^{\circ}$ W; and Tavistock 43.26420$^{\circ}$ N, 80.77209$^{\circ}$ W), both located in Southwestern Ontario and separated by $\sim45$~km. The EMCCDs are binned $2\times2$ resulting in a video resolution of $512 \times 512$~px. They are operated at 32 frames per second and are capable of detecting faint meteors with a limiting magnitude of +7 and a per-point detection limit typically of +8 \citep{vida2020new}. As shown in Figure \ref{img:EMCCDcamera}, the camera pair labeled "F" is positioned at a $70^{\circ}$ elevation to capture meteors above 90 km, while the "G" set, angled at $40^{\circ}$, targets meteors between 70 km and 120 km in altitude. 

All meteors used in this study were also observed with the Canadian Automated Meteor Observatory's mirror tracking system (CAMO). The CAMO system consists of two high frame rate CCD cameras (Prosilica GX1050 digital) coupled with an 18 mm diameter ITT FS9910 series Generation 3 image intensifier \citep{vida2021high}. One serves as the wide-field camera with a 30$^{\circ}$ field of view, capable of detecting meteors with magnitudes up to +4.5 as a cuing system. Once a meteor is detected by the wide-field camera, a pair of mirrors are positioned to track the meteor in real-time across the sky with response time of order 50 ms. CAMO provides ultra-precise spatial resolution (~3–4 m) and high temporal resolution (10 ms exposures) for the latter half of meteor trajectories \citep{vida2021high}. As shown in Figure \ref{img:CAMOcamera}, the CAMO system's unique design makes it the only operational telescopic tracking system for meteor observations \citep{borovivcka2019physical}. The camera details for each of the two systems are summarized in Table \ref{tab:cameras}.

\begin{table}[h!] \label{tab:cameras}
\caption{Comparison of EMCCD and CAMO mirror tracking cameras. The specifications for the CAMO mirror tracking camera are from \cite{vida2021high} and for the EMCCD from \cite{gural2022development}.}
\centering
\resizebox{\textwidth}{!}{
\begin{tabular}{lp{6cm}p{6cm}}
\hline
\textbf{Characteristic} & \textbf{EMCCD Camera} & \textbf{CAMO narrow-field} \\
\hline
\textbf{Camera Model} & HN\"u1024 by N\"uv\"u Cameras & Prosilica GX1050 digital progressive scan CCD \\
\textbf{Sensor Resolution} & 512 x 512 pixels (2x2 binned) & 1024 x 1024 pixels \\ 
\textbf{Frame Rate} & 32 FPS & 80 FPS \\ 
\textbf{Field of View (FOV)} & 14.7$^{\circ}$ $\times$ 14.7$^{\circ}$ & 1.5$^{\circ}$ $\times$ 1.5$^{\circ}$ \\ 
\textbf{Angular Resolution} & 1.72 arcminutes per pixel & 6 arcseconds per pixel \\ 
\textbf{Lens} & 50 mm f/1.2 Nikkor & 80 mm aperture APO Orion refractor with 545 mm focal length \\ 
\textbf{Sensitivity} &  +8 & +7 \\ 
\hline
\end{tabular}
}
\end{table}

Operationally, the EMCCD system record data in rolling video buffers every night when sky conditions are good (dark, no clouds, no moon). A hybrid threshold/matched filter meteor detection algorithm \citet{gural2022development} is applied to the data to produce automated detections and measurements. An initial trajectory solution is produced based on these measurements using the \citet{Vida2020theory} meteor trajectory solver.

For our case study we selected meteors that were both detected by EMCCD cameras and the CAMO tracking system. In addition, we removed all meteors that did not meet strict geometric constraints, ensuring a convergence angle greater than 10$^{\circ}$ between the two stations and perspective angles exceeding 20$^{\circ}$ for each individual station. Under these conditions, the radiants of all ORI and CAP meteors clustered in the same region and were consistent with those reported by GMN over the years, as shown in ~\ref{sec:Apx radiants}.
Finally, we selected 15 meteors from two distinct showers with known physical characteristics: the Orionids (ORI) and the $\alpha$ Capricornids (CAP). For the ORI sample, we analyzed 9 meteors observed with both EMCCD and CAMO, while for the CAP sample, we analyzed 6 meteors. All events were detected between 2019 and 2022.

All selected meteors were manually reduced using the SkyFit2 software which is part of the Raspberry Pi Meteor Station (RMS) library\footnote{https://github.com/CroatianMeteorNetwork/RMS}. Manual reduction consisted of estimating the leading edge position of the meteor for each observed frame and then manually selecting the visible meteor region for photometric measurement. The astrometric/photometric calibration process is described in \citet{vida2021high}.

For this study, we relied solely on EMCCD data derive meteor light curves and used CAMO narrow-field observations for velocity and deceleration measurements.

\subsubsection{Synthetic Data Generation and Considerations Regarding the Measurement Reference Point}


During the development of the simulations, we noticed an effect that impacts the observed heights of meteors. Due to the relatively small fields of view of EMCCDs, their comparatively long exposure time, and fast meteor angular speeds, all meteors appear as streaks on each video frame. The measurements of meteor light curve are done by integrating all the light the meteor producing on each frame, and assigning the leading edge as the reference point.

\begin{figure}[ht]
    \centering
    \includegraphics[width=0.9\linewidth]{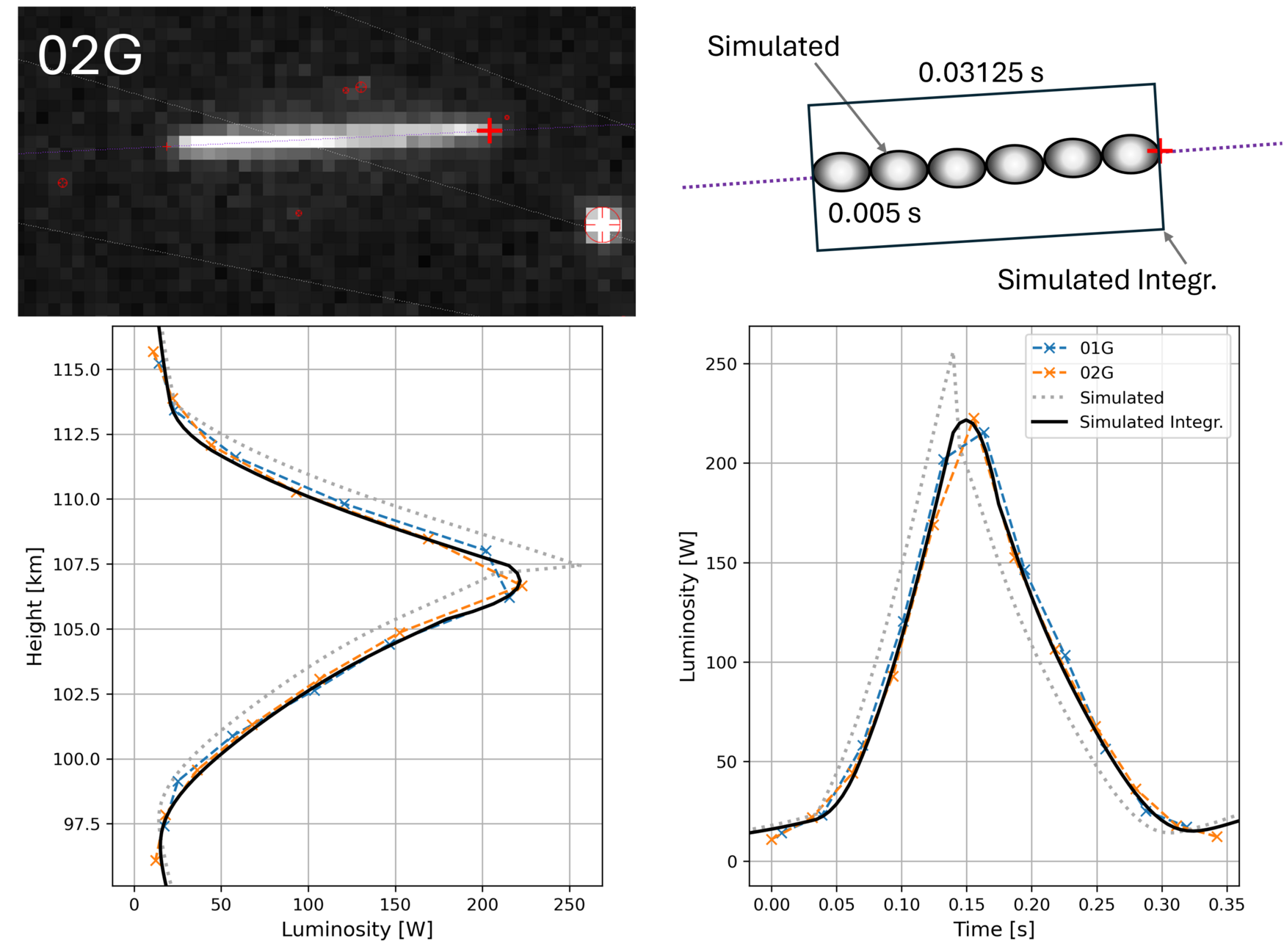}
    \caption{Smearing effect for an EMCCD recorded Orionid meteor. The dashed gray line shows the raw simulation, while the solid black line includes temporal integration over 0.03125 seconds. The top panels display real EMCCD data (left) detected by camera 02G with the chosen leading edge pick point as a red cross and the simulated integrated trail (right). Integration shifts the peak and smooths the light curve.}
    \label{img:example_ORI_integ}
\end{figure}

Without careful consideration, one may want to use these measurements sampled at $\sim31$~ms and compare them directly to simulations that are sampled at much higher rates of around 5~ms. Due to long exposure times and fast meteor speeds, EMCCD images show meteors as streaks. This introduces a height error of up to 1 km if not corrected. Furthermore, due to the higher fidelity of the simulation, any sharp changes in the light curve are more pronounced and may bias the data if they are not averaged to the same cadence as the camera sampling.

To account for these effects, we bin the simulated data and compute per-point luminosity over the camera’s exposure duration and use the last simulated height within the bin as reference. The integration shifts the apparent position of the meteor slightly downward in height (due to the trailing effect of the leading edge) and delays the peak brightness in time. This produces a smoothed light curve that more accurately reflects observational data.

\begin{figure}[ht]
    \centering
    \includegraphics[width=0.9\linewidth]{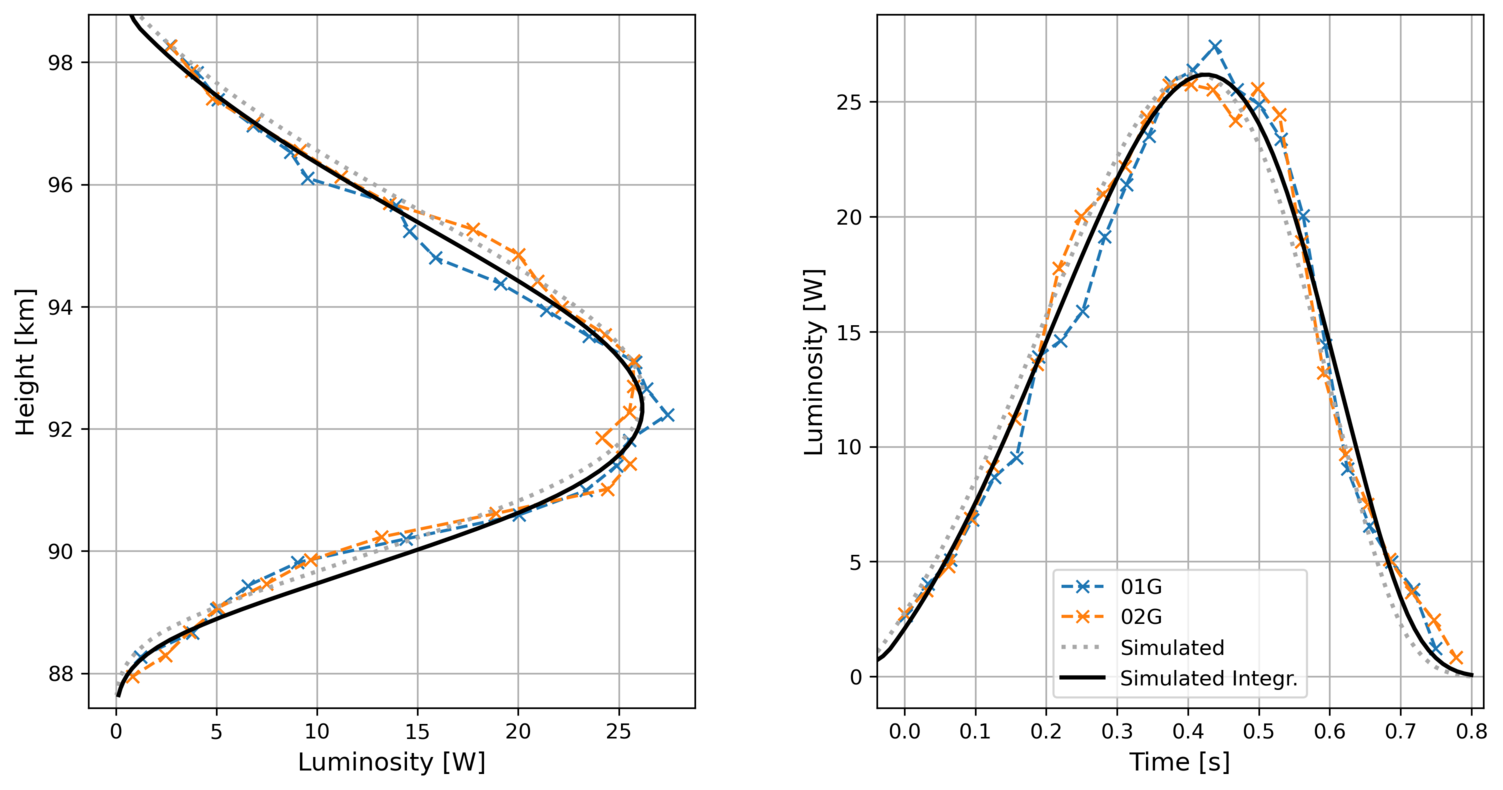}
    \caption{Effect of temporal integration on a CAP simulation. The integrated and raw simulations are nearly identical, reflecting the lower velocity and reduced smearing of the CAP meteor.}
    \label{img:example_CAP_integ}
\end{figure}

Figure~\ref{img:example_ORI_integ} illustrates this effect for a fast ORI meteor, where the difference between the raw simulation and the time-integrated curve is substantial due to the high speed. Conversely, as shown in Figure~\ref{img:example_CAP_integ}, the impact on slower CAP meteors is minimal, demonstrating that this correction becomes increasingly important for faster events.

\subsection{Application of Dynamic Nested Sampling to Automate Erosion Model Fitting}

Having adopted an ablation model to fit our photometric and metric data (described in detail in Section \ref{erosion_param}), we next need to pair it with an optimization method to invert the physical properties of meteoroids consistent with measurements. 

Dynamic nested sampling (DNS) is an extension of the original static nested sampling algorithm, which was designed to compute the Bayesian evidence \citep{higson2019dynamic}. Bayesian evidence quantifies how well a model $M$ with parameters $\mathbf{\theta}$ explains the data $D$.
\begin{equation}
\label{eq:bayes_evidence} 
\mathcal{Z} \;=\; P(D \mid M) \;=\; \int P(D \mid \mathbf{\theta}, M)\,P(\mathbf{\theta}\mid M)\,d\mathbf{\theta}.
\end{equation}
\noindent The evidence $\mathcal{Z}$ is the probability that given the measurements D the adopted model M explains the data. 

The power of this method comes from its ability to adaptively allocate computational resources to the most critical regions of the parameter space, such as those with high likelihood, strong degeneracies, or multimodal structures, thereby refining the evidence estimation and yielding robust posterior distributions. We use the implementation of the algorithm available in the \texttt{dynesty} Python package\footnote{https://dynesty.readthedocs.io/en/stable/index.html}.


We tested different sampling methods within the nested sampling \texttt{dynesty} software package, including random walks and random slice sampling, along with various bounding strategies and different numbers of live points \citep{speagle2020dynesty}. Random slice sampling provided better accuracy at the cost of almost double the computation time of random walks, which suffered from local minima. Varying the bounding strategies did not have a significant impact, and increasing the number of live points from 500 to 1000 drastically increased the computation time without notably improving the results. Thus, we chose random slice sampling (\texttt{`rslice'}), multiple bounding ellipsoids (\texttt{`multi'}), and 500 live points. 

\subsubsection{Likelihood Model}

We assume Gaussian errors and compute the total likelihood by summing contributions from both light curve and lag data. To illustrate how we compute the likelihood for a single meteor (e.g., luminosity~vs.~height), suppose we have $N$ observations $\{x_i\}$ with corresponding model predictions $\{\mu_i(\mathbf{\theta})\}$ using physical parameters $\mathbf{\theta}$ within the erosion model. If the measurement errors are independent and distributed following a Gaussian with known variance $\sigma^2$, the likelihood for each data point is also a Gaussian distribution

\begin{equation}
\label{eq:single_likelihood}
P\bigl(x_i \mid \mu_i(\mathbf{\theta}), \sigma\bigr) 
\;=\;
\frac{1}{\sqrt{2\pi}\,\sigma} \exp\!\Bigl[-\tfrac{(x_i - \mu_i(\mathbf{\theta}))^2}{2\,\sigma^2}\Bigr].
\end{equation}

\noindent Working in log-space, the log-likelihood for each data point is then
\begin{equation}
\label{eq:single_loglike}
\log P\bigl(x_i \mid \mu_i(\mathbf{\theta}), \sigma\bigr)
\;=\;
-\tfrac{1}{2}\,\log\bigl(2\pi\sigma^2\bigr)
\;-\;\tfrac{\bigl(x_i - \mu_i(\mathbf{\theta})\bigr)^2}{2\,\sigma^2}.
\end{equation}

\noindent For $N$ data points, the total log-likelihood becomes
\begin{equation}
\label{eq:loglike_sum}
\log \mathcal{L}_{\text{(all data)}}(\mathbf{\theta})
\;=\; \sum_{i=1}^N \log P\bigl(x_i \mid \mu_i(\mathbf{\theta}), \sigma\bigr).
\end{equation}

We adopt meteor luminosity (in Watts) as the constraint instead of the most commonly used magnitudes. Luminosity is a linear unit which is expected by the algorithm, while the magnitudes are logarithmic. In practice, we found that luminosity cast in linear rather than log form produces better results. The dynamics are represented by the lag which is defined as the cumulative distance a meteor falls behind a hypothetical meteor moving at a constant speed. The lag between the observations and the model is normalized to the observed initial speed, making the algorithm able to invert the speed without considering the traditional time vs. distance measurements. 

We estimate the measurement uncertainty for each input ($\sigma^2$) and use it to normalize them to be equal and allow them to be combined into a single objective function by simple summation
\begin{equation}
\label{eq:combined_ll}
\log \mathcal{L}(\mathbf{\theta})
\;=\; \log \mathcal{L}_{\text{lum}}(\mathbf{\theta})
\;+\; \log \mathcal{L}_{\text{lag}}(\mathbf{\theta}).
\end{equation}
\noindent This sum appropriately captures all information from both datasets under the assumption that their measurement errors are independent. 

The innovation of the dynamical nested sampling algorithm is that the quality of the inversion does not depend on the fixed value of the uncertainty estimate. An approximate measurement uncertainty is needed for both the luminosity and the lag to initialize the algorithm, however, the uncertainties are included as parameters in the optimization process and they are further refined during the process to best match the data.

\subsection{Estimating Luminosity Uncertainty}

\begin{figure}[ht]
    \centering
    \includegraphics[width=0.8\linewidth]{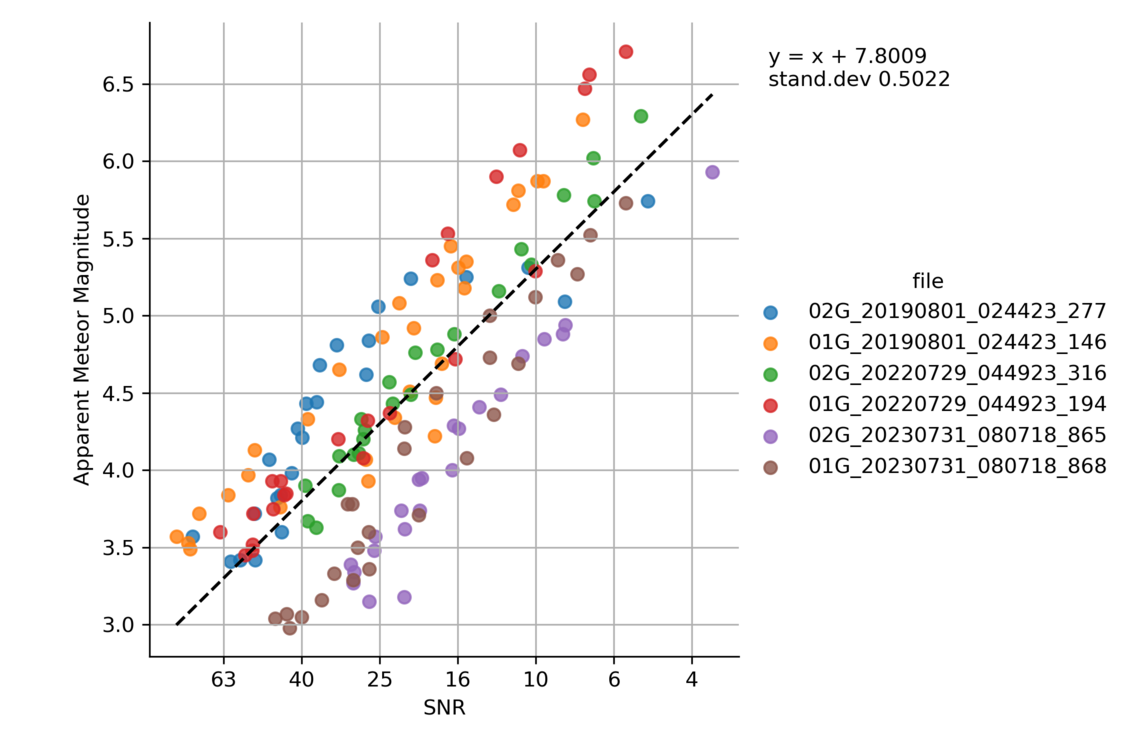}
    \caption{Scatter plot of apparent meteor magnitude vs.~SNR for three meteors, each observed by two cameras (six datasets total). Note that the X axis is in log scale. Each color indicates a separate meteor-camera pair - the designations indicate observations from different sites (01 - Tavistock; 02 - Elginfield), the camera pair (F - high altitude pointing, G lower altitude pointing cameras) followed by the date and time of the recorded event. The dashed black line (slope $m=1$) is the best-fit relation for CAP data, used to assign an initial noise estimate for the first detection frames. The systematic scatter above and below the ideal line reflects variations in real-world observing conditions.}
    \label{img:CAPsnr_vs_mag}
\end{figure}

We estimate the initial value of the luminosity uncertainty $\sigma_{\text{lum}}$ by leveraging the signal-to-noise ratio (SNR) measurements derived from EMCCD observations. To establish empirical estimates, we analyzed three meteors from two meteor showers across a range of speeds (Capricornids and Orionids), observed by two EMCCDs. The resulting measurements for three Capricornids are shown in Figure \ref{img:CAPsnr_vs_mag}. They include the apparent meteor magnitudes $m$ at each frame and the SNR values obtained by comparing the meteor’s flux against the background noise following the ``CCD equation'' \citep{howell1989two}. These data show a fairly large scatter and systematic offsets between meteors, however, the magnitudes follow a linear trend against log(SNR). Most of the offsets are driven by the apparent speed of the meteor and by the observing conditions (as shown by the Photometric offset curves in \ref{sec:Apx photometry}). Our aim is not to have an accurate noise model of the cameras, but to obtain a realistic average value of the noise to kickstart the optimization process. Thus, we fit a linear relationship between $m$ and $-2.5\,\log_{10}(\mathrm{SNR})$, forcing a slope of unity which comes from the standard definition of astronomical magnitude. This fit captures the mapping from the simulated apparent magnitudes in our instrumental bandpass to the SNR, which can be used to compute the luminosity error in the simulation.

This process entails selecting the luminosity $L_{\min}$ of faintest observable point on the meteor produced in the simulation (limiting magnitude is set by the observation), computing the SNR following the relationship derived above, and using the following equation to define the luminosity noise floor
\begin{equation} 
  \label{eq:sigma_from_snr}
  \sigma_{\text{lum}} 
  \;=\; \frac{L_{\min}}{\mathrm{SNR}} \,. 
\end{equation}
\noindent This value serves only as an \emph{initial} noise estimate, since \texttt{dynesty} adaptively refines $\sigma_{\text{lum}}$ during the optimization process.

\subsection{Estimating Lag Uncertainty}

To estimate the initial measurement noise $\sigma_{\text{lag}}$ in the meteoroid's dynamics, We fit a cubic polynomial to the lag data, constrained to ensure deceleration, and use the RMS of residuals as the initial lag uncertainty: 

\begin{equation}
\ell(t) = 
    \begin{cases}
        0 & \text{if } t < t_{0} \\
        -|a| (t - t_{0})^3 - |b| (t - t_{0})^2 & \text{if } t \geq t_{0} \,.
    \end{cases}
\end{equation}
    
\noindent where $t_0$ is the onset time of deceleration, and $a$ and $b$ are polynomial coefficients constrained to be negative to force the deceleration assumption. By definition, $\ell(t)$ remains zero (no deceleration) until $t = t_0$, then turns into a cubic polynomial without producing unphysical accelerations \citep[see also][]{Vida2020theory}. We solve for $(t_0, a, b)$ by minimizing the residuals of function:
  \[
    \Delta \ell_i \;=\; \ell_{\mathrm{obs}}(t_i)\;-\;\ell(t_i),
  \]
where $\ell_{\mathrm{obs}}(t_i)$ is the lag measured at time $t_i$. The root mean square difference (RMSD) of these residuals directly gives our estimate for the positional (lag) noise.

\begin{figure}[ht]
    \centering
    \includegraphics[width=0.65\linewidth]{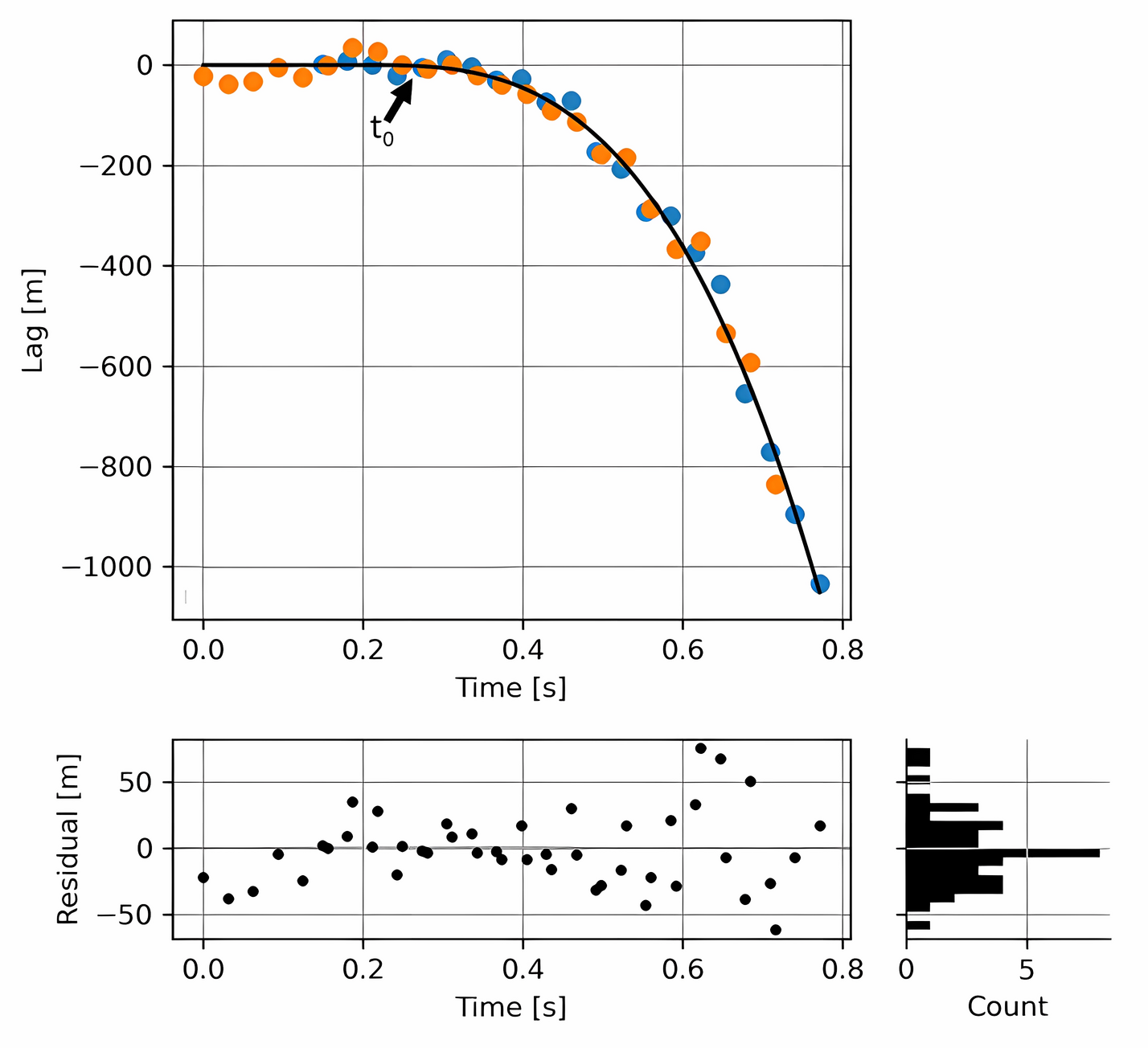}
    \caption{Top: The lag of a Capricornid meteor (points) and the dervied fit (black) with annotated time t$_0$. Each color indicates measurements from a separate camera. Bottom: Fit residuals and a histogram of residuals.}
    \label{img:example_stdv}
\end{figure}

Figure~\ref{img:example_stdv} is an example of a typical fit for an observed CAP meteor, showing that the residuals are approximately Gaussian in distribution.

Upon receiving the initial lag noise estimate, the \texttt{dynesty} algorithm simultaneously explores the physical parameters of the fragmentation model and refines the noise. As part of the log-likelihood calculation $P\bigl(x_i \mid \mu_i(\mathbf{\theta}), \sigma\bigr)$, the code effectively \emph{weights} each dataset according to its noise level, ensuring that more precise measurements have a stronger influence on the parameter inference. Contrary to the adoption of fixed noise, this strategy allows the code to adaptively tune the noise parameters for each meteor, finding the most appropriate weighting between the luminosity and lag datasets in the likelihood function. The noise estimate thus becomes an additional degree of freedom in the fit.




\subsection{Erosion-Fragmentation Model and Assumed Priors}\label{erosion_param}

We adopt the formulation of the meteoroid erosion-fragmentation model by \cite{borovivcka2007atmospheric} as the basis of our work. The model assumes that at a set altitude, a meteoroid begins to shed constituent $\mu$m-sized particles from its surface. This process is called erosion. The grain masses follow a power-law distribution and ablate independently, $dN/dm \propto m^{-s}$, where $s$ is the grain mass index controlling the relative abundance of small versus large grains, and the grains ablate independently while the main body also continues to ablate. Our implementation of the model allows for changing the intensity of erosion along the trajectory. For the purposes of this paper, the number of changes to the erosion has been limited to two discrete heights. The model captures both gradual erosion and sudden fragmentation, making it suitable for a range of meteoroid behaviors \citep{borovivcka2007atmospheric}.

For simplicity, we first introduce the model assuming only a single erosion. The parameters we fit within the framework of the erosion model are as follows,
\[
  \boldsymbol{\theta}_\text{single} \;=\; 
  \bigl(
  v_{0},\, m_{0},\, \rho,\,
  \sigma,\,
  h_{e},\,
  \eta,\,
  s,\,
  m_{l},\,
  m_{u},\,
  \sigma_{lag},\,
  \sigma_{lum}
  \bigr),
\]
\noindent where $v_0$ is the initial velocity, $m_0$ the initial meteoroid mass, $\rho$ is meteoroid bulk density and $\sigma$ is the ablation coefficient. We also fit the erosion coefficient $\eta$, the lower and upper erosion mass grain limits $m_l$ and $m_u$ and the power-law mass distribution index $s$. Finally, the model computes the start height of erosion $h_e$ and the noise parameters $\sigma_{lag}$ and $\sigma_{lum}$.

The choice of priors for each parameter is motivated by a combination of physical constraints and data-informed assumptions. We drew heavily on the previous works of \citet{borovivcka2007atmospheric, vojavcek2017properties, vida2024first, buccongello2024physical}, which applied the model to a suite of meteors from various meteor showers and sporadic sources and assumed similar priors.

A Gaussian prior is applied to the initial velocity $v_0$, computed in the model at the height of detection, with a fixed standard deviation of 500~m/s which is well above our typical trajectory precision. Since the simulations begin at an altitude of 180 km, while the first observed velocity is measured at a lower altitude, we apply a correction of +100 m/s to the velocity in the first frame. This adjustment compensates for the minimal but non-negligible deceleration that occurs before the initial observation point. The prior for the initial mass $m_0$ is uniform within bounds defined by an order-of-magnitude estimate of the photometric mass, $\mathcal{O}(m_{\text{photometric}})$, employing the luminous efficiency model of \citet{vida2024first} based on the area under the lightcurve. The order of magnitude mass uncertainty estimate is completely driven by the uncertainties in the luminous efficiency \citep{subasinghe2018luminous}.

Log-uniform priors are used for the bulk density $\rho$, erosion coefficient $\eta$, and the erosion mass limits $m_l$ and $m_u$, enabling efficient sampling across several orders of magnitude. For the noise parameters $\sigma_{lag}$ and $\sigma_{lum}$, we adopt inverse gamma distributions. The gamma distribution priors provide tight concentration near the expected fit residuals, while their heavy-tailed nature ensures sufficient flexibility to explore larger uncertainty values when warranted by the data \citep{hoff2009first}. Table~\ref{tab:priors_1frg} summarizes the adopted priors.

\begin{table}[h!]
\centering
\caption{Adopted prior distributions used as inputs for the erosion model where only a single-fragmentation is permitted within the \texttt{dynesty} software package. For many of these parameters, the range of values was adopted in part from the findings of the earlier study of \citet{vojavcek2017properties}.}
\resizebox{\textwidth}{!}{
\begin{tabular}{lllp{7cm}}
\hline
\textbf{Variable} & \textbf{Name} & \textbf{Distribution} & \textbf{Range or Formula} \\
\hline
$v_{0}$ [km/s] & Initial velocity at 180 km altitude & Gaussian & Centered at 100 m/s over first-frame speed, $\sigma = 500$ m/s \\
$m_{0}$ [kg] & Initial mass & Uniform & $0.1\ \mathcal{O}(m_{\text{photometric}})$ -- $20\ \mathcal{O}(m_{\text{photometric}})$ \\
$\rho$ [kg/m$^{3}$] & Bulk density & Log-uniform & 100 -- 4000 \\
$\sigma$ [kg/MJ] & Ablation coefficient & Uniform & 0.001 -- 0.05 \\
$h_{e}$ [km] & Erosion start height & Uniform & ($\frac{1}{2}h_{\text{beg}} + \frac{1}{2}h_{\text{peak}}$) -- ($\frac{3}{2}h_{\text{beg}} - \frac{1}{2}h_{\text{peak}}$) \\
$\eta$ [kg/MJ] & Erosion coefficient & Log-uniform & 0 -- 1 \\
$s$ & Mass distribution index & Uniform & 1 -- 3 \\
$m_{l}$ [kg] & Lower erosion mass limit & Log-uniform & $5\times10^{-12}$ -- $1\times10^{-9}$ \\
$m_{u}$ [kg] & Upper erosion mass limit & Log-uniform & $1\times10^{-10}$ -- $1\times10^{-7}$ \\ 
$\sigma_{lag}$ [m] & Lag noise standard deviation & Inverse Gamma & Mode from lag fit, $\alpha = 10$ \\
$\sigma_{lum}$ [W] & Luminosity noise standard deviation & Inverse Gamma & Mode from lum fit, $\alpha = 5$ \\
\hline
\end{tabular}}
\label{tab:priors_1frg}
\end{table}

Together, these priors impose realistic and physically motivated bounds, enabling robust exploration of the parameter space. In addition to the fitted parameters, several quantities are fixed based on standard assumptions in meteoroid physics. The one trajectory parameter that is kept fixed is the meteoroid zenith angle $z_c$, which is normally measured to high accuracy and has a low impact on the results. Second, following previous work, we fix the grain density $\rho_g$ to 3000 kg/m$^3$, which we assume matches chondritic grain densities, and enforce $\rho \le \rho_g$ by setting $\rho=\rho_g$ whenever $\rho$ is sampled above $\rho_g$, corresponding to a non-porous meteoroid. Furthermore, we adopt a drag coefficient $\Gamma$ of unity as appropriate to free molecular flow \citet{popova2019modelling}, and a fixed shape factor, A of 1.21 (appropriate for a sphere), as summarized in Table~\ref{tab:fixed_params}.

\begin{table}[h!]
\centering
\caption{Fixed parameters used in the erosion-fragmentation model.}
\begin{tabular}{lll}
\hline
\textbf{Variable} & \textbf{Name} & \textbf{Value} \\
\hline
$z_c$ & Zenith angle & Fixed based on the trajectory solution \\
$\rho_g$ [kg/m$^3$] & Grain density & 3000 \\
$\Gamma$ & Drag coefficient & 1 \\
$A$ & Shape factor & 1.21 \\
\hline
\end{tabular}
\label{tab:fixed_params}
\end{table}

For cases where we permit a second fragmentation, we extend the same set of parameters by introducing a second fragmentation event at a different altitude $h_{e2}$. This adds four additional parameters $h_{e2}, \rho_2, \sigma_2, \eta_2$ to the overall parameter space.
The added constraints are that the second erosion height $h_{e2}$ is lower than the first ($h_{e2} < h_{e}$) and that the bulk density of the second fragment is higher than the first, $\rho_{2} > \rho$. Any parameters shared with the single fragmentation model have the same priors shown in Table \ref{tab:priors_1frg}, while the priors for the four new parameters are shown in Table \ref{tab:priors_2frg}. 

\begin{table}[h!]
\centering
\caption{Prior distributions for the parameters unique to the second fragmentation event. The second erosion height uses the complete range of detected heights of the meteor, from $h_{\text{beg}}$ to $h_{\text{end}}$.}
\resizebox{\textwidth}{!}{
\begin{tabular}{llll}
\hline
\textbf{Variable} & \textbf{Name} & \textbf{Distribution} & \textbf{Range or Formula} \\
\hline
$h_{e2}$ [km] & Second erosion start height & Uniform & $h_{\text{end}}$ -- $\min\left(h_e,\,h_{beg}\right)$ \\
$\rho_{2}$ [kg/m$^{3}$] & Bulk density of the second fragment & Log-uniform & $\rho$ -- 4000 \\
$\sigma_{2}$ [kg/MJ] & Ablation coefficient of the second fragment & Uniform & 0.001 -- 0.05 \\
$\eta_{2}$ [kg/MJ] & Erosion coefficient of the second fragment & Log-uniform & 0 -- 1 \\
\hline
\end{tabular}}
\label{tab:priors_2frg}
\end{table}

\subsubsection{The Importance of Bayesian Evidence for Model Selection} \label{subsec:evidence importance}

To assess the utility of Bayesian evidence in determining the optimal number of fragmentation events, we applied DNS to synthetic and real meteor data. Simpler models with a single onset of erosion are favored unless additional complexity provides a statistically significant improvement in fit quality. In two illustrative cases, we show that higher evidence corresponds to physically supported and visually confirmed fragmentation behavior, while models with excessive complexity lead to unconstrained parameters and reduced evidence. A full description of the synthetic example, Bayesian evidence values, and corresponding model fits is provided in \ref{sec:Apx bayes-evidence}.

\subsection{Validation Test Cases}\label{validation-test}



For each meteor considered shower (Orionids and Capricornids), we generated two synthetic meteors based on our estimated physical parameters: one using the mean of the parameters and the other using the mode. The specific values for the four cases are listed in Table \ref{tab:test_parameters}.

\begin{table}[h!]
\centering
\caption{The four synthetic meteors generated based on the mode and the mean inverted physical parameters of all manually reduced Orionids and Capricornids.}
\resizebox{\textwidth}{!}{
\begin{tabular}{lrrrr}
\hline
\textbf{Parameter} & \textbf{Orionid Mode} & \textbf{Orionid Mean} & \textbf{Capricornid Mode} & \textbf{Capricornid Mean} \\
\hline
$v_{\rm 0}$ [km/s]    & 67.72      & 67.38      & 25.12  & 25.35  \\
$z_{\rm c}$ [deg]        & 33.27        & 28.13        & 54.06    & 49.83    \\
$m_{\rm 0}$ [kg]              & 3.78$\times10^{-6}$ & 1.69$\times10^{-6}$ & 1.03$\times10^{-5}$ & 3.67$\times10^{-6}$ \\
$\rho$ [kg/m$^{3}$]            & 195     & 134     & 875       & 1125     \\
$\sigma$ [kg/MJ]        & 2.03$\times10^{-2}$ & 2.17$\times10^{-2}$ & 1.84$\times10^{-2}$ & 2.15$\times10^{-2}$ \\
$h_{e}$ [km]         & 116.19     & 115.74     & 98.49  & 97.52  \\
$\eta$ [kg/MJ]  & 0.22        & 0.17        & 0.24    & 0.13    \\
$s$                       & 2.13      & 2.14      & 2.09  & 2.05  \\
$m_{l}$ [kg]           & 1.39$\times10^{-11}$ & 1.33$\times10^{-11}$ & 7.44$\times10^{-11}$ & 3.04$\times10^{-11}$ \\
$m_{u}$ [kg]              & 1.54$\times10^{-8}$  & 5.08$\times10^{-9}$  & 1.82$\times10^{-8}$  & 5.90$\times10^{-9}$  \\
\hline
\end{tabular}
}
\label{tab:test_parameters}
\end{table}

Each synthetic meteor was modelled using the integration time step of 0.005\,s for the Orionids and 0.01\,s for the Capricornids. To make the computation traceable, the longer time step was necessary due to the lower speed of the Capricornids. Tens of thousands of \texttt{dynesty} test runs showed that some parameter combinations caused numerical instability at shorter time steps. All other parameters used in the forward modeling were fixed to the values listed in Table~\ref{tab:fixed_params}.

Next, we sampled the model-generated luminosity and magnitude data at 32 FPS, mimicking the EMCCD cameras, and the lag and velocity data at 80 FPS, corresponding to the CAMO frame rate, to produce the synthetic measurements. To simulate CAMO observations, we introduced a delay to the synthetic lag and velocity profiles to mimic the tracking lag caused by the mirror system \citep{vida2021high}. This latency is caused by the requirement that meteor detection occurs in at least 6 frames of the wide field camera before a tracking solution is created and mirrors moved to center on the meteor. For each event, CAMO data were generated starting at a random point between 30\% and 60\% of the total EMCCD observation time. This reproduces the typical delay in CAMO's acquisition of early meteor trajectory data. 

The synthetic meteors are assumed to have been captured at a second station to emulate the two station EMCCD/CAMO measurements. The simulation also includes a random phase shift in time sampling of frames, as the cameras at the two sites are not phase-locked. Based on real observations, measurement noise corresponding to the average uncertainties of the CAMO and EMCCD data were added to the synthetic data. For the Orionid CAMO data, the average lag error was $\sigma_{\rm lag}$ = 7.5 m, while the average luminosity error for the EMCCD data was $\sigma_{\rm lum}$ = 2.5 W. Similarly, for the Capricornids, the CAMO lag error was 6.1 m, and the EMCCD luminosity error was 1.7 W. Next, all data points where the magnitude was fainter than +8 were removed, as that was imposed as the instrumental limiting magnitude. Finally, the we used \texttt{dynesty} to recover the physical parameters from the synthetic data and compared the results to the inputs.



\begin{figure}[ht]
    \centering
    \includegraphics[width=\linewidth]{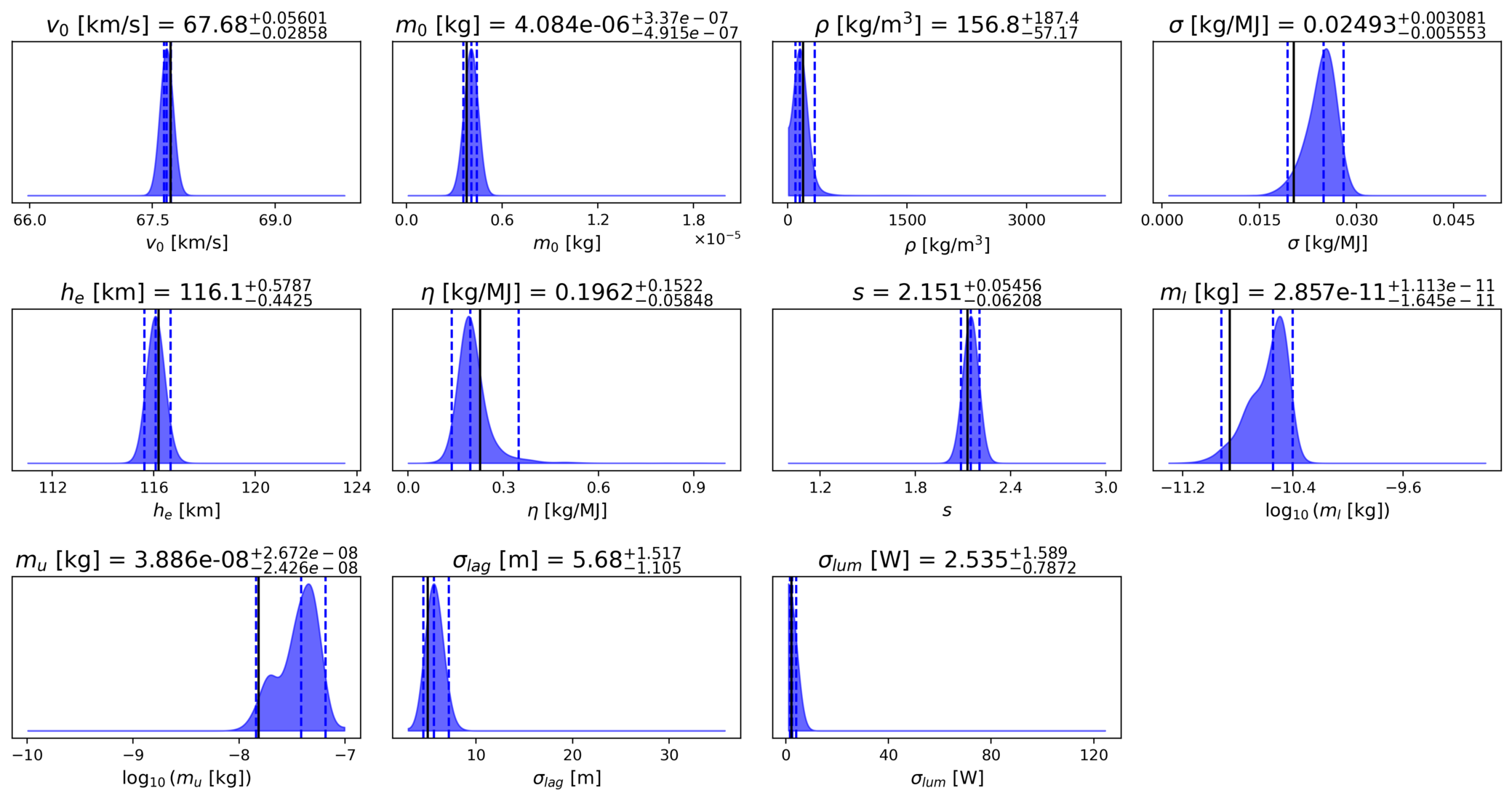}
    \caption{Posterior distributions for the Orionid mean test case with EMCCD light-curve noise and CAMO lag noise.
    The thick black line marks the true input value. Blue dashed lines denote the $95\%$ credible intervals, the dashed vertical line indicates the posterior median. Numerical summaries are annotated in each panel.}
    \label{img:ORI_mean_istib}
\end{figure}

In Figure \ref{img:ORI_mean_istib} we show one representative example to demonstrate that the true inputs lie within the $95\%$ credible intervals. Consistent with expectations, credible intervals tighten for longer more massive meteors that yield more data points. In all cases, the Best Guess (maximum log-likelihood) is statistically indistinguishable from the noiseless simulation at the plotted scales, indicating accurate recovery of both the physical parameters and the observational noise. Complete best-fit plots and posterior tables for all four test cases are provided in \ref{sec:Apx EMCCD validat}.

The two synthetic Orionids behave similarly, but differences in initial mass and erosion height lead to slightly different durations. The {mean} case is $\sim$0.05\,s longer, supplying more samples and yielding tighter posteriors for several parameters. 


Compared to the Orionids, the synthetic Capricornids are substantially longer and produce more measurements, resulting in narrower credible intervals. In our simulations, the {mode} Capricornid lasts $\sim$0.75\,s on EMCCD and $\sim$0.375\,s on CAMO, whereas the {mean} case lasts $\sim$1.125\,s (EMCCD) and $\sim$0.8\,s (CAMO). The increased sampling in the mean case produces better-constrained estimates for most physical parameters than in the mode case (see Tables~\ref{tab:posterior_summary_CAP_mode} and \ref{tab:posterior_summary_CAP_mean}). 

In summary, the results from the four validation test cases confirm the robustness of the method. The nested-sampling approach produces uncertainties in the synthetically-generated physical parameters compatible with the known true model input values while also correctly recovering the introduced noise from both the EMCCD and CAMO data.

\subsubsection{Validation Using EMCCD-Only Observations} \label{sec:EMCCD_validation}

We additionally tested how the inference performs when relying solely on EMCCD observations, simulating cases where no CAMO narrow-field data are available. For these tests, we produce synthetic meteors with noise characteristics appropriate to the EMCCD cameras. Based on observations, the average root-mean-square deviation (RMSD) for the lag was 24.5\,m for the simulated Capricornids and 20\,m for the simulated Orionids.

\begin{figure}[h!]
    \centering
    \includegraphics[width=\linewidth]{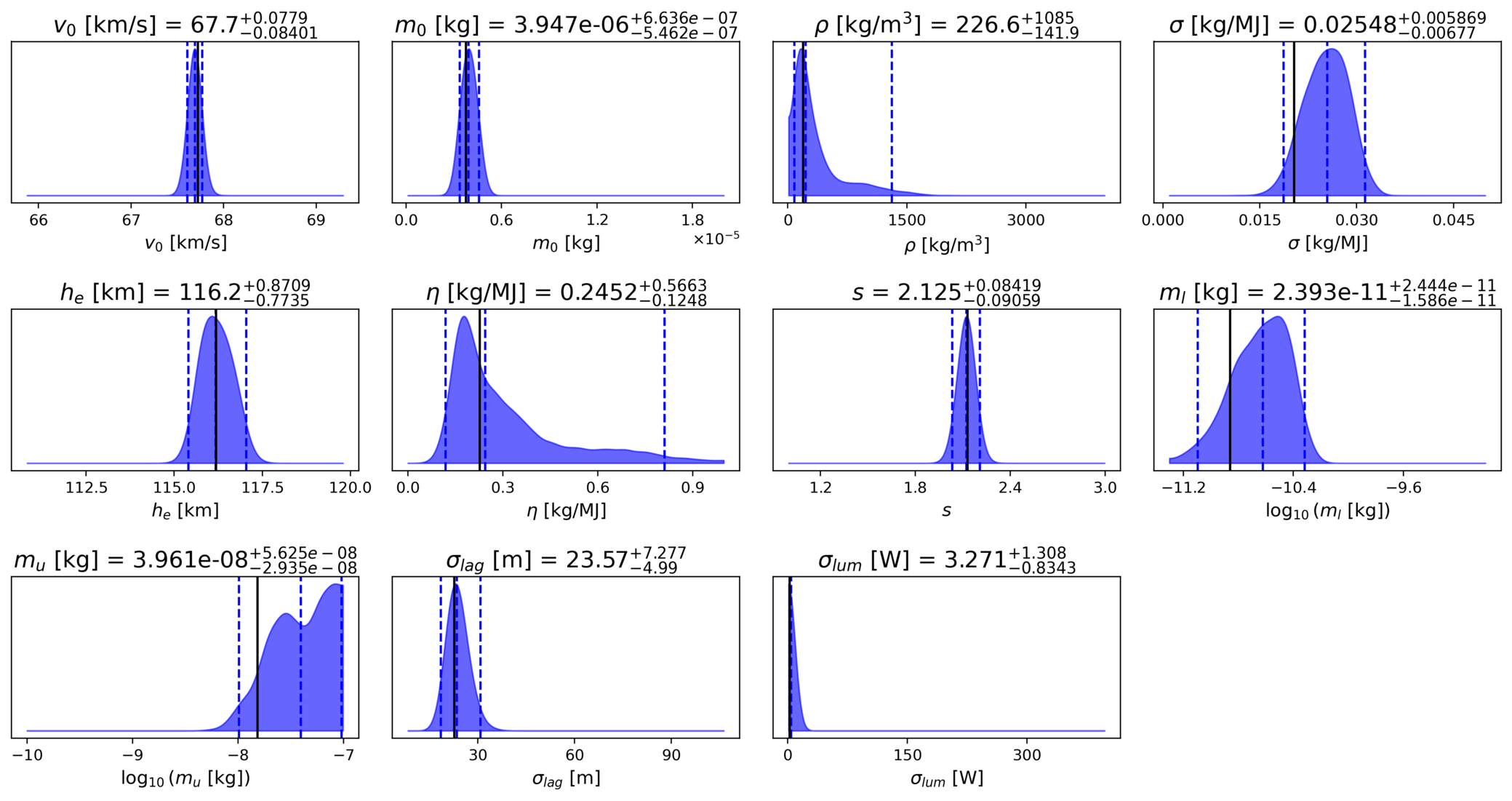}
    \caption{Posterior distribution statistics for the Orionid test case with mean physical parameters only using EMCCD noise. The true (input) solution is indicated by a thick black line and is still found in the 95\% credible intervals bounded by the two blue opposite dashed lines, while the dashed vertical line in between marks the posterior median. The region are broader compared to the other validation results with CAMO noise in lag.}
    \label{img:ORI_mean_istib_EMMCCD}
\end{figure}

All four validation test cases using EMCCD-only noise for both the light curve and deceleration are presented in ~\ref{sec:Apx EMCCD validat}. These tests reveal an important trade-off: while EMCCD-derived lag has higher noise compared to CAMO narrow-field measurements, it avoids the loss of early trajectory segments caused by the finite time taken for CAMO’s mirror to begin tracking. Overall, we found that the tightest parameter constraints were generally achieved when combining high-quality EMCCD light curves with high-accuracy CAMO deceleration data, as can be seen by comparing the mean Orionid with EMCCD only noise in Figure \ref{img:ORI_mean_istib_EMMCCD} and the synthetic meteor with CAMO lag noise in Figure \ref{img:ORI_mean_istib}. However, for the Orionid test case with the mode parameters, a synthetic meteors with the fewest number of data point (only 25 in total), the ablation coefficient $\sigma$ and the lower erosion mass limit $m_l$, showed slightly narrower credible intervals in the EMCCD-only configuration. This indicates that for some meteors the important information about their structure is contained in the initial part, i.e. the amount of early deceleration.

Importantly, across all EMCCD-only validation runs, the known input values were always contained within the 95\% credible intervals, just as in the CAMO validation. The posterior distributions for key physical parameters showed similar overall trends between the two approaches. While initial mass and velocity estimates remained comparably well constrained, the resulting bulk density distributions tended to be broader if only the EMCCD cameras are used, though they still clustered closely around the known input values, strongly showing that the quality of deceleration measurements is critical for quality measurements of the bulk density. Furthermore, among all parameters, the ablation $\sigma$ and erosion $\eta$ coefficients and the lower and upper grain mass limits ($m_l$, $m_u$) were the most sensitive to lag noise. These showed the largest broadening in the posterior distributions as the lag noise was increased. 

These findings confirm that, although EMCCD-only analyses can provide meaningful parameter constraints, combining EMCCD light curves with high-precision CAMO deceleration data consistently delivers tighter and more robust inversion results.

\section{Observational Results} \label{sec:results}







We applied our novel dynamic nested sampling method to a total of 15 meteors, 9 Orionids and 6 Capricornids. All meteors were observed by both the EMCCCD and CAMO systems. For each observed meteor, we explored two scenarios: one assuming a single erosion episode and another with two erosion episodes. The choice between these models was in part guided by manual assessment of the meteoroid morphology using high-resolution CAMO data, which allowed us to directly observe whether a meteor underwent complex fragmentation.


To validate that the investigated meteoroids truly belong to the two showers and are not sporadics, in ~\ref{sec:Apx radiants} we compare the meteors' radiants to the radiant distributions observed by the Global Meteor Network \citep[GMN;][]{vida2021global}. The GMN dataset contains thousands of meteors from each shower, allowing to effectively reconstruct their dispersion profiles. All considered meteors were within the 95\% confidence interval of the shower radiants and speeds observed by the GMN and were within the observed shower activity period.

The following subsections detail the derived physical properties for each shower based on the dynesty inversions of the erosion model, highlighting both the range of inferred values and notable parameter correlations.


\subsection{Orionid Results}

\begin{figure}[p] 
    \centering
    \includegraphics[height=0.85\textheight]{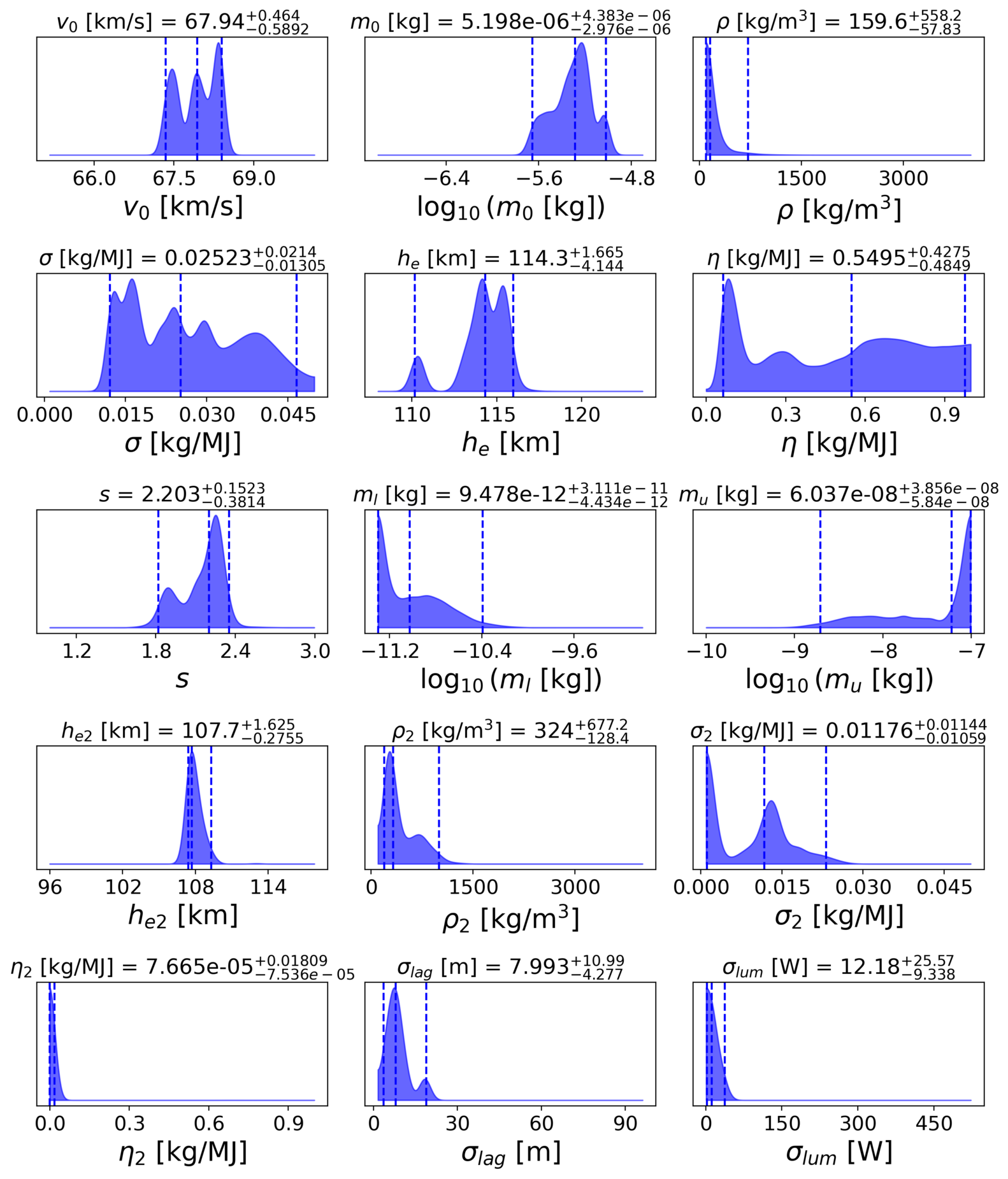}
    \caption{The posterior physical parameter distribution for the 9 modelled Orionids based on the weights from the dynamic nested sampling. The central dashed blue line show the median and the 95\% confidence interval bounds.}
    \label{img:ORI_distrib}
\end{figure}

The physical parameter distributions for the Orionids reflect their well-established cometary origin. Our analysis, illustrated in Figure~\ref{img:ORI_distrib} and summarized in Table~\ref{tab:overall_summary_ori}, covers a range of initial velocities ($v_0$) from 67.35 to 68.4~km/s and initial masses ($m_0$) from $2 \times 10^{-6}$~kg to $1 \times 10^{-5}$~kg. The bulk densities ($\rho$) are predominantly low, clustering around 150~kg/m$^3$, with only a few events approaching values near 700~kg/m$^3$.

For Figure~\ref{img:ORI_distrib}, we used the same plotting routine as for the single-meteor distribution plots, but with a combined (``stacked'') posterior constructed from all Orionid meteors. Specifically, for each meteor we take the posterior samples returned by \texttt{dynesty} together with their weights and first renormalize the weights so they sum to 1 (removing any arbitrary weight scaling between runs). We then concatenate the samples and weights across all meteors and finally renormalize the full set of weights to sum to 1. The resulting distribution is therefore an equal-per-meteor mixture of posteriors (not dominated by any single run due to weight scaling or sample count), and the confidence-interval marginals shown in Figure~\ref{img:ORI_distrib} are computed from this combined, weighted sample set.

\begin{table}[htbp]
    \centering
    \renewcommand{\arraystretch}{1.2}
    \setlength{\tabcolsep}{4pt}
    \caption{Overall posterior summary statistics for the 9 Orionids.}
    \resizebox{\textwidth}{!}{
    \begin{tabular}{lrrrrr}
    \hline
    \textbf{Parameter} & \textbf{2.5 percentile} & \textbf{Mode} & \textbf{Mean} & \textbf{Median} & \textbf{97.5 percentile} \\
    \hline
    $v_0$ [km/s] & 67.35 & 67.34 & 67.93 & 67.94 & 68.4 \\
    $m_0$ [kg] & 2.222$\times10^{-6}$ & 5.41$\times10^{-6}$ & 5.245$\times10^{-6}$ & 5.198$\times10^{-6}$ & 9.581$\times10^{-6}$ \\
    $\rho$ [kg/m$^3$] & 101.8 & 236.9 & 181.9 & 159.6 & 717.8 \\
    $\sigma$ [kg/MJ] & 0.012 & 0.015 & 0.026 & 0.025 & 0.046 \\
    $h_e$ [km] & 110.2 & 115.5 & 114.1 & 114.3 & 116 \\
    $\eta$ [kg/MJ] & 0.064 & 0.906 & 0.377 & 0.549 & 0.977 \\
    $s$ & 1.821 & 2.256 & 2.145 & 2.203 & 2.355 \\
    $m_{l}$ [kg] & 5.044$\times10^{-12}$ & 1.3$\times10^{-11}$ & 1.037$\times10^{-11}$ & 9.478$\times10^{-12}$ & 4.058$\times10^{-11}$ \\
    $m_{u}$ [kg] & 1.97$\times10^{-9}$ & 7.583$\times10^{-8}$ & 2.942$\times10^{-8}$ & 6.037$\times10^{-8}$ & 9.893$\times10^{-8}$ \\
    $h_{e2}$ [km] & 107.4 & 107.4 & 108 & 107.7 & 109.3 \\
    $\rho_{2}$ [kg/m$^3$] & 195.7 & 330.1 & 379.9 & 324 & 1001 \\
    $\sigma_{2}$ [kg/MJ] & 0.001 & 0.023 & 0.01 & 0.011 & 0.023 \\
    $\eta_{2}$ [kg/MJ] & 1.293$\times10^{-6}$ & 9.161$\times10^{-6}$ & 1.32$\times10^{-4}$ & 7.665$\times10^{-5}$ & 0.018 \\
    $\sigma_{lag}$ [m] & 3.71 & 19.25 & 8.76 & 7.99 & 18.98 \\
    $\sigma_{lum}$ [W] & 2.845 & 14.95 & 15.44 & 12.18 & 37.75 \\
    \hline 
    \end{tabular}}    \label{tab:overall_summary_ori}
\end{table}

We manually evaluated the fragmentation behavior using CAMO narrow-field camera data. This data, with of order 3~m resolution and 10~ms temporal cadence, provides high-resolution imagery allowing details of the meteor morphology to be directly observed. Based on a visual inspection, three meteors were have clear evidence of a second fragmentation episode, as listed in Table~\ref{tab:frag_logz_summary_ORI}. For these events, CAMO narrow-field imagery reveals a distinct fragment that emerges after the main erosion phase removes most of the mass from the main body. This remaining fragment is likely a denser, more compact core that survives the initial erosion, as was previously also found by \citet{vida2024first} for the Orionids and \citet{Pinhas2024QuantifyingObservatory} for the South Delta Aquariids.


This physical parameter details of the two-stage erosion behavior are shown in Table \ref{tab:frag_logz_summary_ORI}. The second erosion coefficient ($\eta_2$) is very low compared to the erosion in the fist stage (nearly four orders of magnitude less), indicating that the erosion virtually ceases around the peak of the light curve, confirming the findings of \citet{vida2024first} that were derived using manual model fits. Our numbers show that the bulk density of this remaining fragment is only a factor of two higher than the bulk density of the entire meteoroid and is not a high-density refractory silicate particle. The limited complexity of our model do not allow probing deeper into the structure of this fragment, but its ablation coefficient is half than of the bulk meteoroid, indicating a different composition. The mass distribution index ($s$) may indicate a bimodal structure, with preferred values clustering around 1.9 and 2.2, although this effect may be due to low number statistics. The grain mass range of the eroded fragments spans four orders of magnitude, from about $10^{-11}$~kg to $10^{-7}$~kg.

Uncertainties in lag and luminosity measurements are consistent with validation results, though photometric errors are higher for brighter meteors, high dynamic-range light curves can be harder to fit, and the fit becomes dominated by the brightest points, increasing the effective residuals in the fainter parts of the light curve.



\subsection{Capricornid Results}
\begin{figure}[p] 
    \centering
    \includegraphics[height=0.85\textheight]{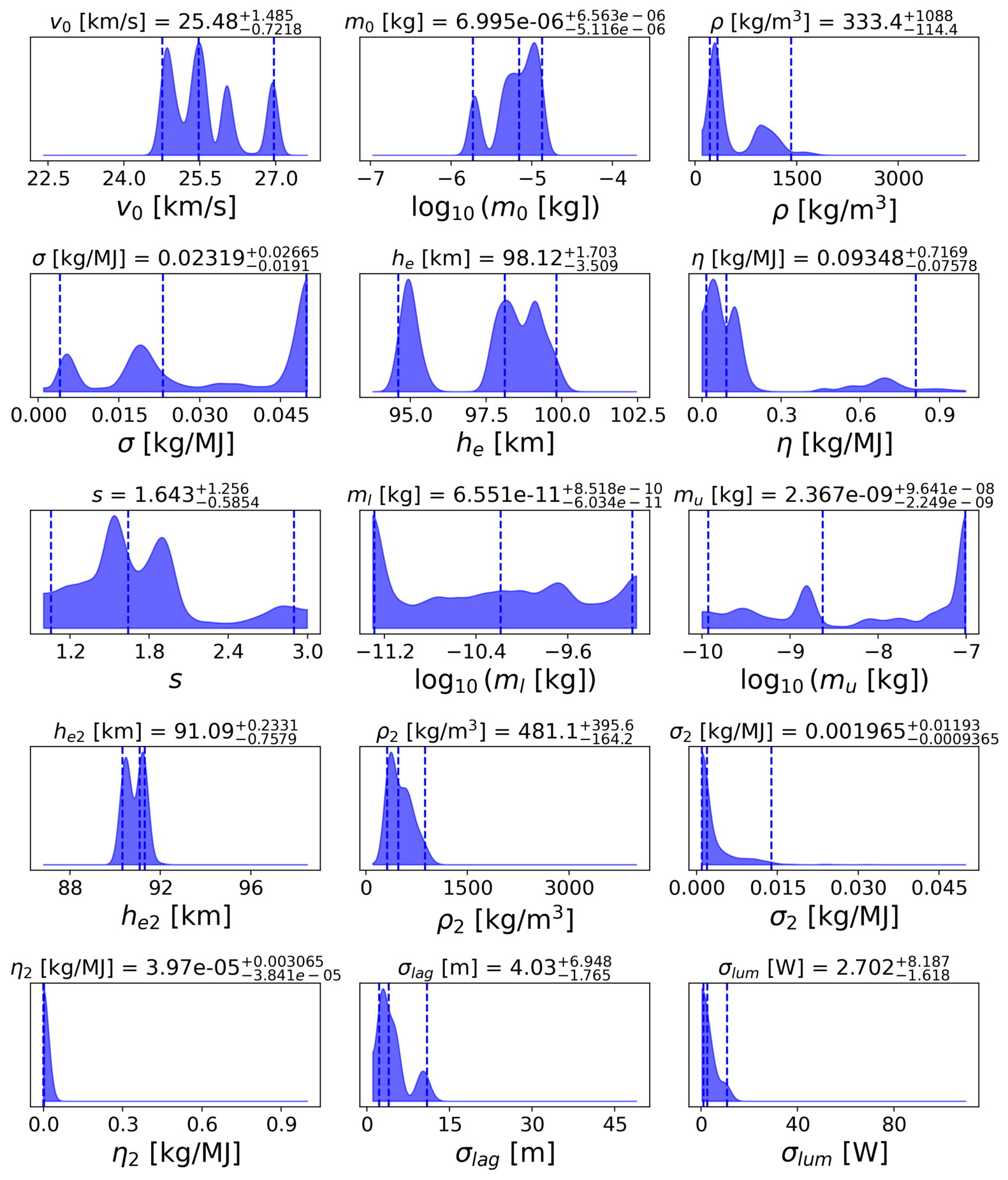}
    \caption{The Capricornid physical parameters distribution based on the weights of the dynamic nested sampling. The dashed blue line shows the median and the 95\% CI.}
    \label{img:CAP_distrib}
\end{figure}

The physical parameter distributions for the Capricornid meteors are summarized in Figure~\ref{img:CAP_distrib} and Table~\ref{tab:overall_summary_cap}. Compared to the Orionids, the Capricornids exhibit markedly lower entry velocities ($v_0$ ranging from 24.7 to 26.9~km/s) and systematically higher material densities. Capricornid bulk densities span a wide range, with clusters around  300~kg/m$^3$ and 1300~kg/m$^3$ with a mean near 455~kg/m$^3$ and a credible interval spanning from 219~kg/m$^3$ to 1422~kg/m$^3$.


\begin{table}[htbp]
    \centering
    \renewcommand{\arraystretch}{1.2}
    \setlength{\tabcolsep}{4pt}
    \caption{Overall posterior summary statistics for 6 Capricornid meteors.}
    \resizebox{\textwidth}{!}{
    \begin{tabular}{lrrrrr}
    \hline
    \textbf{Parameter} & \textbf{2.5CI} & \textbf{Mode} & \textbf{Mean} & \textbf{Median} & \textbf{97.5CI} \\
    \hline
    $v_0$ [km/s] & 24.76 & 24.82 & 25.63 & 25.48 & 26.97 \\ 

    $m_0$ [kg] & 1.879$\times10^{-6}$ & 7.612$\times10^{-6}$ & 7.364$\times10^{-6}$ & 6.995$\times10^{-6}$ & 1.356$\times10^{-5}$ \\

    $\rho$ [kg/m$^3$] & 219 & 325.7 & 455.5 & 333.4 & 1422 \\ 

    $\sigma$ [kg/MJ] & 0.004 & 0.046 & 0.028 & 0.023 & 0.049 \\

    $h_e$ [km] & 94.61 & 95.44 & 97.46 & 98.12 & 99.82 \\

    $\eta$ [kg/MJ] & 0.017 & 0.132 & 0.093 & 0.093 & 0.810 \\

    $s$ & 1.057 & 1.526 & 1.736 & 1.643 & 2.898 \\

    $m_{l}$ [kg] & 5.176$\times10^{-12}$ & 3.618$\times10^{-11}$ & 6.14$\times10^{-11}$ & 6.551$\times10^{-11}$ & 9.173$\times10^{-10}$ \\

    $m_{u}$ [kg] & 1.184$\times10^{-10}$ & 9.668$\times10^{-8}$ & 4.971$\times10^{-9}$ & 2.367$\times10^{-9}$ & 9.878$\times10^{-8}$ \\

    $h_{e2}$ [km] & 90.33 & 91.3 & 90.86 & 91.09 & 91.32 \\

    $\rho_{2}$ [kg/m$^3$] & 316.9 & 356.4 & 486 & 481.1 & 876.7 \\

    $\sigma_{2}$ [kg/MJ] & 0.001 & 0.003 & 0.003 & 0.002 & 0.013 \\

    $\eta_{2}$ [kg/MJ] & 1.285$\times10^{-6}$ & 2.479$\times10^{-6}$ & 4.19$\times10^{-5}$ & 3.97$\times10^{-5}$ & 0.003 \\

    $\sigma_{lag}$ [m] & 2.26 & 8.72 & 4.81 & 4.03 & 10.98 \\

    $\sigma_{lum}$ [W] & 1.08 & 5.25 & 3.82 & 2.70 & 10.89 \\
    \hline
    \end{tabular}}
    \label{tab:overall_summary_cap}
\end{table}

The erosion coefficient ($\eta$) is generally lower than in cometary meteoroids but still shows considerable variation, suggesting variability in structural properties. Notably, for the two meteors exhibiting clear secondary fragmentation in CAMO narrow-field imagery (see Table~\ref{tab:frag_logz_summary_CAP}), the very low erosion coefficient the second-stage indicates a sharp near cessation of erosion. While most Capricornids disintegrate into a line of discrete fragments near the end of their luminous phase, these two cases showed a brighter fragment that persisted after the end of erosion. For event 20200726\_060419, the bulk density remained roughly constant, whereas in 20220729\_044924, it increased from 300 to 600\,kg/m$^{3}$, indicating exposure of a more compact core. The ablation coefficient after the erosion change ($\sigma_2$) is on average half of the initial value, further strengthening the argument that the core is composed of a different material. The mass distribution index ($s$) displays a wide spread, with values ranging from around 1 to 2.9. The range of grain masses on average spans three orders of magnitude, revealing that the inverted mass index is meaningful and the large observed reange is physical.

The inverted measurement uncertainties remain within expected limits. The lag error ($\sigma_{lag}$) shows a tight dispersion, typically under 4 meters, while luminosity errors ($\sigma_{lum}$) range from 1 to 10~W, as expected from the validation cases.

\subsection{Weighted Correlation of Sampled Parameters}
\label{sec:siglemeter_weighted-corr}

For each individual meteor, we computed the pairwise weighted correlation matrix of all model parameters using posterior samples. Each correlation value therefore reflects the internal relationships among parameters within the posterior distribution of a single meteor. It is important to note that these correlations arise primarily from the structure of the adopted physical model and the fitting process itself, rather than representing direct physical dependencies. In other words, they indicate how the DNS algorithm compensates for parameter degeneracies when reproducing a meteor’s observed light curve and deceleration, rather than intrinsic causal links between meteoroid properties.

Across the majority of meteors, a consistently strong positive correlation is found between the erosion coefficient and the bulk density, also previously found by \citet{borovivcka2009material} and \citet{vovk_PER_2025}. This trend reflects a model-driven trade-off: denser meteoroids require higher erosion rates to reproduce the same luminosity and amount of deceleration. Similarly, both parameters typically show a pronounced negative correlation with the entry velocity, as the model balances kinetic energy input against the erosion coefficient (defined in units of kg/MJ). Meteors modeled with double fragmentation generally display weaker correlations among these key parameter pairs, owing to the introduction of additional free parameters that increase posterior uncertainty and dilute simple linear dependencies.

For most Orionids, a robust negative correlation is also detected between bulk density and initial mass, as well as between initial mass and the erosion coefficient. Capricornids exhibit the same qualitative behavior but with smaller magnitudes. These relationships arise from how the DNS sampler adjusts parameter combinations to match observed light-curve amplitudes and slopes under different entry velocities, rather than indicating a physical anti-correlation between mass and density.

All correlations involving the observational noise parameters, both for luminosity and lag, remain close to zero, confirming that the estimated instrumental noise does not couple with physical parameters and that the two noise terms are mutually uncorrelated.

\section{Discussion} \label{sec:discussion}



This study presents a novel statistically robust framework for inferring physical properties and the associated uncertainties of meteoroids from optical observations. Using the meteoroid erosion model in combination with dynamic nested sampling (DNS), we derived full posterior distributions of meteoroid physical parameters by fitting the observed meteor light curves and deceleration profiles, without explicitly constraining the meteor wake.

Despite this, the simulations corresponding to the highest posterior probabilities generally reproduce the observed wakes profiles directly observed by the CAMO narrow-field system (see Section~\ref{sec:Apx wake}). This agreement indicates that the fitted lag and luminosity alone are often sufficient to recover physically consistent solutions. Notable examples with excellent correspondence include events 20191023\_084916 and 20200726\_032722, while a few cases (e.g. 20191028\_050616, 20221022\_075829) show partial mismatches. Incorporating wake profiles in the model fits, which are computationally expensive to produce, might help narrow down the range of physical uncertainties. This will be explored in a future publication.

\subsection{Sensitivity Analysis}
\label{sec:model-assumptions}

In all simulations, we made fundamental fixed assumptions about the physical processes of meteoroid ablation. The drag factor \(\Gamma\) was fixed to unity, consistent with the free-molecular flow regime applicable to small meteoroids \citep{popova2019modelling}. The shape coefficient \(A\) was set to 1.21, appropriate for spheres or rapidly rotating bodies whose effective cross sections approach near-spherical values \citep{Opik1958}. Beyond these fairly certain assumptions, we made two additional key assumptions: the value of the grain density was set to \(3000~\mathrm{kg\,m^{-3}}\) to follow past publications using the erosion model, and the luminous efficiency was assumed to follow the empirical model of \citet{vida2024first}. 

To evaluate the robustness of the inferred parameters, we further examined how the results change when adopting different luminous efficiency models, varying the assumed grain bulk density, and when only lower-resolution EMCCD data is used for both the light curve and deceleration measurements. These tests allow us to assess how the credible intervals respond to different necessary assumptions or worse measurement uncertainty.

\subsubsection{Luminous Efficiency}

Luminous efficiency \(\tau\) was sampled following the \citet{vida2024first} empirical model, which expresses \(\tau\) as a function of object mass and initial velocity. Using the DNS-derived physical parameters for each meteor, we obtained a grain luminosity weighted nominal values of \(\tau \approx 0.4^{+0.2}_{-0.2}\%\) for Orionids and \(\tau \approx 1.4^{+1.8}_{-0.9}\%\) for Capricornids. To test the sensitivity of our results to this assumption, we adopted the highest and lowest velocity-dependent formulations from the literature as summarized by \citet{subasinghe2018luminous}, selecting the two extremes among published models: the upper limits from \citet{weryk2013simultaneous} (\(3\%\) for Orionids, \(3.5\%\) for Capricornids) and the lower limits from \citet{saidov1989luminous} (\(0.05\%\) for Orionids, \(0.12\%\) for Capricornids).

\begin{figure}[h!]
    \centering
    \includegraphics[width=0.9\linewidth]{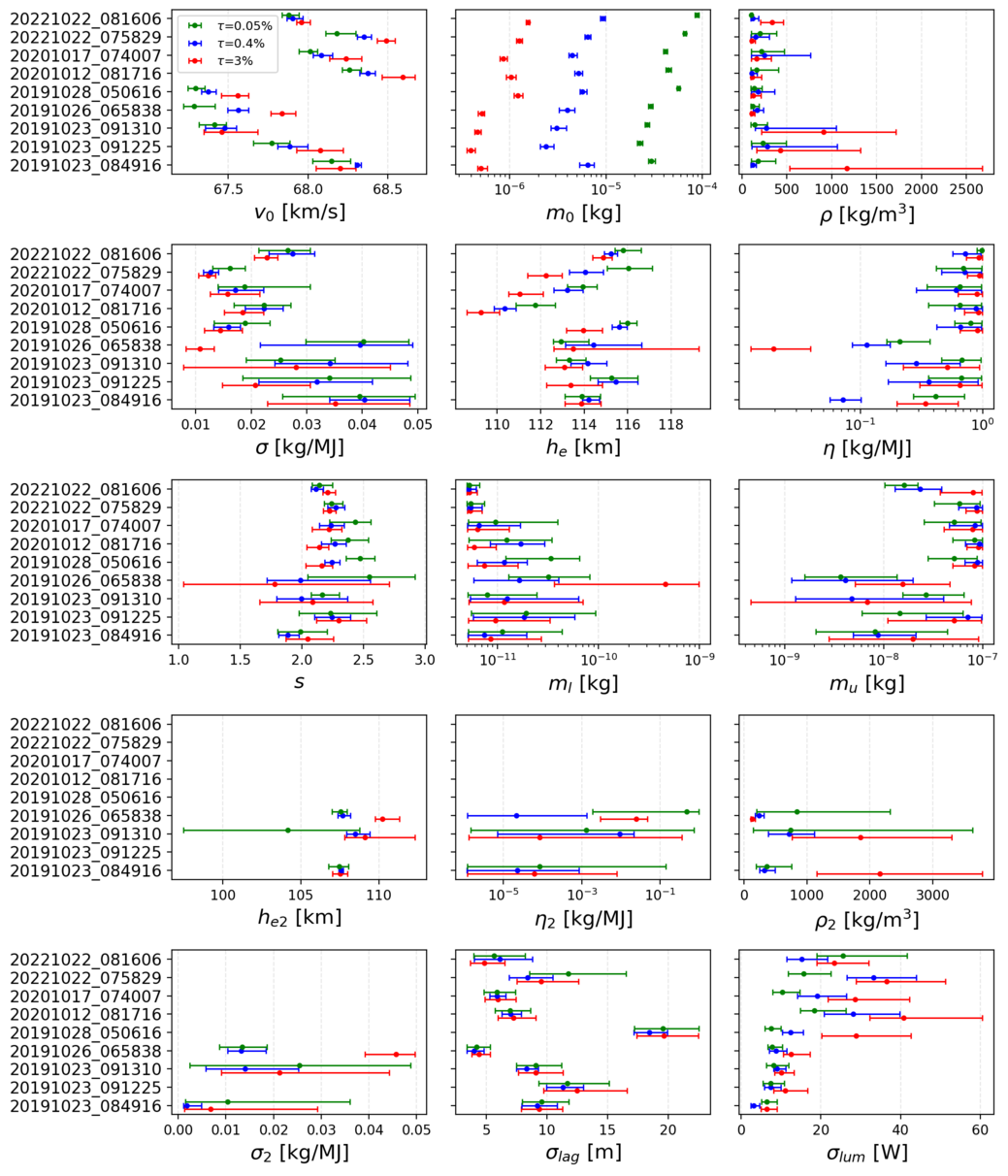}
    \caption{ORI sensitivity to luminous efficiency \(\tau\). Colors: red 3\%, blue 0.4\% (nominal), green 0.05\%.}
    \label{img:ORI_taucheck}
\end{figure}

\begin{figure}[h!]
    \centering
    \includegraphics[width=0.9\linewidth]{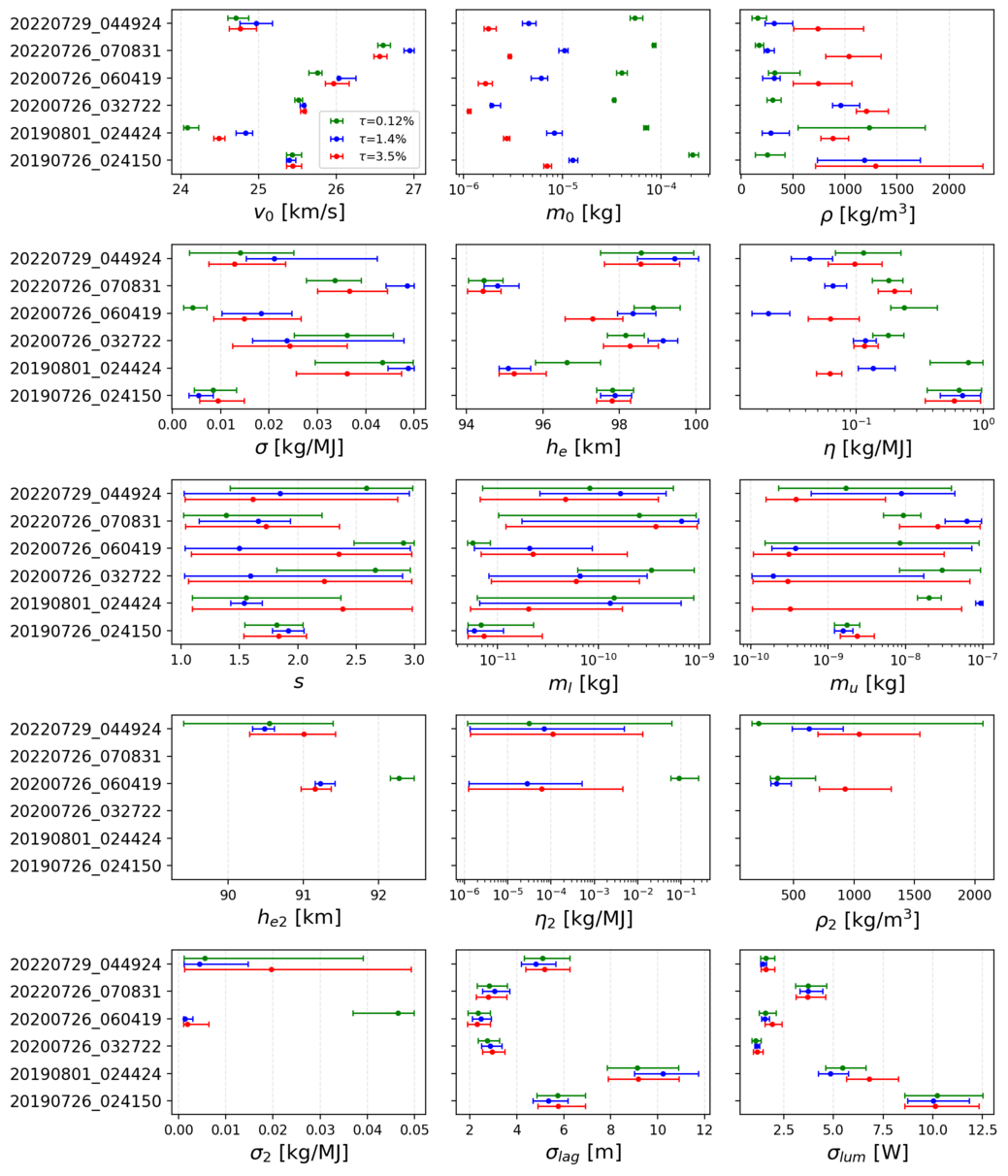}
    \caption{CAP sensitivity to luminous efficiency \(\tau\).}
    \label{img:CAP_tauheck}
\end{figure}

As shown in Figures Figures~\ref{img:ORI_taucheck} and \ref{img:CAP_tauheck}, the simulations with modified luminous efficiency reproduce observed data across all meteors, but higher luminous efficiencies consistently yield lower initial masses and higher bulk densities, while lower efficiencies produce the opposite trend. For the Orionids, the effect on density is modest and for most remains within credible intervals. For Capricornids, the variation is substantial, with median bulk densities increasing from \(333~\mathrm{kg\,m^{-3}}\) to nearly \(950~\mathrm{kg\,m^{-3}}\) at \(\tau = 3.5\%\). Other parameters show no systematic shifts across the tested range. This sensitivity may also help explain why early single-body ablation studies, such as \citet{Verniani1969}, reported unusually low bulk densities - his adoption of a much lower luminous efficiency than contemporary values likely led to systematically lower densities.

These tests confirm that luminous efficiency has a significant impact on the inferred bulk densities and the initial masses. Accurate constraints on \(\tau\) are therefore critical for physically meaningful meteoroid characterization. 



\subsubsection{Grain density change}

The grain density was assumed to be \(3000~\mathrm{kg\,m^{-3}}\), consistent with mixed silicate organic compositions, as assumed in the literature. Tests with \(3500~\mathrm{kg\,m^{-3}}\), closer to meteoritic values \citep{flynn2018physical}, produced results slightly higher in bulk density that kept the same porosity. The results are generally well within the uncertainty limits of the main runs, as shown in \ref{sec:Apx sensitivity rhoGr plots}, showing a low sensitivity of the model to this parameter.

\subsubsection{Derived Physical Properties Based Only on EMCCD Data}
\label{sec:EMCCD-results}

The CAMO tracking system records an order of magnitude fewer meteors than the EMCCD systems. Furthermore, manual analysis of CAMO data is time-consuming. In contrast, measurements from EMCCD data can be made much more quickly, partly because the EMCCD framerate is three times lower.

To assess the impact of using these lower-precision EMCCD deceleration measurements on derived physical properties and their associated uncertainties, we reanalyzed the 15 meteors in our sample. This reanalysis used DNS but relied solely on the luminosity and lag measurements derived from the EMCCD data.

\begin{figure}[ht]
    \centering
    \includegraphics[width=\linewidth]{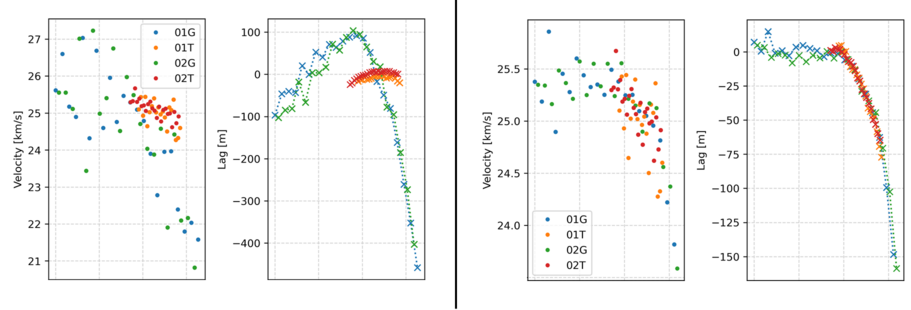}
    \caption{Comparison of lag data for two Capricornid meteors. Left: An event where the EMCCD-derived lag (01G, 02G) is inconsistent with the CAMO narrow-field lag (01T, 02T). Right: An event showing agreement between the CAMO narrow-field and EMCCD measurements.}
    \label{img:EMCCDvsCAMO_lag}
\end{figure}

While EMCCD lag can align well with CAMO narrow-field lag under certain conditions, discrepancies often arise. These stem mainly from biases introduced during manual leading-edge selection, especially for fast faint meteors where the low EMCCD resolution makes it difficult to resolve fine morphological details. As shown in Figure~\ref{img:EMCCDvsCAMO_lag} for some meteors EMCCD and CAMO lag data agree well. The slower-moving Capricornid meteors tend to show better agreement, whereas the faster Orionids often diverge. This divergence may not only be a function of velocity but also reflect differences in meteor morphology as the Orionids exhibit leading fragments which are too faint to be observed with the EMCCDs. In this case, the CAMO measurements track the faster and slower decelerating fragment, while the EMCCD measurements track the rapidly decelerating cloud of grains. Both can be accomoated in the model with the correct choice of the reference point - for CAMO, the leading fragment should be chosen as the reference point, while for EMCCD measurements the location of the brightest grain bin should be used.

Overall, validation test cases show that when the leading edge is clearly resolved, EMCCD-only analyses yield comparable physical solutions to those incorporating CAMO lag, though with broader uncertainties (see section \ref{sec:EMCCD_validation}). However, when the leading edge is less well defined for the real meteor data detected only with EMCCD, uncertainties in the lag measurement propagate strongly into the secondary fragmentation parameters, particularly the second-stage erosion coefficients. This is due to the higher noise and lower temporal precision of EMCCD data compared to CAMO narrow-field measurements as shown in ~\ref{sec:Apx sensitivity cam plots}.

We suggest that an automated algorithm to measure EMCCD data in a statistically robust way is to be developed which will reduce subjective biases introduced by manual picking. In addition, such an automated algorithm will be able to produce direct measurement uncertainties that could be included in the DNS process.

\subsection{Distribution of Physical Parameters}
\label{sec:phys_dist_parameters}

The posterior distributions recovered for Orionid and Capricornid meteoroids in the $10^{-6}$ to $10^{-5}$~kg mass range are consistent with the expected physical nature of their respective parent bodies. The Orionids, originating from comet 1P/Halley, exhibit low bulk densities ($159^{+558}_{-57}$~kg/m$^3$), characteristic of porous, fragile material from an active, volatile-rich nucleus. In contrast, the Capricornids, associated with the low-activity or dormant comet 169P/NEAT, display significantly higher densities ($333^{+1089}_{-114}$~kg/m$^3$, with credible intervals extending up to $1400$~kg/m$^3$). This suggests a more compact structure, likely shaped by space weathering and thermal processing.

\subsubsection{Mass Index and Grain Mass}
\label{sec:mass_index}

In-situ measurements from 1P/Halley (DIDSPY on Giotto; SP-2 on Vega 1 and 2) show a grain mass index $s$ ranging from $1.5$ to $2.27$, and $s=1.85$ when summed over the Giotto trajectory for $10^{-13}$--$10^{-8}$~kg meteoroids \citep{hughes1988p, mcdonnell1987dust}. A value of $s$ around 2 means that the number of particles increases rapidly toward smaller masses, so the population is dominated by small grains. Our Orionid results ($s = 1.8$--$2.4$, clustering near $s \approx 2$) therefore imply a size distribution similar to Halley's dust, consistent with fresh cometary ejecta composed of small, fragile aggregates.

We also note that the inferred minimum and maximum grain masses often approach the prior bounds, indicating that the mass index strongly influences both the light-curve shape and the deceleration behavior.

No direct in-situ data exist for 169P/NEAT. However, 67P/Churyumov–Gerasimenko, another Jupiter-family comet (JFC) with available in-situ measurements, shows $s = 1.76\text{--}1.9$ for grains above $10^{-13}$~kg \citep{agarwal2007dust}. This is consistent with JFC meteors studied by \citet{vojavcek2019properties}, which have $s=1.8$. Our Capricornid sample yields similar values, clustering around $s=1.5$ and $s=2.0$. These results have broader uncertainties, reflecting weaker constraints, particularly for events where the inferred upper and lower grain mass bounds were close.

\begin{figure}[h!]
    \centering
    \includegraphics[height=0.4\textwidth]{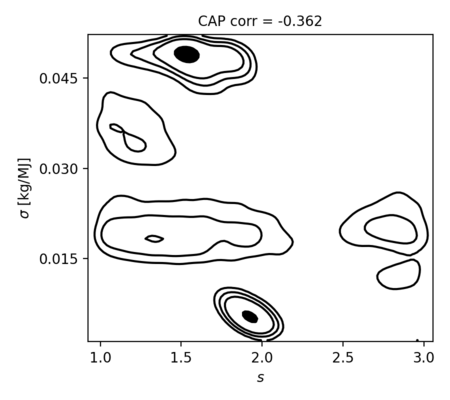}
    \includegraphics[height=0.4\textwidth]{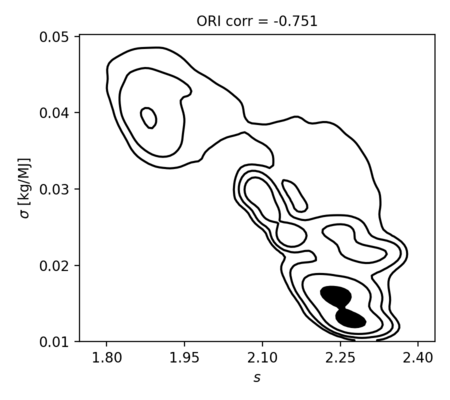}
    \caption{Inverse correlation between mass index \(s\) and ablation coefficient \(\sigma\) for representative Capricornid (left) and Orionid meteors (right).}
    \label{img:ORI_CAPsigmas}
\end{figure}

Across our meteor data, we find a strong inverse correlation for the Orionids between the mass index $s$ and the ablation coefficient $\sigma$ (Figure~\ref{img:ORI_CAPsigmas}). This suggests that populations dominated by smaller fragments (high $s$) are composed of highly refractory particles, as they ablate less readily (low $\sigma$). Conversely, populations with a lower mass index (dominated by larger particles) exhibit higher ablation coefficients, suggesting they are rich in volatile material, similar to the CHON (carbon, hydrogen, oxygen and nitrogen) particles found by Giotto at 1P/Halley \citep{clark1987systematics}.

\begin{figure}[h!]
    \centering
    \includegraphics[height=0.4\textwidth]{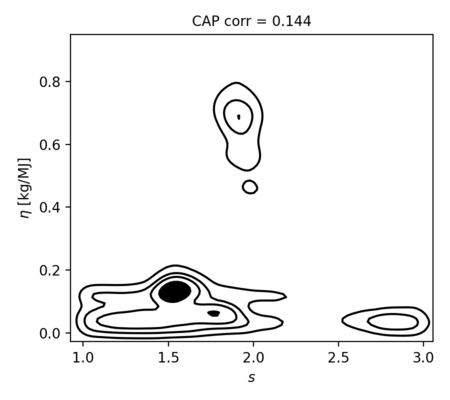}
    \includegraphics[height=0.4\textwidth]{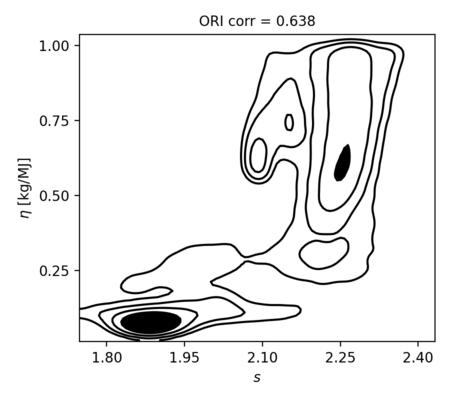}
    \caption{Positive correlation between erosion coefficient \(\eta\) and ablation coefficient \(\sigma\) in selected Capricornid (left) and Orionid  meteors (right).}
    \label{img:ORI_CAPetas}
\end{figure}

An additional positive correlation appears for Orionids between the mass index and the erosion coefficient $\eta$ (Figure~\ref{img:ORI_CAPetas}). This implies that populations with a high mass index (dominated by small grains) are less resistant to thermal and pressure stresses and fragment more efficiently. Conversely, populations dominated by larger particles (low $s$) appear to be more resistant, suggesting a stronger matrix holding the grains together. This trend is less evident for the Capricornids, where the correlations are weaker and are influenced by individual outliers.

While these relationships offer plausible physical interpretations, caution is warranted. Given the limited sample size, some observed correlations may be statistical artifacts or model-driven, rather than reflecting intrinsic material differences.

\subsubsection{Erosion energy}
 
We examined the average energy per unit mass and per unit cross section required to initiate erosion, following the formulation of \citet{borovivcka2007atmospheric}. Compared to other studies the first detection was comparatively higher thanks to the more sensitive EMCCD cameras used for this study. The majority of Orionid meteoroids show erosion onset energies per unit cross section (\(E_s\)) clustered between {1 and 2~MJ\,m\(^{-2}\)}, while the Capricornid sample reaches higher values in the range of {1.5–4~MJ\,m\(^{-2}\)}. When expressed per unit mass, the Orionids display a broader distribution, from 2 to 6~MJ\,kg\(^{-1}\), whereas the Capricornids range from 1 to 5.5~MJ\,kg\(^{-1}\). These differences likely reflect variations in grain binding strength and internal cohesion rather than bulk density alone.

\subsubsection{Ablation and Erosion Coefficients}

Comparing the ablation coefficient $\sigma$ and the erosion coefficient $\eta$ provides insight into the balance between ablation and fragmentation energies \citep{vojavcek2017properties}. Orionids show a higher median erosion coefficient ($\eta \approx 0.55~\mathrm{kg\,MJ^{-1}}$) than Capricornids ($\eta \approx 0.09~\mathrm{kg\,MJ^{-1}}$), consistent with a more fragile, porous structure. Conversely, Capricornids exhibit slightly lower ablation coefficients ($\sigma_{\mathrm{CAP}} \approx 0.023~\mathrm{kg\,MJ^{-1}}$) than Orionids ($\sigma_{\mathrm{ORI}} \approx 0.025~\mathrm{kg\,MJ^{-1}}$), reflecting their more compact and cohesive composition.

\begin{figure}[h!]
    \centering
    \includegraphics[height=0.4\textwidth]{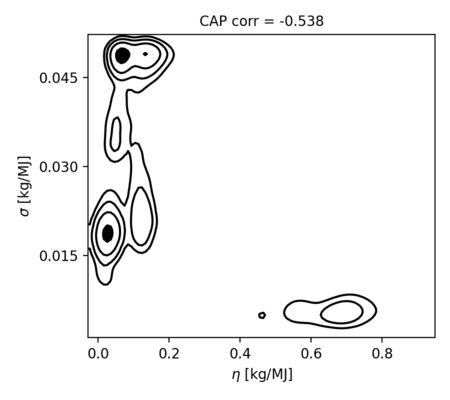}
    \includegraphics[height=0.4\textwidth]{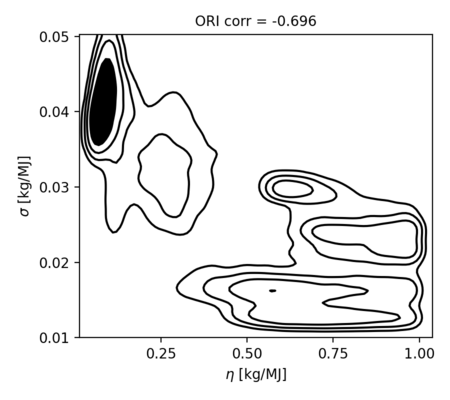}
    \caption{Inverse correlation between ablation coefficient \(\sigma\) and erosion coefficient \(\eta\) for representative CAP (left) and ORI (right) meteors.}
    \label{img:ORI_CAPsigmaeta}
\end{figure}

A clear anti-correlation between $\sigma$ and $\eta$ is found for Orionids (Figure~\ref{img:ORI_CAPsigmaeta}), indicating that when ablation is less efficient ($\sigma$ is small), erosion dominates ($\eta$ is large). This is expected for loosely bound aggregates that shed many small grains. For Capricornids, this trend is not significant and appears to be influenced by a single outlier.

Considering $\sigma$, $\eta$, and the mass index $s$ together, populations rich in small grains ($s$ high) appear more prone to fragmentation, while populations dominated by larger grains ($s$ low) ablate more efficiently.

\subsubsection{Bulk Density}
\label{sec:bulk_density}

The measured bulk densities for Orionids ($159^{+558}_{-57}\,\mathrm{kg\,m^{-3}}$) and Capricornids ($333^{+1089}_{-114}\,\mathrm{kg\,m^{-3}}$) align broadly with previous studies and in-situ data, though variations arise from differences in instrumentation and modeling. \citet{buccongello2024physical} reported mean values near $830\,\mathrm{kg\,m^{-3}}$ for Capricornids and slightly higher masses for Orionids, consistent with the upper tails of our posteriors. Similarly, \citet{vida2024first} derived Orionid bulk densities of $\sim300$ using CAMO data and a fixed $\sigma=0.3\,\mathrm{kg\,MJ^{-1}}$. This contrasts with our results, which show a lower median but have overlapping credible intervals. In contrast to these works, our DNS approach uses both EMCCD light curves and CAMO lag data, capturing early ablation phases and yielding tighter constraints on initial erosion heights.

Broader context from \citet{vojavcek2019properties} shows median densities of $400\,\mathrm{kg\,m^{-3}}$ for Halley-type and $695\,\mathrm{kg\,m^{-3}}$ for Jupiter-family meteoroids. Our Orionid results agree well with the Halley-type population, while the Capricornid densities, though lower, still reflect a predominantly cometary composition with possible contributions from more compact fragments.

\begin{figure}[htbp] 
    \centering
    \includegraphics[height=0.4\textwidth]{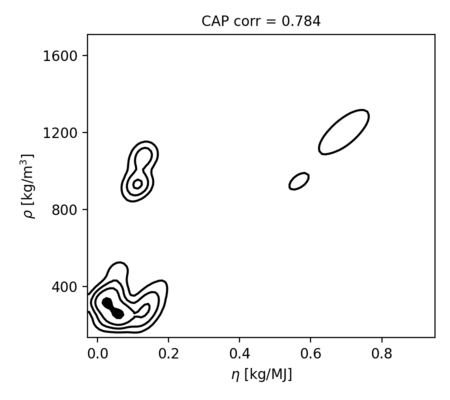}
    \includegraphics[height=0.4\textwidth]{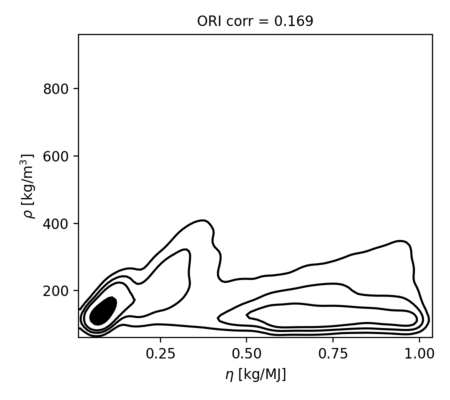}
    \caption{Posterior distributions showing the positive correlation between bulk density and erosion coefficient for representative Capricornids and Orionids.}
    \label{img:ORI_CAPerosionRho}
\end{figure}

A consistent positive correlation emerges between bulk density and erosion coefficient across both showers (Figure~\ref{img:ORI_CAPerosionRho}), and the same trend appears within individual meteor solutions (see Section~\ref{sec:siglemeter_weighted-corr}). Its presence at both the single-meteor and population levels suggests that this relationship is can simply be explained physically - denser meteoroids penetrate deeper into the atmosphere, experience higher dynamic pressures and temperatures, and thus fragment more rapidly. The correlation is strongest among Capricornids, while Orionids show a weaker relationship due to their narrower density range. Other correlations with density are less pronounced and may reflect the limited sample size rather than intrinsic dependencies.

In-situ measurements from the three-channel spectrometer aboard Vega-2 at Comet 1P/Halley indicate bulk densities near $300~\mathrm{kg\,m^{-3}}$ for small grains ($10^{-13}$--$10^{-9}\,\mathrm{kg}$), though with uncertainties up to a factor of two \citep{krasnopolsky1988properties}. \citet{divine1986comethalley} further suggested that bulk density decreases with particle size, as larger aggregates are expected to be more porous. This trend could explain the slightly lower densities retrieved for our Orionid meteoroids ($10^{-5}$--$10^{-6}\,\mathrm{kg}$), which likely represent more loosely bound aggregates of similar material. 

A complementary analysis based on coma modeling and data from the Optical Probe Experiment (OPE) and Dust Impact Detection System (DIDSY) on Giotto estimated Halley dust bulk densities could range from $50$ to $500~\mathrm{kg\,m^{-3}}$, with $100~\mathrm{kg\,m^{-3}}$ favored, for particles between $10^{-12}$ and $10^{-3}\,\mathrm{kg}$ \citep{fulle2000situ}. Our estimted range of ($159^{+558}_{-57}\,\mathrm{kg\,m^{-3}}$ almost exactly matches this estimate. This reinforces that the physical properties inferred from our DNS analysis are consistent with the composition and porosity observed in Halley-type cometary dust.

\section{Conclusions}\label{sec:conclusions}



This study presents a statistically rigorous framework to infer the physical properties of meteoroids---specifically mass, bulk density, and fragmentation behavior---from optical observations. These parameters are essential for evaluating spacecraft impact risks using Ballistic Limit Equations (BLEs), with the bulk density being a classical challenging to estimate, particularly for the sub-millimeter to millimeter-sized meteoroids.

We applied the erosion-fragmentation model of \citet{borovivcka2007atmospheric} within a dynamic nested sampling (DNS) framework to fit meteor light curves and deceleration profiles, providing automated model fits and statistically sound uncertainties. Observational data were collected using the Canadian Automated Meteor Observatory (CAMO) for high-precision deceleration measurements and Electron Multiplying CCD (EMCCD) cameras for sensitive photometry.

The method was validated using four synthetic meteors and then applied to 15 real events: 9 Orionids and 6 Capricornids. The key results are:

\begin{itemize}
    \item A median bulk density of $159^{+558}_{-57}\,\mathrm{kg\,m^{-3}}$ for Orionids, consistent with porous cometary material.
    
    \item A median bulk density of $333^{+1089}_{-114}\,\mathrm{kg\,m^{-3}}$ for Capricornids, suggesting a more compact cometary material.
    
    \item Posterior distributions and credible intervals for all key physical parameters are in good agreement with previous studies that used similar models and luminous efficiency assumptions.
    
    \item Sensitivity tests show that the luminous efficiency ($\tau$) is the dominant fixed parameter influencing the inferred bulk density. Higher $\tau$ values increase the estimated density and decrease the initial mass, and vice versa. Accurate constraints on $\tau$ are therefore critical for physically meaningful interpretations.
    
\end{itemize}

Beyond bulk density, the recovered parameter distributions reveal distinct structural and compositional differences between the two showers. Orionid meteoroids (from 1P/Halley) exhibit mass indices $s \approx 2$, high erosion coefficients, and lower ablation efficiencies. This is consistent with loosely bound, porous aggregates composed of large, fragile grains that fragment easily. In contrast, Capricornids (from 169P/NEAT) show $s \approx 1.6$, lower erosion coefficients, and slightly higher ablation efficiencies, reflecting a more compact, cohesive structure. The higher erosion onset energies found for Capricornids (2--4~MJ\,m$^{-2}$, compared to 1--2~MJ\,m$^{-2}$ for Orionids) further suggest stronger grain binding and internal cohesion. 

While consistent trends are observed between parameters (e.g., bulk density and erosion coefficient), their presence in single-meteor posteriors suggests some correlations may be partly model-driven. These should be interpreted with caution given the limited sample size.

The DNS framework enables an automated, uncertainty-aware characterization of meteoroids and is scalable to larger datasets. By integrating multi-instrument data, it enhances the resolution of early ablation behavior---a phase critical for accurate density estimation and spacecraft shielding design.

Looking ahead, this approach will be extended to sporadic meteors, which originate from a more diverse set of parent bodies. This will allow for a broader mapping of meteoroid material properties and fragmentation behaviors across orbital classes, advancing both scientific understanding and practical risk assessment.

\section*{Acknowledgments}

Funding for this work was provided by the NASA Meteoroid Environment Office under cooperative agreement 80NSSC24M0060 and the European Space Agency (ESA) through Contract Number 4000145350, as part of the ESA Initial Support for Innovation (EISI) program.
The authors would like to thank Dr. Bill Cooke and Mark Millinger for providing insight and expertise that assisted the research. We thank Z. Krzeminski for help in optical data reduction.

\section*{Note on Code Availability}

Implementation of all methods used in this work is published as open source on the following GitHub web page and the dataset used for this work is available in the following mendeley link:
\begin{itemize}
    \item WesternMeteorPyLib: 
    \\\url{https://github.com/wmpg/WesternMeteorPyLib}
    \item dataset:
    \\\url{https://data.mendeley.com/datasets/ckv4jnp2rr}
\end{itemize}
Readers are encouraged to contact the lead author in the event they are not able to obtain the code on-line.

\section*{Declaration of generative AI and AI-assisted technologies in the writing process}

During the preparation of this work the authors used ChatGPT in order to improve language and readability. After using this tool/service, the authors reviewed and edited the content as needed and take full responsibility for the content of the publication.





\newpage

\appendix\section{Radiants}\label{sec:Apx radiants}

\begin{figure}[h!]
    \centering
    \includegraphics[width=\linewidth]{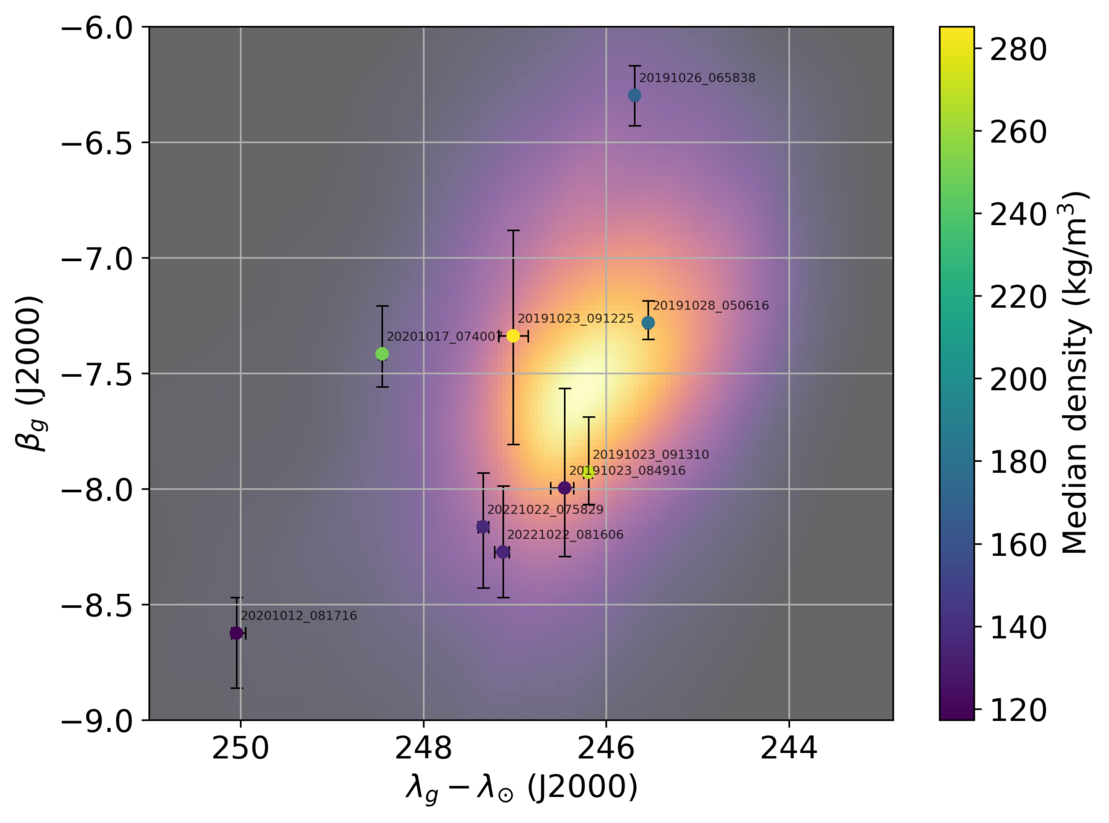}
    \caption{The sun-centered ecliptic radiant distribution for all 9 analyzed Orionid meteors. The background density map shows the radiant distribution for 10,316 Orionids from GMN 2024 data, while the individual points with error bars represent the 1 sigma level of uncertainty for the meteors analyzed in this study with the median bulk density from the posterior distribution of the meteor.}
    \label{fig:ORI_radiants}
\end{figure}

\begin{figure}[h!]
    \centering
    \includegraphics[width=\linewidth]{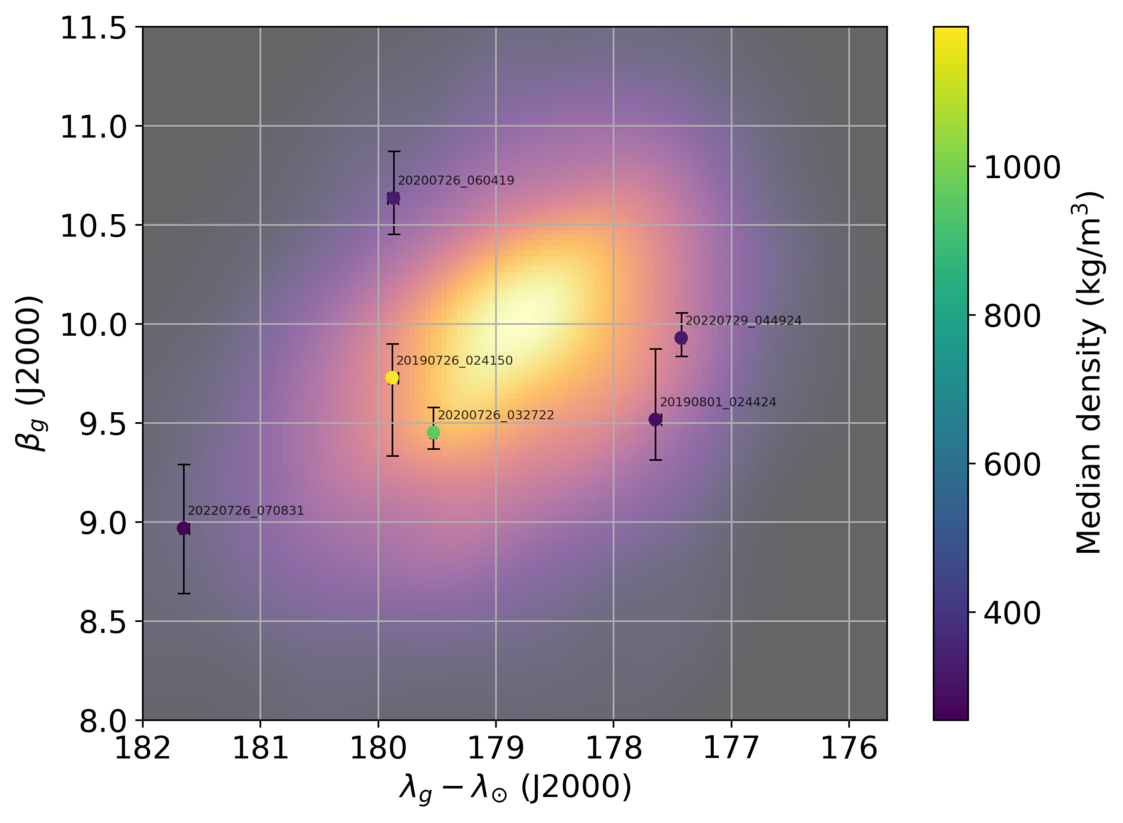}
    \caption{Sun-centered ecliptic radiant distribution for the 6 analyzed Alpha Capricornid meteors. The background density map shows the 3845 radiants from the GMN 2024 data, while the individual points with error bars represent the 1 sigma level of uncertainty for the meteors analyzed in this study with the median bulk density from the posterior distribution of the meteor.}
    \label{fig:CAP_radiants}
\end{figure}

This section presents the radiant locations of all EMCCD/CAMO observed Orionid and Capricornid meteors used in our study in Sun-centered ecliptic coordinates. Figures~\ref{fig:ORI_radiants} and~\ref{fig:CAP_radiants} display the Sun-centered ecliptic longitude and ecliptic latitude for each meteor radiant in our sample in comparison with the density of radiants from the GMN for the same showers. 

In each plot, the radiant locations derived from our observations are compared against the corresponding shower populations identified by the Global Meteor Network (GMN) in 2024. This comparison serves to validate that the meteors analyzed in this study are consistent in their orbital characteristics.

The plotted points for each meteor show their measured radiant positions with the error and their median bulk density, while the surrounding GMN reference data provide context for the typical spread and clustering of the respective showers.


\subsection{Orionids}

\newpage

\subsection{Capricornids}

\newpage

\section{SNR and photometric offset}\label{sec:Apx photometry}

Appendix A provides a comprehensive list of the SNR line fits for the three meteors recorded by two EMCCDs. The apparent shift in some events is attributed to the photometric offset at the specific moment the meteor was detected. The Capricornids and the Orionids were recorded in different years to capture potential variations over time. Because these meteors were detected on different dates, the photometric offset plot is displayed with dashed lines in two colors, indicating which meteor each offset refers to. The SNR fit curves and photometric offset plots are presented here, except for the Capricornid SNR line fit curve, which appears in the main text as Figure~\ref{img:CAPsnr_vs_mag}.

\subsection{Capricornids}

\begin{figure}[h!]
\centering

\begin{minipage}{0.8\linewidth}
    \centering
    \includegraphics[width=\linewidth]{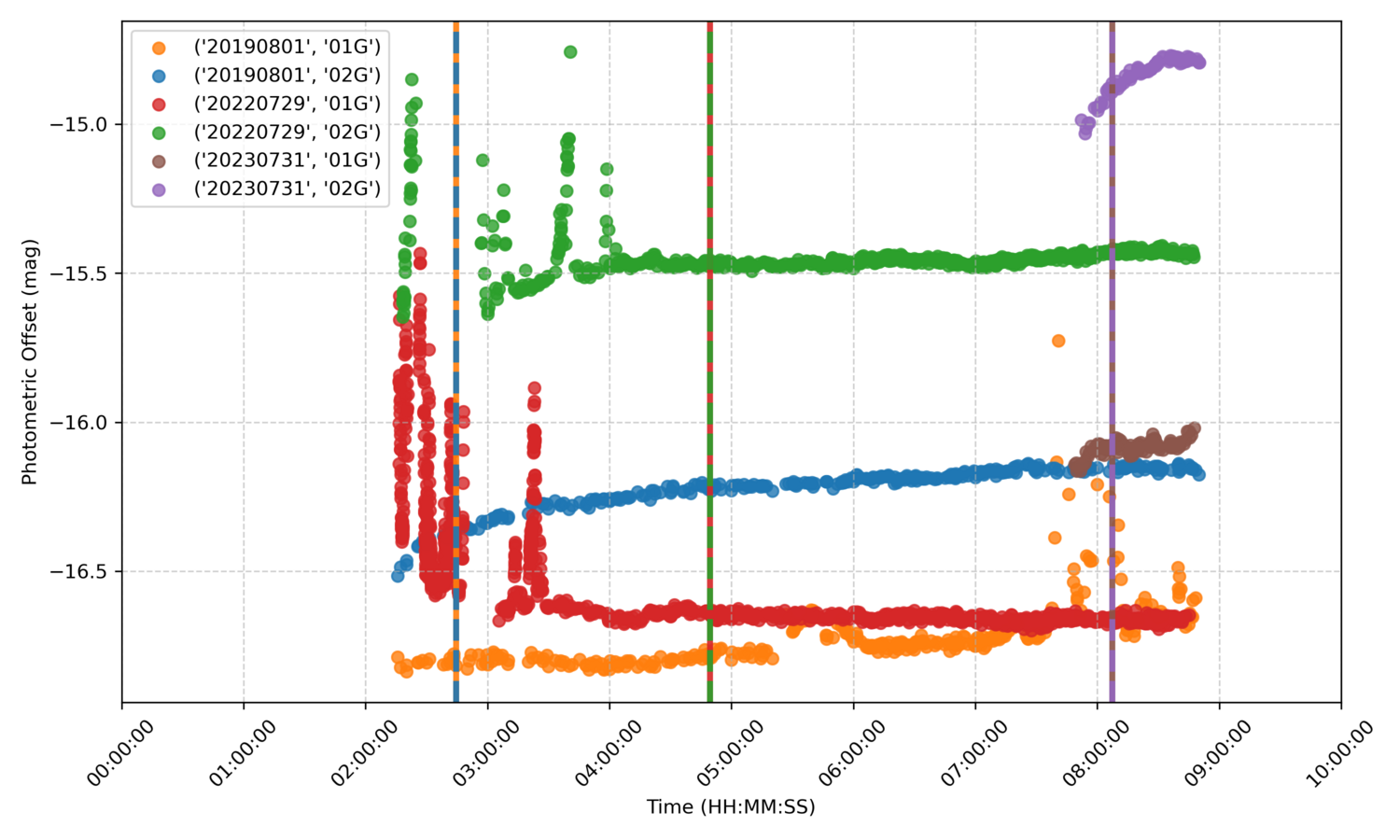}
\end{minipage}

\caption{Photometric offset curve using dashed lines in two colors to indicate the respective curves. These meteors were recorded in a different year to assess temporal variations.}
\label{img:DRAonlyphotometric_offset}
\end{figure}

\newpage

\subsection{Orionids}

\begin{figure}[h!]
\centering
\begin{minipage}{0.8\linewidth}
    \centering
    \includegraphics[width=\linewidth]{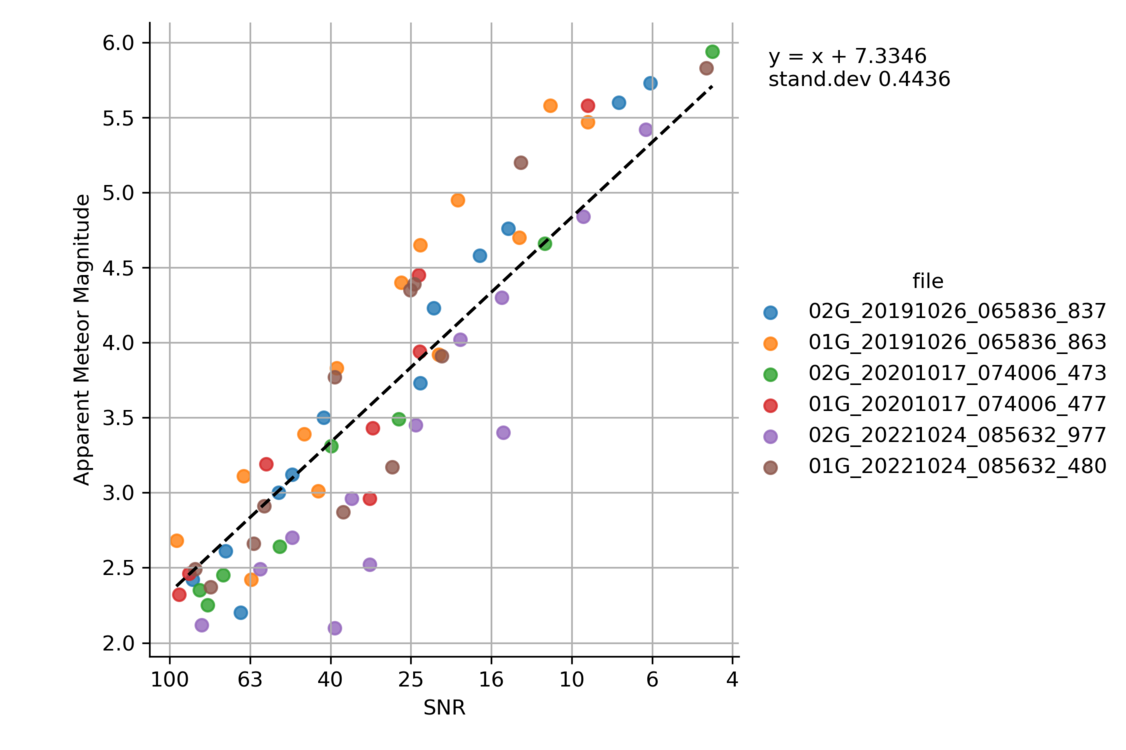}
\end{minipage}

\begin{minipage}{0.8\linewidth}
    \centering
    \includegraphics[width=\linewidth]{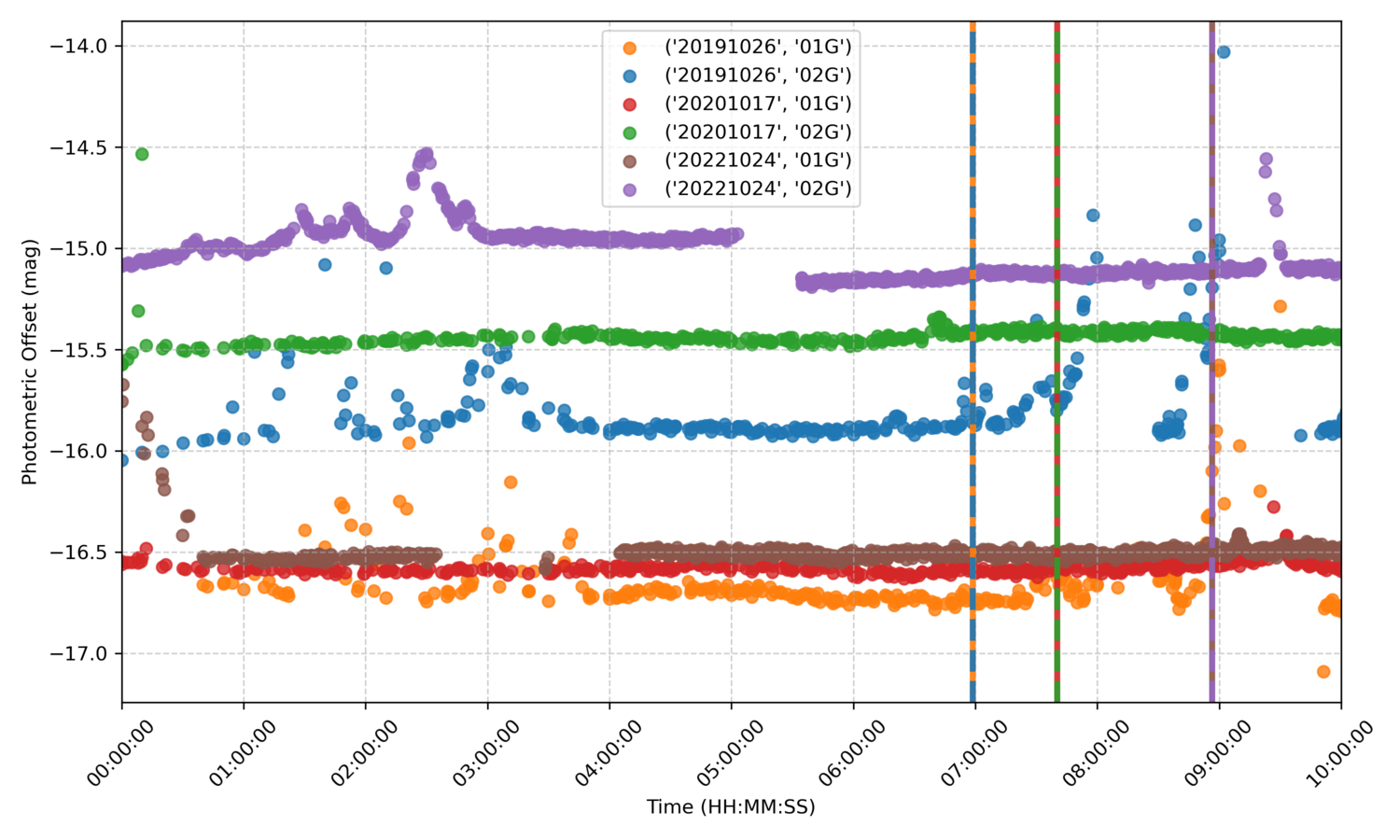}
\end{minipage}

\caption{Top: SNR line fit for the Orionids. Bottom: Photometric offset curve using dashed lines in two colors to indicate the respective curves. These meteors were recorded in a different year to capture interannual differences.}
\label{img:ORIsnr_vs_mag}
\end{figure}

\newpage

\section{Bayesian Evidence and Fragmentation Model Selection}
\label{sec:Apx bayes-evidence}

This appendix presents how the Bayesian evidence computed via dynamic nested sampling varies based on and single or double fragmentation model.

In the Bayesian approach, the model with higher evidence corresponds to a better physical explanation of the observed data (for instance, confirming the presence of multiple fragmentation events). Conversely, if the single-fragmentation and two-fragmentation models yield comparable evidences, it would suggest that the data quality is insufficient to justify the introduction of additional an additional fragmentation and model complexity. In this case, the posterior distributions of the second fragmentation parameters (i.e. $h_{e2},\rho_{2},\sigma_{2},\,\eta_{2}$), appear broad and poorly localized, often spanning the entire prior range—indicating that these additional parameters are effectively unnecessary.

\begin{figure}[h!]
\centering
\includegraphics[width=\linewidth]{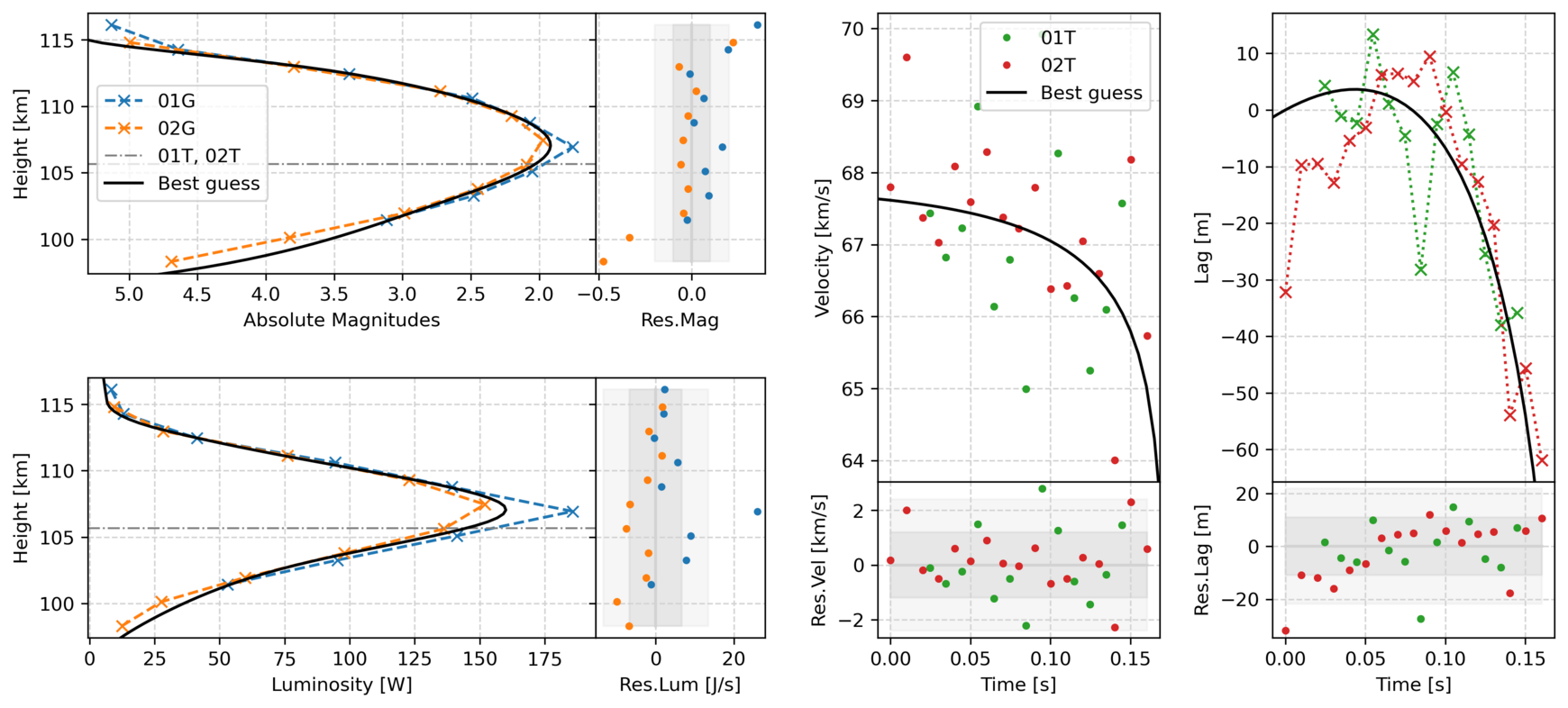}
\caption{Best guess simulation for meteor 20191023\_091225 (ORI) using a single-fragmentation model. This solution yielded the highest posterior probability in the dynamic nested sampling run.}
\label{fig:1frg_2019}
\end{figure}

As an example, Figure~\ref{fig:1frg_2019} shows the specific simulation of brightness and lag for the individual model run which \texttt{dynesty} identified as the highest-posterior-probability in matching meteor 20191023\_091225 using a single-fragmentation model. Note that while this represents the combination of parameters with the best evidence in fit, it is not necessarily the correct solution -  any set of erosion model parameters which match measurements within uncertainty are viable solutions.

\begin{figure}[h!]
\centering
\includegraphics[width=\linewidth]{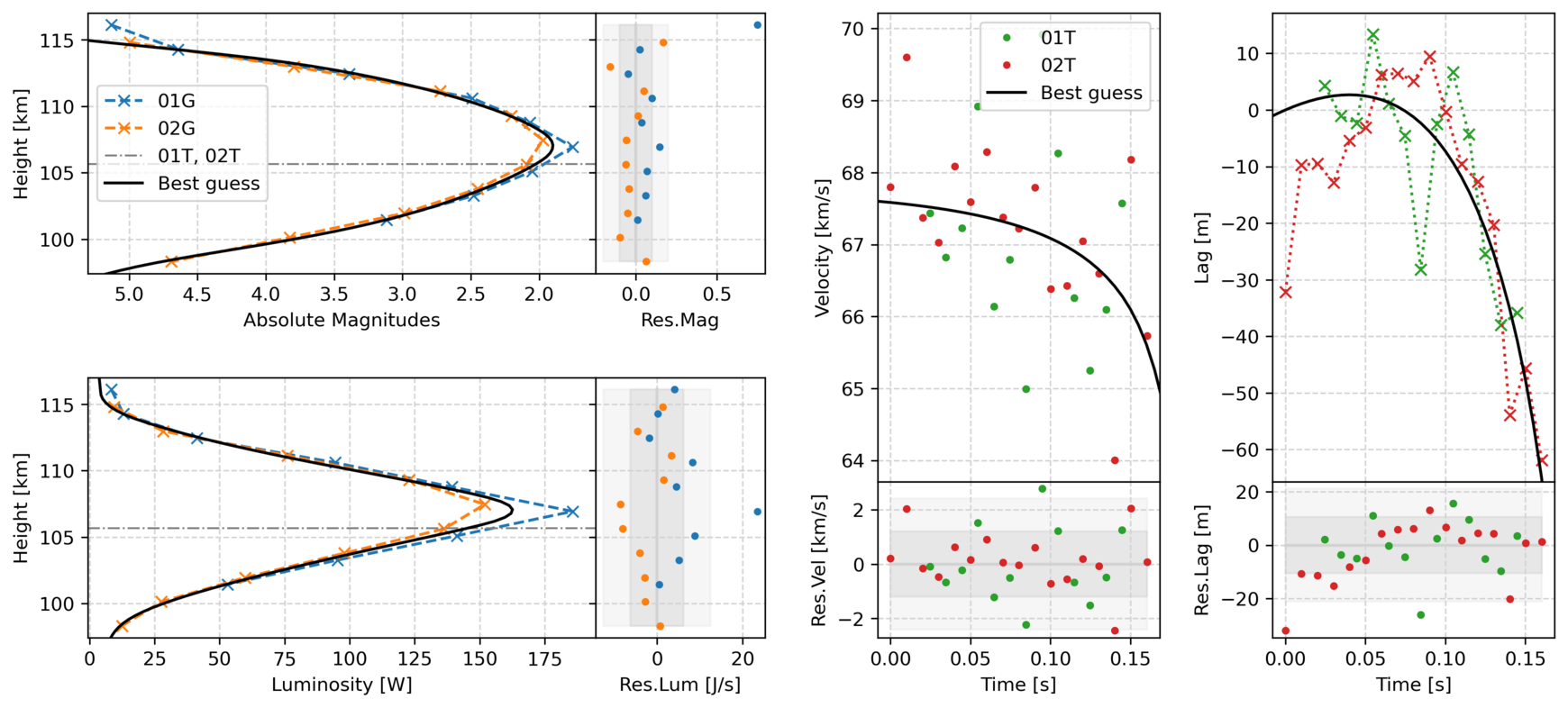}
\caption{The same meteor as  Figure~\ref{fig:1frg_2019}, but allowing a double-fragmentation. The added complexity results in broader posteriors and lower Bayesian evidence.}
\label{fig:2frg_2019}
\end{figure}

In contrast, Figure~\ref{fig:2frg_2019} presents the same meteor but allowing for a two-fragmentation solution. 

Dynamic nested sampling yields a Bayesian evidence of
\[
\log \mathcal{Z}_\text{1frg} = -521.10 \;\pm\; 0.15
\quad \text{vs.} \quad
\log \mathcal{Z}_\text{2frg} = -521.89 \;\pm\; 0.17.
\]

The lower evidence for the double-fragmentation model, despite its added complexity, indicates no significant gain in explanatory power. 

\begin{figure}[h!]
\centering
\begin{minipage}{0.3\linewidth}
\centering
\textbf{105 km}\\
\includegraphics[width=\linewidth]{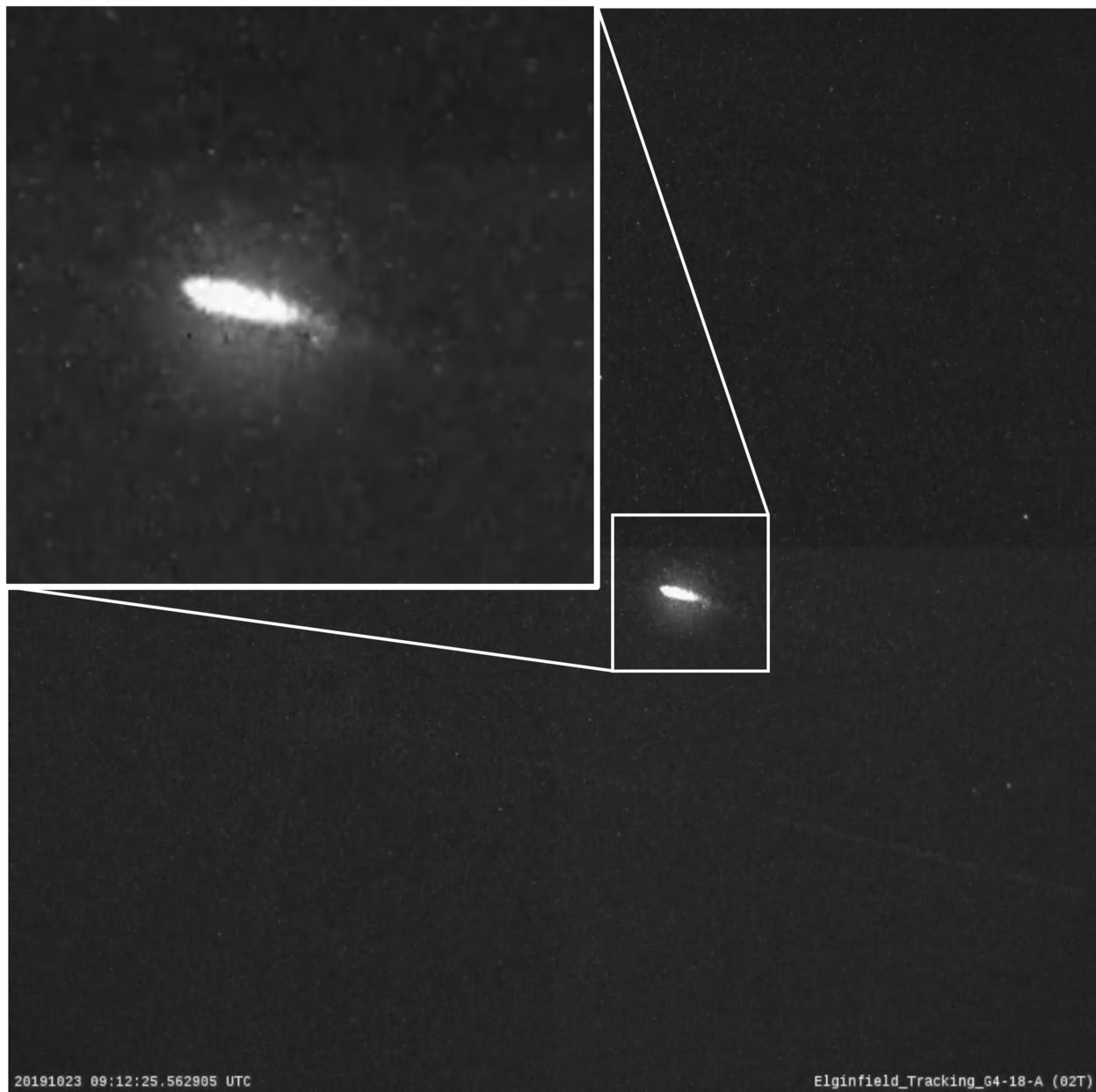}
\end{minipage}
\begin{minipage}{0.3\linewidth}
\centering
\textbf{103 km}\\
\includegraphics[width=\linewidth]{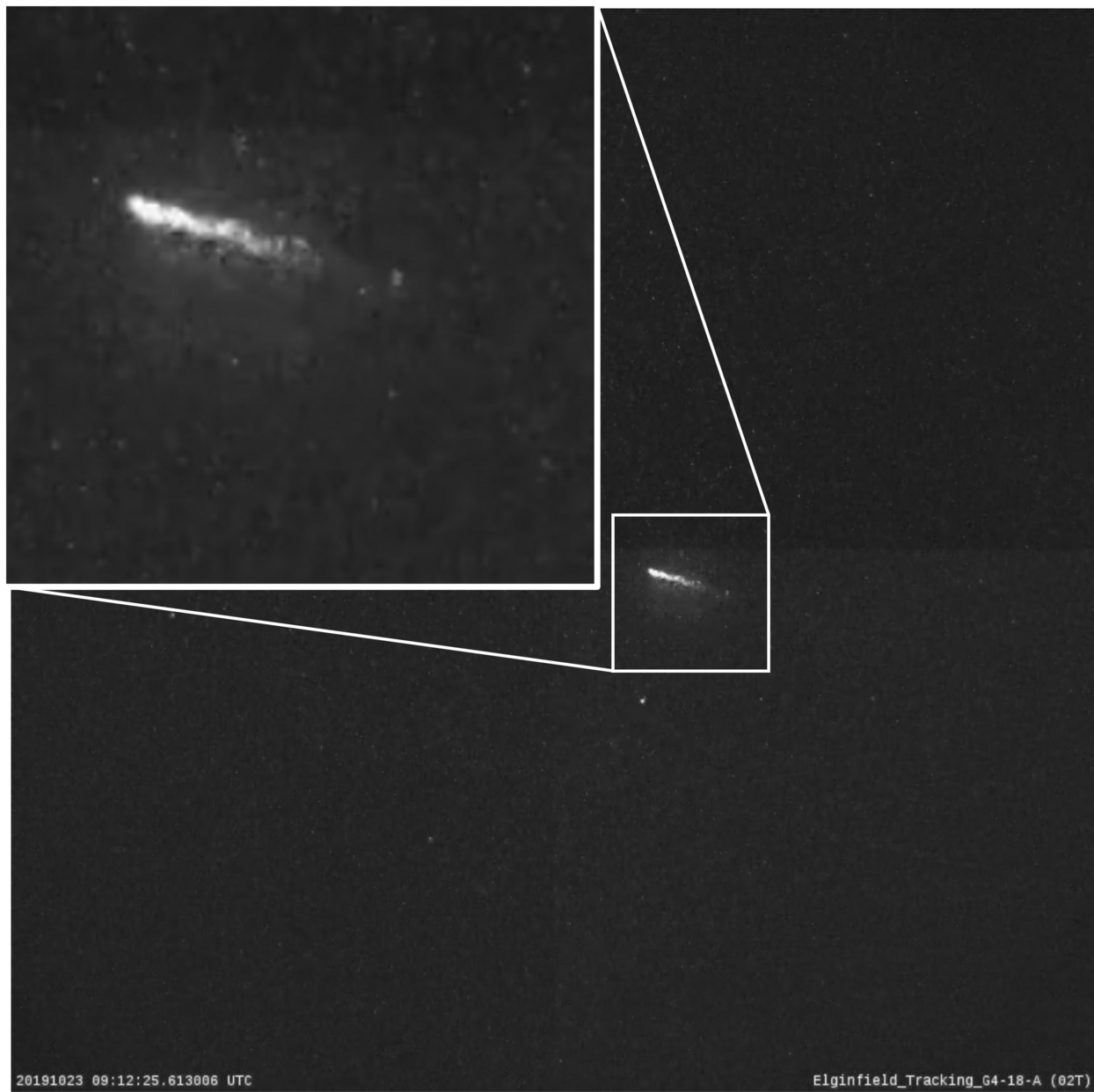}
\end{minipage}
\begin{minipage}{0.3\linewidth}
\centering
\textbf{101 km}\\
\includegraphics[width=\linewidth]{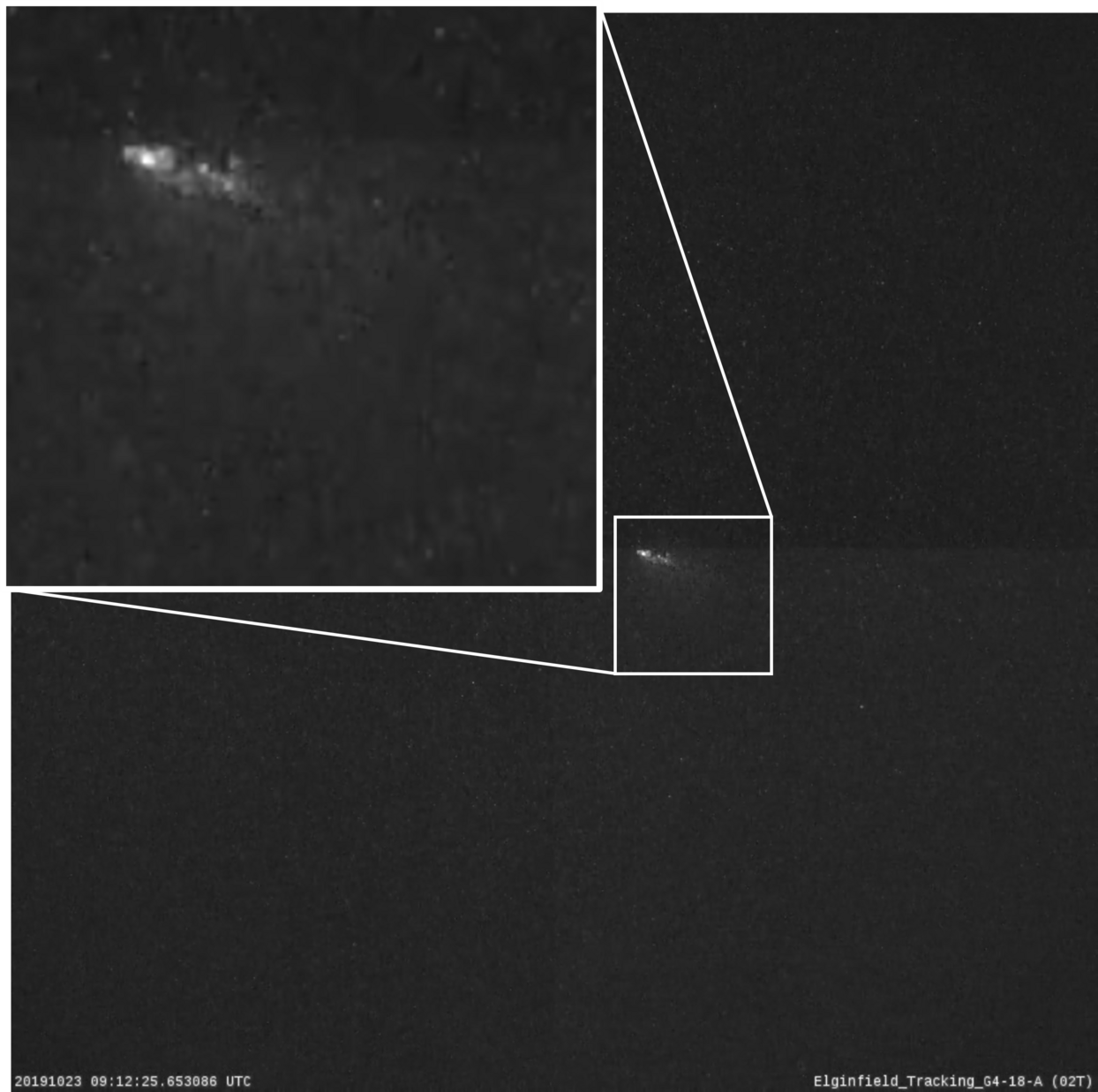}
\end{minipage}
\caption{Three consecutive frames showing meteor 20191023\_091225 (ORI) as imaged by CAMO station 02T (Elginfield). The images show the primary luminous mass fading rapidly, with no surviving fragment visible beyond the main ablation phase. The meteor had a speed of 67.7 km/s, a peak brightness of $+1.7$ absolute magnitude, a photometric mass of $6.5\times10^{-6}$ kg, and a luminous trajectory spanning altitudes from 116 km to 98 km.}
\label{fig:1frg_CAMO}
\end{figure}

For this case, the CAMO observations clearly support a single fragmentation event: the high resolution imagery shows no distinct secondary fragment as shown in Figure~\ref{fig:1frg_CAMO}. The one-fragmentation scenario thus provides both an observationally consistent and statistically preferred solution.

\begin{figure}[h!]
\centering
\includegraphics[width=\linewidth]{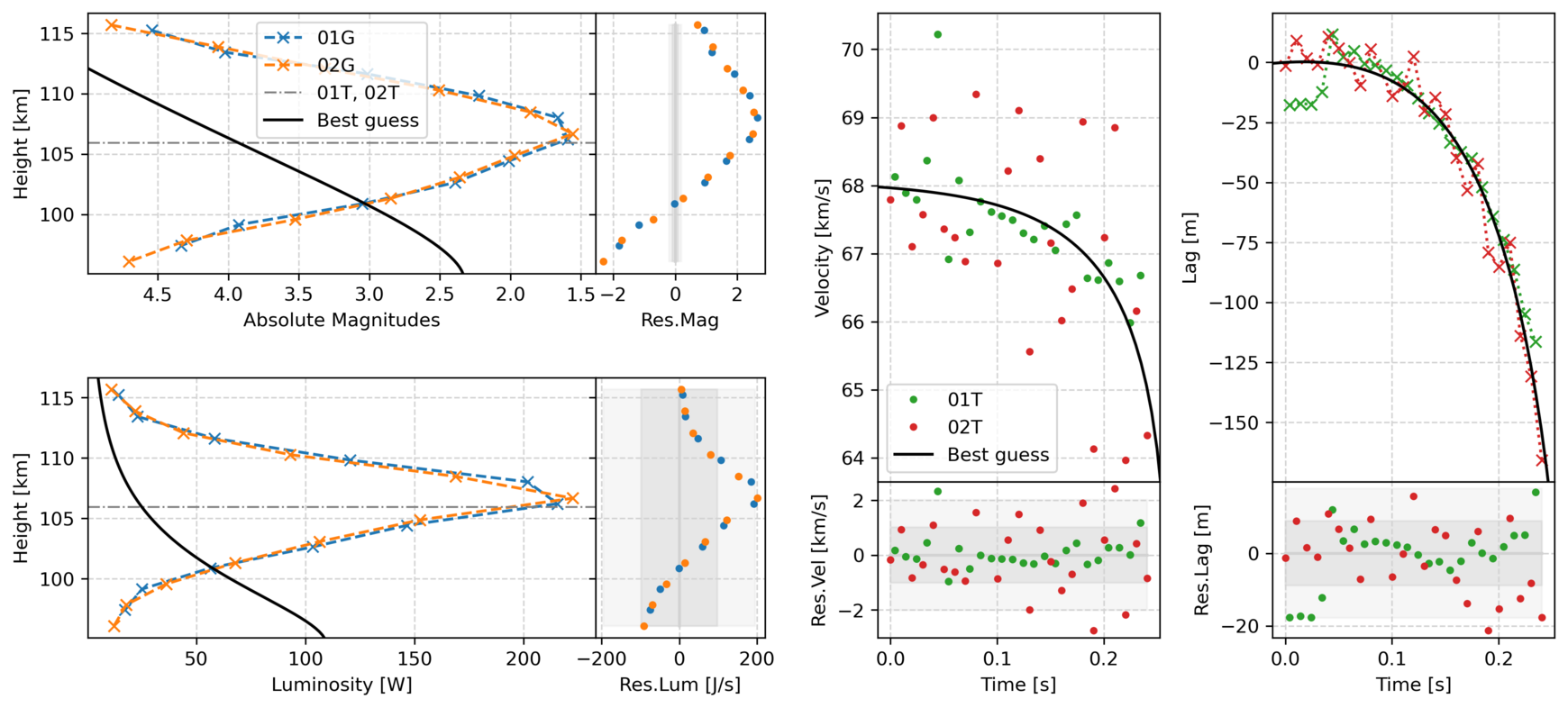}
\caption{Best-fit simulation for meteor 20191023\_084916 (ORI) using a single-fragmentation model. The fit fails to simultaneously match both luminosity and lag, resulting in a lower Bayesian evidence.}
\label{fig:1frg_2019b}
\end{figure}

\begin{figure}[htbp]
\centering
\includegraphics[width=\linewidth]{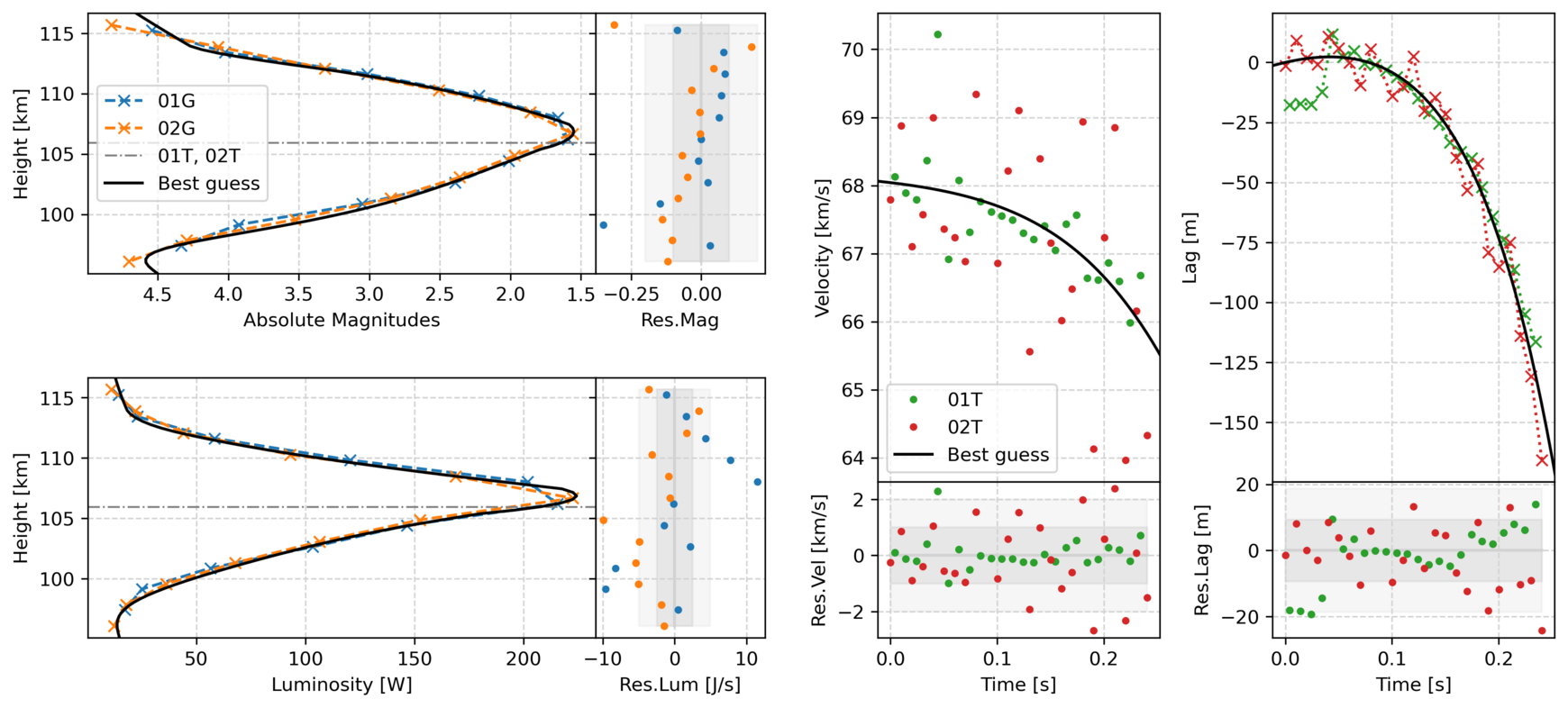}
\caption{Same meteor as in Fig.~\ref{fig:1frg_2019b}, but using a two-fragmentation model. The fit aligns well with both luminosity and lag observations and yields significantly higher evidence.}
\label{fig:2frg_2019b}
\end{figure}

A contrasting case is shown in Figures~\ref{fig:1frg_2019b} and \ref{fig:2frg_2019b}, corresponding to meteor 20191023\_084916. In this example, the single-fragmentation model completely fails to simultaneously reproduce the luminosity and lag. Because the log-likelihood combines contributions from both lag and luminosity, and the model can adaptively adjust the noise parameters based on the data as it is part of the priors, the dynamic nested sampling algorithm favors fitting the better-constrained lag curve. This results in an effective down-weighting of the poorly matched luminosity data. The Bayesian evidence for this fit is:

\[
\log \mathcal{Z}_\text{1frg} = -495.17 \;\pm\; 0.14.
\]

In contrast, the two-fragmentation model achieves a much better match to the observed data in both luminosity and lag. The fit quality is reflected in a substantially higher Bayesian evidence:
\[
\log \mathcal{Z}_\text{2frg} = -396.83 \;\pm\; 0.25.
\]

This large increase in evidence, almost a factor of 100,demonstrates that the added complexity is not only justified but essential to explain the data. 

\begin{figure}[h!]
\centering
\begin{minipage}{0.3\linewidth}
\centering
\textbf{106 km}\\
\includegraphics[width=\linewidth]{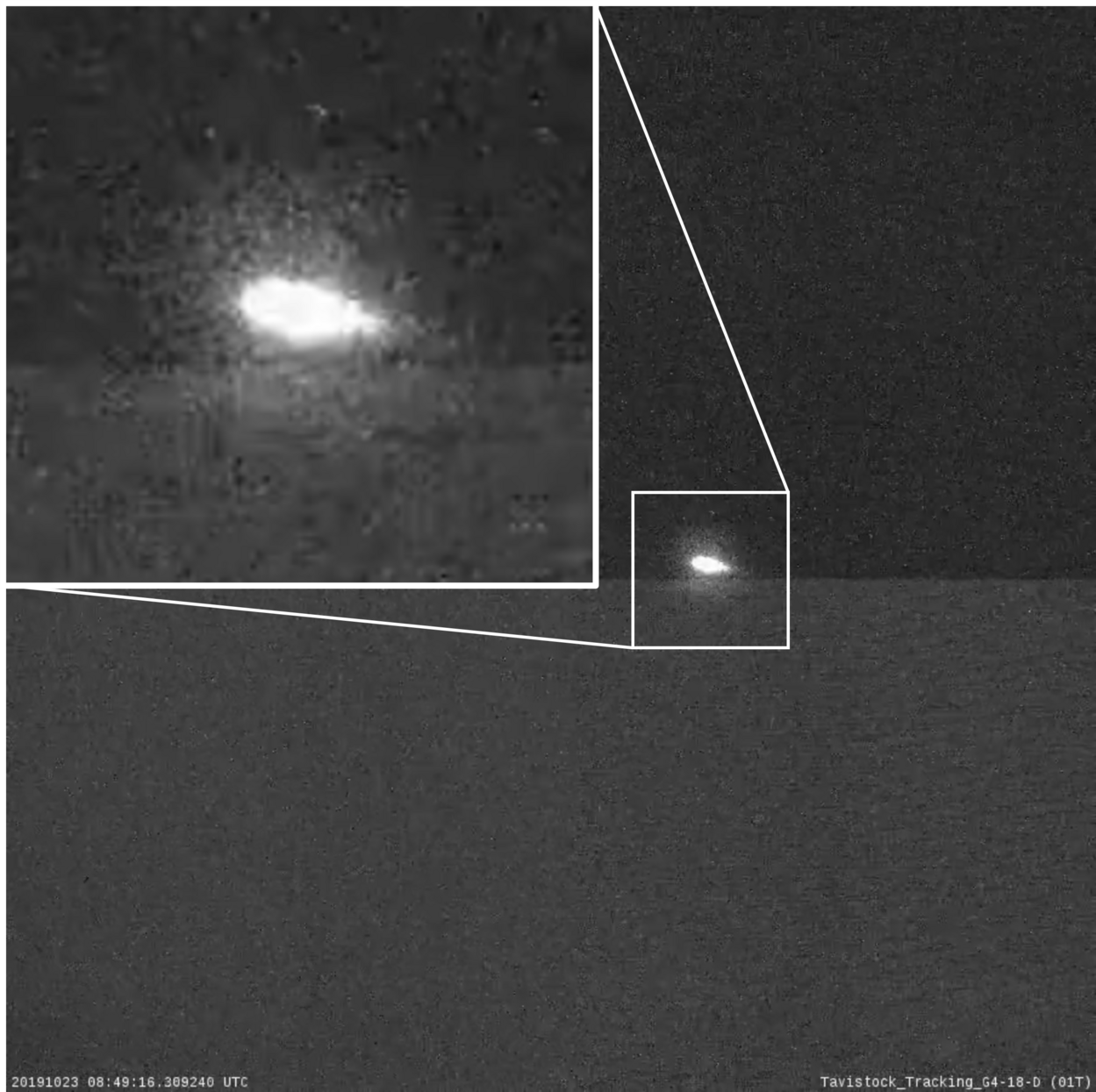}
\end{minipage}
\begin{minipage}{0.3\linewidth}
\centering
\textbf{102 km}\\
\includegraphics[width=\linewidth]{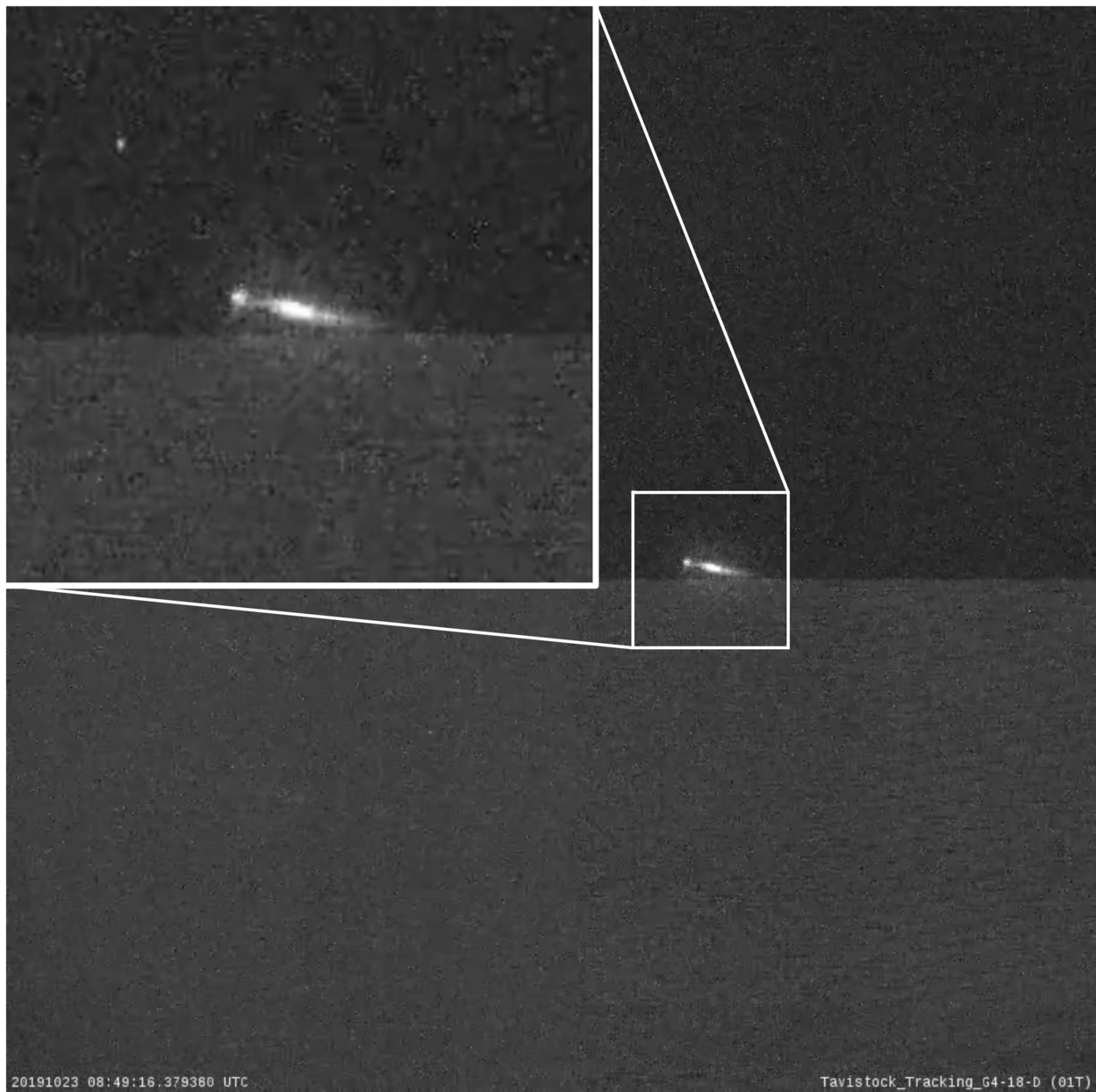}
\end{minipage}
\begin{minipage}{0.3\linewidth}
\centering
\textbf{99 km}\\
\includegraphics[width=\linewidth]{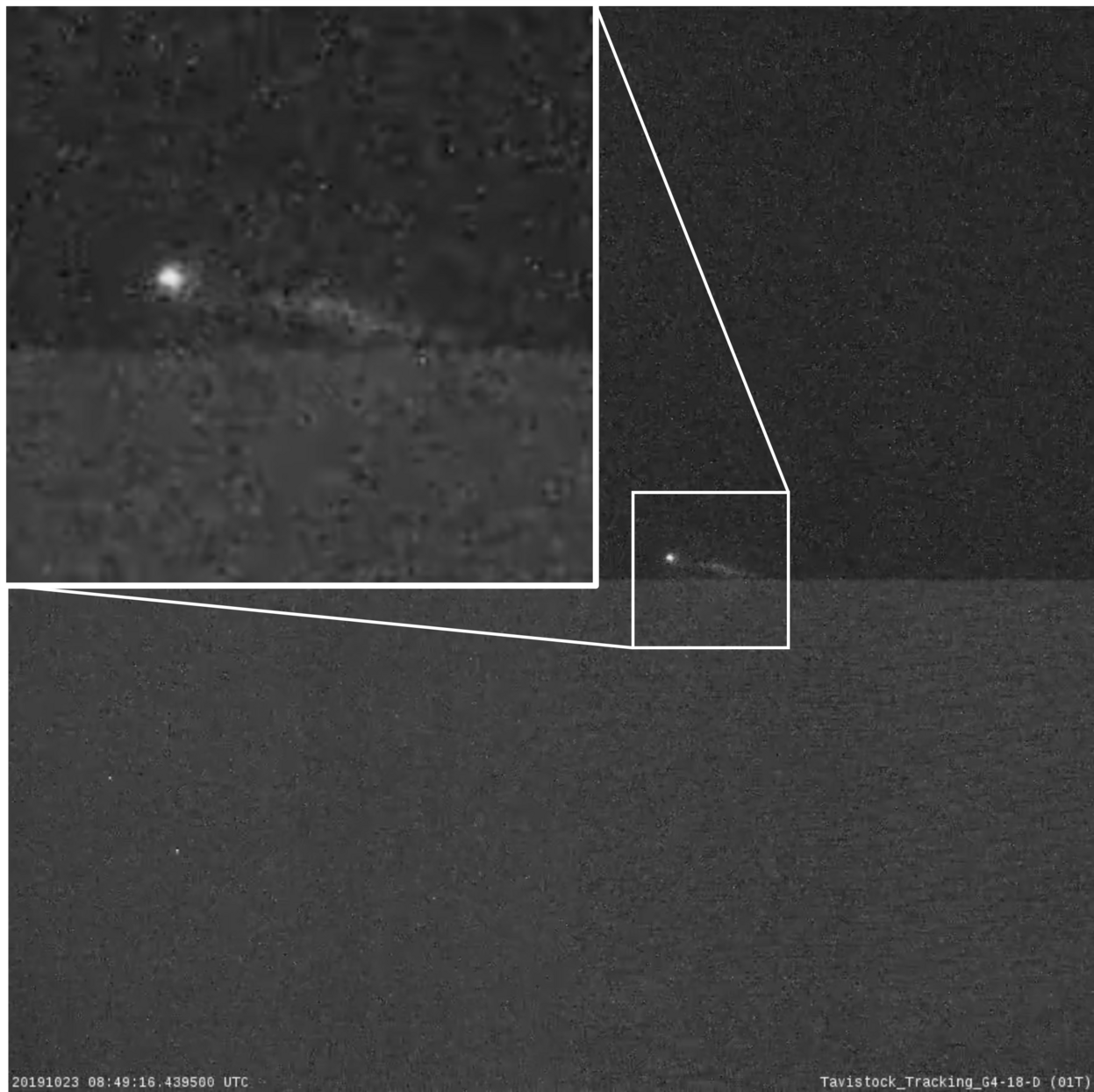}
\end{minipage}
\caption{Three consecutive frames showing meteor 20191023\_084916 (ORI) as imaged by CAMO station 01T (Tavistock). The images reveal the primary, diffuse luminous mass fading rapidly, while a smaller fragment with no wake continues along the trajectory. This is consistent with a second, denser core surviving after the main ablation phase. The meteor had a speed of 68.1 km/s, a peak brightness of $+1.5$ absolute magnitude, a photometric mass of $2.3\times10^{-6}$ kg, and a luminous trajectory spanning altitudes from 115 km to 96 km. Here the meteor is moving from the lower right to upper left, though it is nearly stationary as mirrors are tracking its motion.}
\label{fig:2frg_CAMO}
\end{figure}

This interpretation is further supported by CAMO observations. The high resolution tracking imagery clearly shows a rapid drop in luminosity following the primary ablation phase, followed by appearance of a smaller, persistent star-like fragment continuing along the trajectory as shown in Figure~\ref{fig:2frg_CAMO}. These observations are consistent with the expected behavior of a second, more refractory fragment that separates and survives longer, reinforcing the two-component structure hypothesis for some meteoroids proposed by \citet{vida2024first}.


Although we selected the fragmentation model for each meteor based on visual inspection of CAMO narrow-field camera data—choosing either a single or double fragmentation scenario accordingly—we also performed full DNS runs for both models across all events. As outlined in Section~\ref{subsec:evidence importance}, the Bayesian evidence ($\log \mathcal{Z}$) provides a statistically grounded metric to evaluate which model better explains the observed data.

\begin{table}[htbp]
    \centering
    \renewcommand{\arraystretch}{1.2}
    \setlength{\tabcolsep}{4pt}
    \caption{Summary of ORI $\log \mathcal{Z}$ values for 1frag and 2frag models. The Difference column shows the difference of the logarithm of the two evidences. The last column indicates whether CAMO clearly showed a second fragmentation event. }
    \label{tab:frag_logz_summary_ORI}
    \resizebox{\textwidth}{!}{%
    \begin{tabular}{lrrrr}
    \hline
    Meteor & $\log \mathcal{Z}_{1frag}$ & $\log \mathcal{Z}_{2frag}$ & $\Delta \log \mathcal{Z}$ & 2fr \\
    \hline
    20191023\_084916 & $-495.165 \pm 0.137$ & $-396.834 \pm 0.249$ & $-98.331$ & Yes\\
    20191023\_091225 & $-521.098 \pm 0.151$ & $-521.890 \pm 0.168$ & $+0.792$ & No\\
    20191023\_091310 & $-664.724 \pm 0.247$ & $-645.544 \pm 0.204$ & $-19.180$ & Yes\\
    20191026\_065838 & $-320.479 \pm 0.188$ & $-306.609 \pm 0.245$ & $-13.870$ & Yes\\
    20191028\_050616 & $-1738.635 \pm 0.190$ & $-1715.955 \pm 0.197$ & $-22.680$ & No\\
    20201012\_081716 & $-632.905 \pm 0.190$ & $-611.436 \pm 0.255$ & $-21.469$ & No\\
    20201017\_074007 & $-581.276 \pm 0.172$ & $-554.619 \pm 0.270$ & $-26.657$ & No\\
    20221022\_075829 & $-387.349 \pm 0.217$ & $-385.494 \pm 0.183$ & $-1.855$ & No\\
    20221022\_081606 & $-299.406 \pm 0.206$ & $-277.511 \pm 0.211$ & $-21.895$ & No\\
    \hline
    \end{tabular}}
\end{table}

\begin{table}[htbp]
    \centering
    \renewcommand{\arraystretch}{1.2}
    \setlength{\tabcolsep}{4pt}
    \caption{Summary of CAP $\log \mathcal{Z}$ values for 1frag and 2frag models. The Difference column shows the difference of the logarithm of the two evidences. The last column indicates whether CAMO clearly showed a second fragmentation event. }
    \label{tab:frag_logz_summary_CAP}
    \resizebox{\textwidth}{!}{%
    \begin{tabular}{lrrrr}
    \hline
    Meteor & $\log \mathcal{Z}_{1frag}$ & $\log \mathcal{Z}_{2frag}$ & $\Delta \log \mathcal{Z}$ & 2fr \\
    \hline
    20190726\_024150 & $-693.243 \pm 0.195$ & $-676.746 \pm 0.305$ & $-16.497$ & No\\
    20190801\_024424 & $-708.588 \pm 0.228$ & $-682.678 \pm 0.214$ & $-25.910$ & No\\
    20200726\_032722 & $-533.255 \pm 0.169$ & $-479.885 \pm 0.208$ & $-53.370$ & No\\
    20200726\_060419 & $-497.086 \pm 0.209$ & $-419.312 \pm 0.188$ & $-77.774$ & Yes\\
    20220726\_070831 & $-359.010 \pm 0.189$ & $-347.187 \pm 0.232$ & $-11.823$ & No\\
    20220729\_044924 & $-598.664 \pm 0.215$ & $-523.599 \pm 0.241$ & $-75.065$ & Yes\\
    \hline
    \end{tabular}}
\end{table}

To assess model adequacy, we examined the difference in log evidence between the single and double fragmentation models, defined as $\Delta \log \mathcal{Z} = \log \mathcal{Z}_{\text{1frag}} - \log \mathcal{Z}_{\text{2frag}}$. The results, summarized in Tables~\ref{tab:frag_logz_summary_ORI} and \ref{tab:frag_logz_summary_CAP}, reveal distinct trends.

For Orionids (ORI), the fragmentation classification was generally straightforward: CAMO imagery frequently revealed distinct primary fragments, aiding model selection. In contrast, the Capricornids (CAP) presented a more challenging case. CAMO narrow-field observations showed that many CAP meteors disintegrated into complex “string of pearls” morphologies, making it difficult to determine whether a second fragmentation occurred, particularly in the final stages of ablation. As a result, the visual classification between single and double fragmentation was more uncertain for the CAP, and the evidence-based model selection became especially valuable in distinguishing cases where the morphological cues were ambiguous.


\newpage

\section{Validation Plots and Tables}\label{sec:Apx EMCCD validat}

This appendix compliments the validation results for four synthetic test cases representative of Orionid (ORI) and Capricornid (CAP) meteors, using both EMCCD ligh-curve and CAMO lag noise configurations and only EMCCD noise. Each test case was generated for two parameter sets: mode and mean, as summarized in Table~\ref{tab:test_parameters}.

For every case, we show:
\begin{itemize}
  \item the best-fit (“Best Guess”) simulation, defined as the synthetic meteor realization corresponding to the highest likelihood obtained from the dynamic nested sampling runs. In the light-curve plots, this is shown as a thick black line;
  \item the true (input) simulation without any introduced observational noise, shown as a thin dotted black line;
  \item the synthetic “camera” data, which correspond to the noisy observations generated by adding EMCCD-like or CAMO-like measurement noise to the true simulation. These are the data fed into the model to test its inversion capability;
  \item the residuals between the noisy synthetic data (camera points) and the model fits, shown as colored dots, while the residuals relative to the noiseless simulation are shown in black;
  \item a summary table listing the 95\% credible intervals, best-guess, mode, mean, median, and true input values, together with the absolute and relative errors between the best-guess solution and the true values.
\end{itemize}

To avoid repetition, captions are abbreviated. “CAMO+EMCCD” refers to test cases using EMCCD light-curve noise and CAMO lag noise to generate the synthetic data, while “EMCCD only” refers to tests where EMCCD noise was applied to both light-curve and lag to generate the synthetic meteors.

\newpage

\subsection{ORI mode EMCCD with CAMO lag test case}

\begin{figure}[ht]
    \centering
    \includegraphics[width=\linewidth]{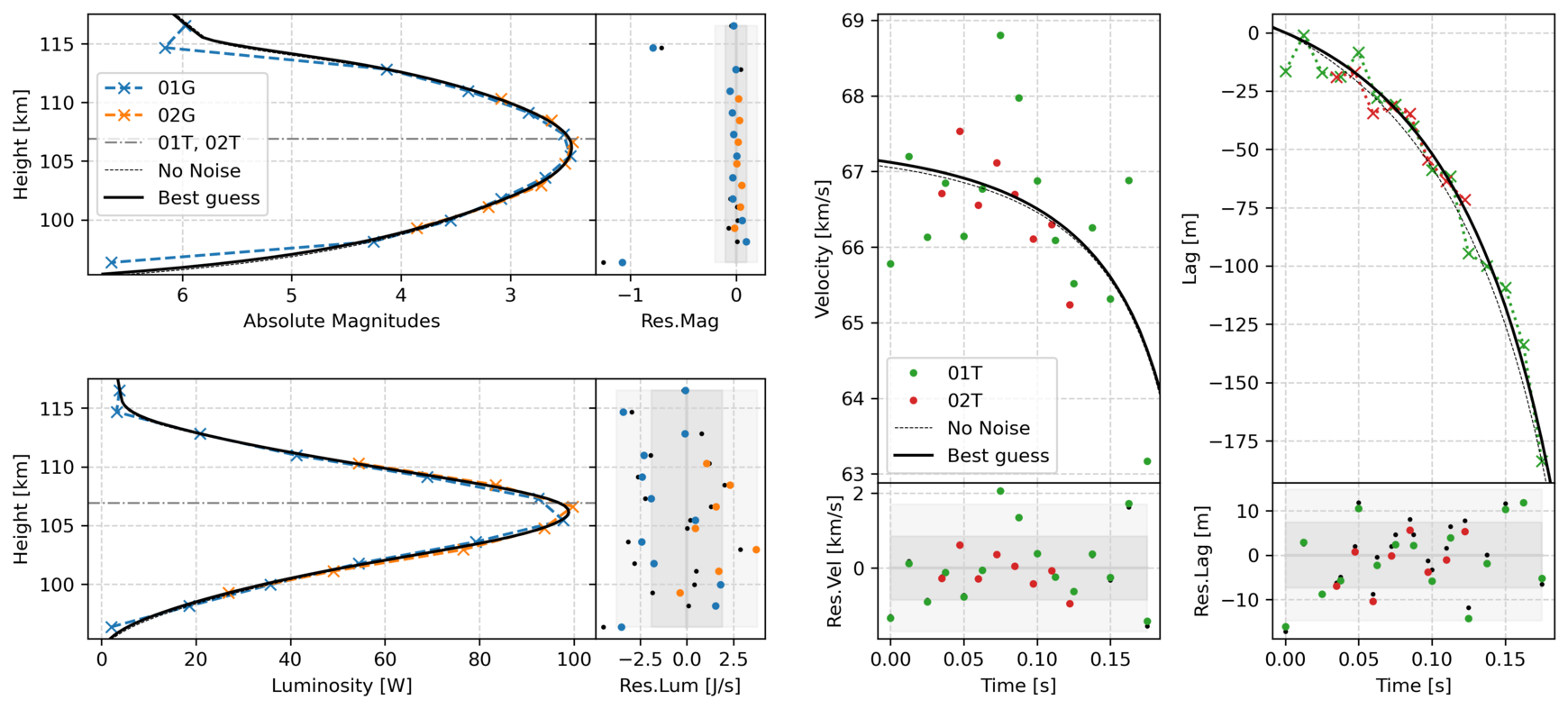}
    \caption{Orionid mode test case — CAMO+EMCCD synthetic data.}
    \label{img:ORI_mode_best}
\end{figure}

\begin{table}[h!]
    \centering
    \renewcommand{\arraystretch}{1.2}
    \setlength{\tabcolsep}{4pt}
    \caption{Posterior summary for Orionid mode — CAMO+EMCCD.}
    \label{tab:posterior_summary_ORI_mode}
    \resizebox{\textwidth}{!}{
    \begin{tabular}{lrrrrrrrrr}
    \hline
    Parameter & 2.5CI & True Value & Best Guess & Mode & Mean & Median & 97.5CI & Abs.Error & Rel.Error\%\\
    \hline
    $v_0$ [km/s]             & 67.12   & 67.38     & 67.47      & 67.20 & 67.25 & 67.23 & 67.47    & 0.082 & 0.12 \\
    $m_0$ [kg]               & 1.435$\times10^{-6}$ & 1.687$\times10^{-6}$ & 1.654$\times10^{-6}$ & 1.592$\times10^{-6}$ & 1.634$\times10^{-6}$ & 1.606$\times10^{-6}$ & 1.976$\times10^{-6}$ & 3.38$\times10^{-8}$ & 2.01 \\
    $\rho$ [kg/m$^3$]        & 101.39   & 134.6     & 122.8      & 536.8 & 469.4 & 491.5 & 1691.0   & 11.79    & 8.76 \\
    $\sigma$ [kg/MJ]         & 0.02012 & 0.02167   & 0.02005    & 0.03478 & 0.03434 & 0.03430 & 0.04831 & 0.00162 & 7.47 \\
    $h_e$ [km]               & 115.1   & 115.7     & 116.1      & 116.6 & 116.1 & 116.1 & 117.1    & 0.38     & 0.33 \\
    $\eta$ [kg/MJ]           & 0.1196  & 0.1725    & 0.1562     & 0.4020 & 0.3837 & 0.3984 & 0.9383  & 0.0163   & 9.45 \\
    $s$                       & 1.785   & 2.143     & 2.140      & 2.146 & 1.995 & 1.993 & 2.225    & 0.0025   & 0.12 \\
    $m_{l}$ [kg]             & 6.17$\times10^{-12}$ & 1.325$\times10^{-11}$ & 1.147$\times10^{-11}$ & 5.325$\times10^{-11}$ & 2.712$\times10^{-11}$ & 2.786$\times10^{-11}$ & 1.149$\times10^{-10}$ & 1.79$\times10^{-12}$ & 13.47 \\
    $m_{u}$ [kg]             & 3.09$\times10^{-9}$ & 5.083$\times10^{-9}$ & 3.184$\times10^{-9}$ & 1.745$\times10^{-8}$ & 1.138$\times10^{-8}$ & 1.038$\times10^{-8}$ & 5.866$\times10^{-8}$ & 1.90$\times10^{-9}$ & 37.36 \\
    $\sigma_{\mathrm{lag}}$ [m] & 6.026   & 7.368     & 7.368      & 9.796 & 7.917 & 7.796 & 10.530   & 0 & 0.00 \\
    $\sigma_{\mathrm{lum}}$ [W] & 1.804   & 1.889     & 1.889      & 3.531 & 2.601 & 2.522 & 3.869    & 0 & 0.00 \\
    \hline
    \end{tabular}}
\end{table}

\newpage

\subsection{ORI mean EMCCD with CAMO lag test case}

\begin{figure}[ht]
    \centering
    \includegraphics[width=\linewidth]{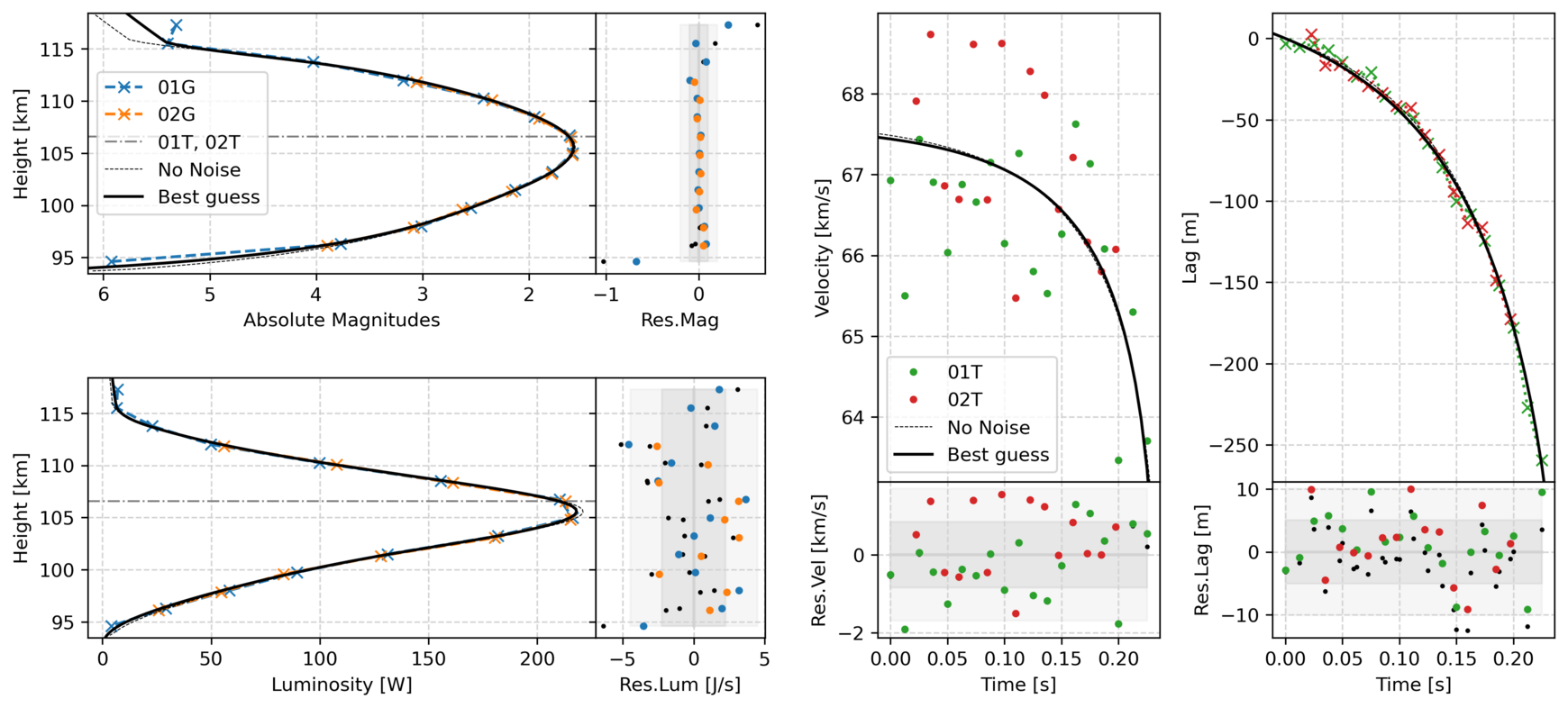}
    \caption{Orionid mean test case — CAMO+EMCCD synthetic data.} 
    \label{img:ORI_mean_best}
\end{figure}

\begin{table}[h!]
    \centering
    \renewcommand{\arraystretch}{1.2}
    \setlength{\tabcolsep}{4pt}
    \caption{Posterior summary for Orionid mean — CAMO+EMCCD.}
    \label{tab:posterior_summary_ORI_mean}
    \resizebox{\textwidth}{!}{
    \begin{tabular}{lrrrrrrrrr}
    \hline
    Parameter & 2.5CI & True Value & Best Guess & Mode & Mean & Median & 97.5CI & Abs.Error & Rel.Error\%\\
    \hline
    $v_0$ [km/s]             & 67.65  & 67.73     & 67.68      & 67.67 & 67.68 & 67.68 & 67.73 & 0.046  & 0.07 \\
    $m_0$ [kg]               & 3.592$\times10^{-6}$ & 3.778$\times10^{-6}$ & 3.969$\times10^{-6}$ & 4.010$\times10^{-6}$ & 4.060$\times10^{-6}$ & 4.084$\times10^{-6}$ & 4.420$\times10^{-6}$ & 1.90$\times10^{-7}$ & 5.04 \\
    $\rho$ [kg/m$^3$]        & 103.58  & 195.2     & 159.7      & 178.2 & 162.1 & 156.8 & 344.1 & 35.5   & 18.18 \\
    $\sigma$ [kg/MJ]         & 0.01938 & 0.02033   & 0.02353    & 0.02492 & 0.02459 & 0.02493 & 0.02801 & 0.00320 & 15.73 \\
    $h_e$ [km]               & 115.6  & 116.2     & 116.1      & 115.9 & 116.1 & 116.1 & 116.7 & 0.11    & 0.09 \\
    $\eta$ [kg/MJ]           & 0.1377 & 0.2281    & 0.2033     & 0.2224 & 0.2008 & 0.1962 & 0.3485  & 0.0248  & 10.89 \\
    $s$                       & 2.089  & 2.131     & 2.125      & 2.182 & 2.150 & 2.151 & 2.206   & 0.00625 & 0.29 \\
    $m_{l}$ [kg]             & 1.212$\times10^{-11}$ & 1.390$\times10^{-11}$ & 2.178$\times10^{-11}$ & 3.248$\times10^{-11}$ & 2.605$\times10^{-11}$ & 2.857$\times10^{-11}$ & 3.970$\times10^{-11}$ & 7.88$\times10^{-12}$ & 56.64 \\
    $m_{u}$ [kg]             & 1.460$\times10^{-8}$ & 1.538$\times10^{-8}$ & 2.075$\times10^{-8}$ & 5.641$\times10^{-8}$ & 3.579$\times10^{-8}$ & 3.886$\times10^{-8}$ & 6.558$\times10^{-8}$ & 5.36$\times10^{-9}$ & 34.88 \\
    $\sigma_{\mathrm{lag}}$ [m] & 4.576  & 5.022     & 5.022      & 5.748 & 5.743 & 5.680 & 7.197    & 0 & 0.00 \\
    $\sigma_{\mathrm{lum}}$ [W] & 1.747  & 2.232     & 2.232      & 3.369 & 2.624 & 2.535 & 4.123    & 0 & 0.00 \\
    \hline
    \end{tabular}}
\end{table}

\newpage

\subsection{CAP mode EMCCD with CAMO lag test case}

\begin{figure}[ht]
    \centering
    \includegraphics[width=\linewidth]{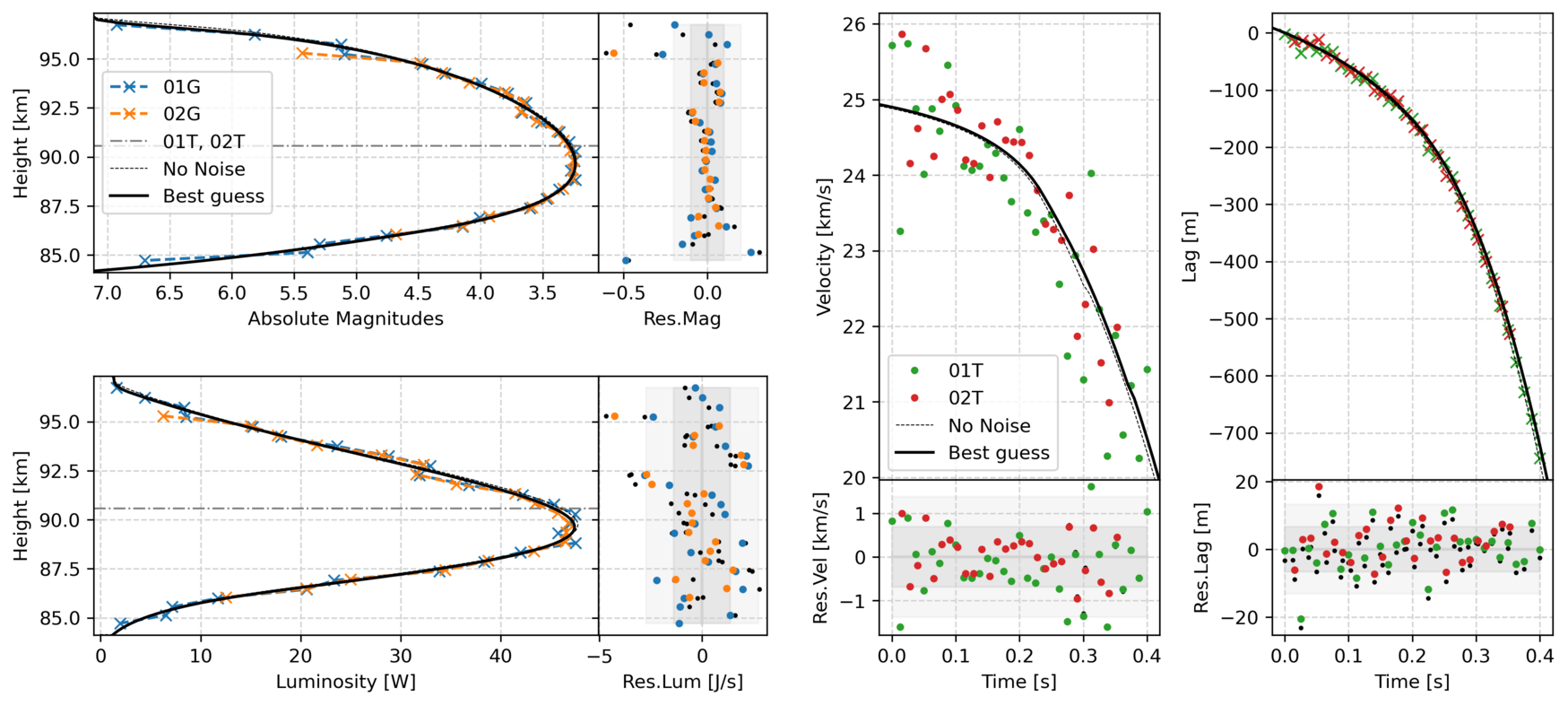}
    \caption{Capricornid mode test case — CAMO+EMCCD synthetic data.} 
    \label{img:CAP_mode_best}
\end{figure}

\begin{table}[h!]
    \caption{Posterior summary for Capricornid mode — CAMO+EMCCD.}
    \label{tab:posterior_summary_CAP_mode}
    \centering
    \renewcommand{\arraystretch}{1.2} 
    \setlength{\tabcolsep}{4pt} 
    \resizebox{\textwidth}{!}{ 
    \begin{tabular}{lrrrrrrrrr}
    \hline
    Parameter & 2.5CI & True Value & Best Guess & Mode & Mean & Median & 97.5CI & Abs.Error & Rel.Error\% \\
    \hline
    $v_0$ [km/s]        & 25.22        & 25.36         & 25.37      & 25.32  & 25.37 & 25.37   & 25.57        & 0.01828    & 0.07  \\
    $m_0$ [kg]          & 3.531$\times10^{-6}$    & 3.667$\times10^{-6}$     & 3.722$\times10^{-6}$  & 3.676$\times10^{-6}$ & 3.718$\times10^{-6}$ & 3.692$\times10^{-6}$ & 4.14$\times10^{-6}$     & 5.463$\times10^{-8}$  & 1.49     \\
    $\rho$ [kg/m$^3$]   & 668.8        & 1125          & 1079       & 1276   & 1092  & 1097    & 1611         & 46.37      & 4.12    \\
    $\sigma$ [kg/MJ]    & 0.01736      & 0.0215        & 0.02286    & 0.02374 & 0.02264 & 0.02255 & 0.02834      & 0.001364   & 6.34    \\
    $h_e$ [km]          & 96.88        & 97.53         & 97.27      & 97.73  & 97.35 & 97.32   & 97.88        & 0.2596     & 0.26   \\
    $\eta$ [kg/MJ]      & 0.07227      & 0.135         & 0.13       & 0.1537 & 0.1287 & 0.1313  & 0.1918       & 0.004957   & 3.67    \\
    $s$                 & 1.502        & 2.05          & 1.715      & 2.059  & 1.996 & 1.978   & 2.56         & 0.3346     & 16.32    \\
    $m_{l}$ [kg]        & 5.475$\times10^{-12}$    & 3.042$\times10^{-11}$     & 6.759$\times10^{-12}$  & 8.092$\times10^{-11}$ & 2.743$\times10^{-11}$ & 2.885$\times10^{-11}$ & 1.681$\times10^{-10}$  & 2.366$\times10^{-11}$  & 77.78    \\
    $m_{u}$ [kg]        & 2.479$\times10^{-9}$    & 5.9$\times10^{-9}$       & 4.16$\times10^{-9}$   & 9.438$\times10^{-9}$ & 6.774$\times10^{-9}$ & 5.194$\times10^{-9}$ & 5.846$\times10^{-8}$  & 1.74$\times10^{-9}$   & 29.49    \\
    $\sigma_{lag}$ [m] & 5.727   & 6.586         & 6.586      & 6.802  & 6.68  & 6.617   & 7.994        & 0          & 0        \\
    $\sigma_{lum}$ [W] & 1.367   & 1.37          & 1.37       & 1.42   & 1.674 & 1.658   & 2.1          & 0          & 0        \\
    \hline
    \end{tabular}}
\end{table}

\newpage

\subsection{CAP mean EMCCD with CAMO lag test case}

\begin{figure}[ht]
    \centering
    \includegraphics[width=\linewidth]{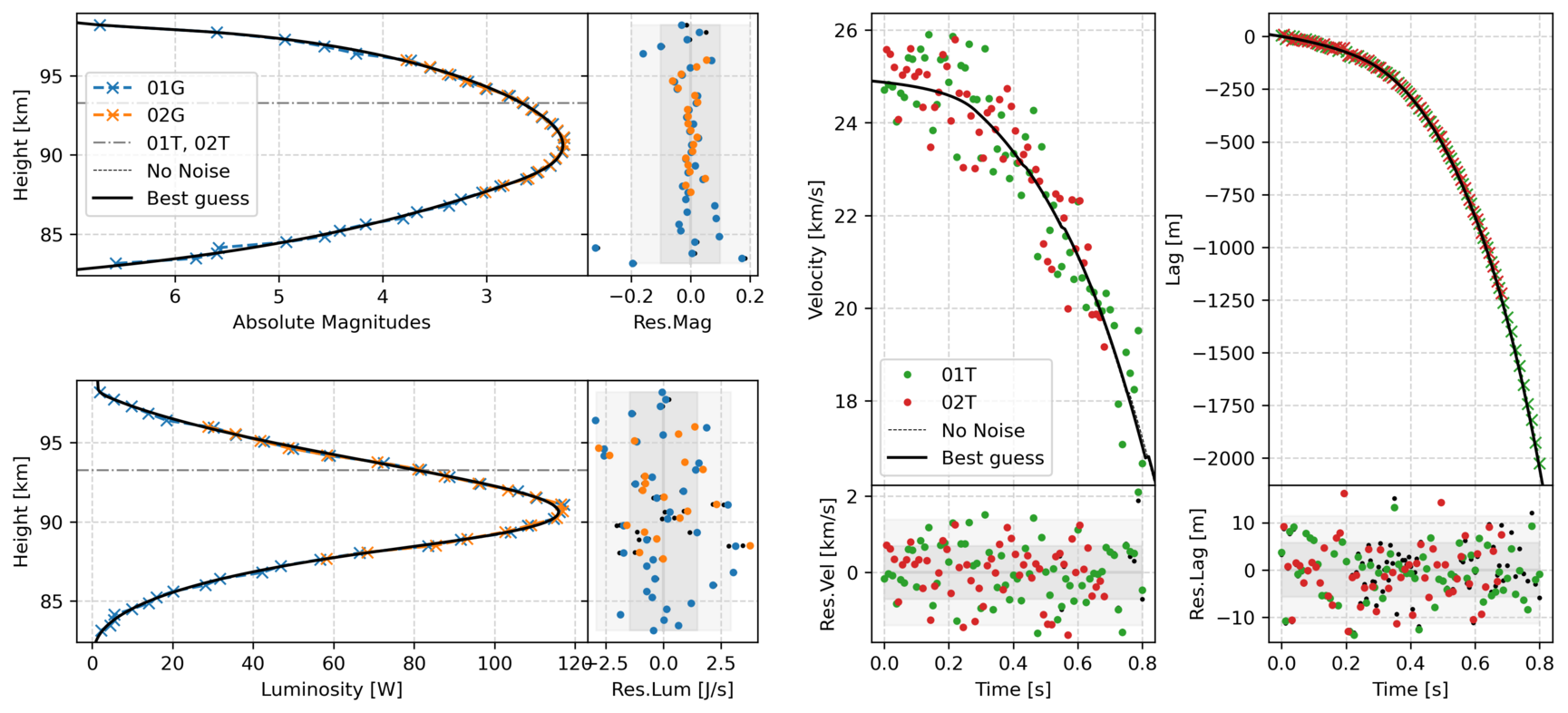}
    \caption{Capricornid mean test case — CAMO+EMCCD synthetic data.}
    \label{img:CAP_mean_best}
\end{figure}

\begin{table}[h!]
    \centering
    \renewcommand{\arraystretch}{1.2} 
    \setlength{\tabcolsep}{4pt} 
    \caption{Posterior summary for Capricornid mean — CAMO+EMCCD.}
    \resizebox{\textwidth}{!}{ 
    \begin{tabular}{lrrrrrrrrr}
    \hline
    Parameter & 2.5CI & True Value & Best Guess & Mode & Mean & Median & 97.5CI & Abs.Error & Rel.Error\% \\
    \hline
    $v_0$ [km/s] & 25.07 & 25.13 & 25.12 & 25.13 & 25.14 & 25.12 & 25.21 & 0.007177 & 0.02 \\
    $m_0$ [kg] & 1.007$\times10^{-5}$ & 1.025$\times10^{-5}$ & 1.021$\times10^{-5}$ & 1.028$\times10^{-5}$ & 1.023$\times10^{-5}$ & 1.022$\times10^{-5}$ & 1.039$\times10^{-5}$ & 3.853$\times10^{-8}$ & 0.37 \\
    $\rho$ [kg/m$^3$] & 664.7 & 875 & 920.6 & 878.8 & 862.5 & 906.8 & 1077 & 45.6 & 5.21 \\
    $\sigma$ [kg/MJ] & 0.01697 & 0.01844 & 0.01839 & 0.01855 & 0.01831 & 0.01833 & 0.0193 & 4.655$\times10^{-5}$ & 0.25 \\
    $h_e$ [km] & 98.29 & 98.5 & 98.58 & 98.45 & 98.48 & 98.48 & 98.63 & 0.08151 & 0.08 \\
    $\eta$ [kg/MJ] & 0.1906 & 0.2444 & 0.2553 & 0.2488 & 0.2408 & 0.2511 & 0.2896 & 0.01097 & 4.48 \\
    $s$ & 1.985 & 2.098 & 2.106 & 2.02 & 2.07 & 2.101 & 2.135 & 0.007965 & 0.37 \\
    $m_{l}$ [kg] & 3.895$\times10^{-11}$ & 7.438$\times10^{-11}$ & 9.003$\times10^{-11}$ & 7.073$\times10^{-11}$ & 6.833$\times10^{-11}$ & 7.558$\times10^{-11}$ & 9.354$\times10^{-11}$ & 1.565$\times10^{-11}$ & 21.04 \\
    $m_{u}$ [kg] & 1.417$\times10^{-8}$ & 1.825$\times10^{-8}$ & 1.902$\times10^{-8}$ & 1.509$\times10^{-8}$ & 1.707$\times10^{-8}$ & 1.819$\times10^{-8}$ & 1.944$\times10^{-8}$ & 7.726$\times10^{-10}$ & 4.23 \\
    $\sigma_{lag}$ [m] & 5.339 & 5.689 & 5.689 & 6.651 & 6.057 & 6.051 & 6.813 & 0 & 0 \\
    $\sigma_{lum}$ [W] & 1.396 & 1.466 & 1.466 & 2.034 & 1.698 & 1.682 & 2.079 & 0 & 0 \\
    \hline

    \end{tabular}} 
    \label{tab:posterior_summary_CAP_mean}
\end{table}

\newpage

\subsection{ORI mode EMCCD only test case}

\begin{figure}[ht]
    \centering
    \includegraphics[width=\linewidth]{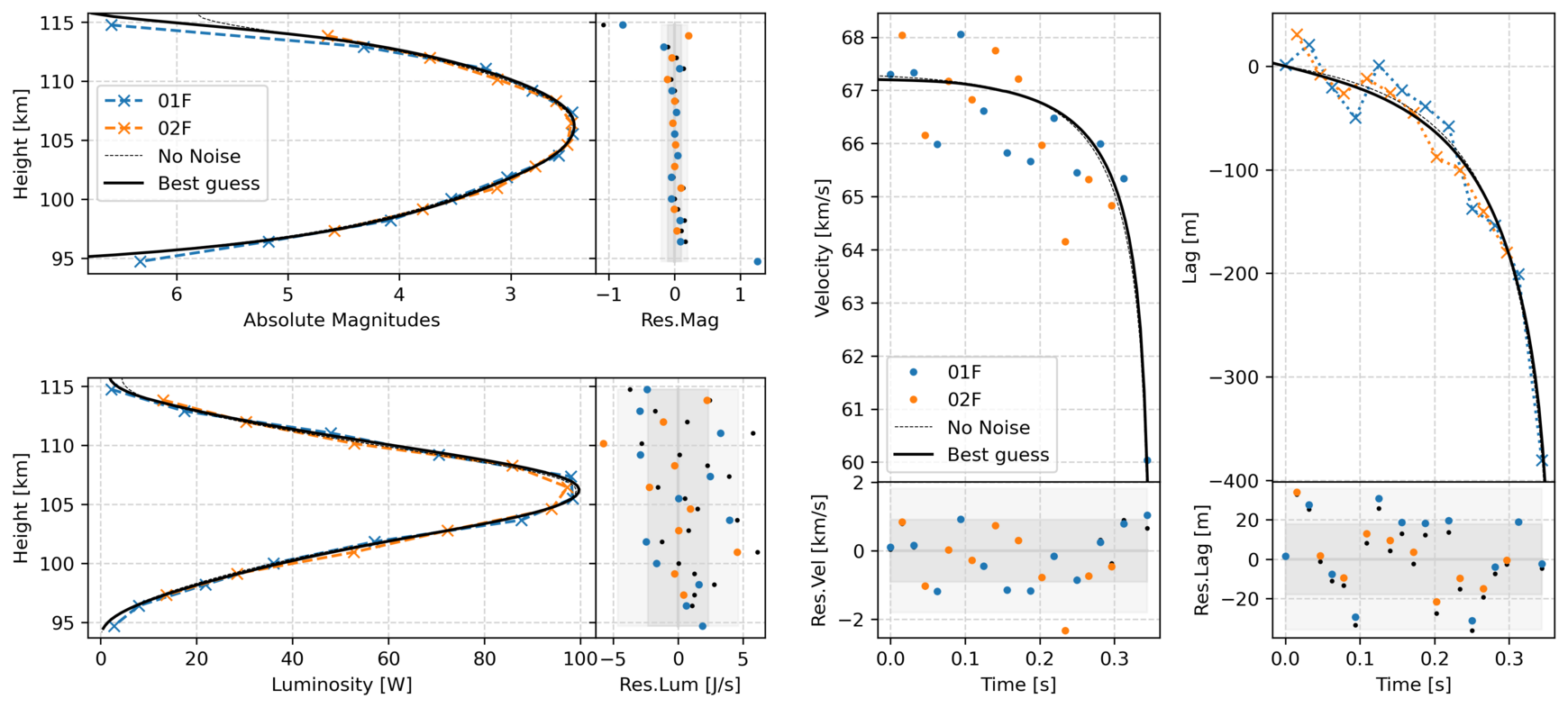}
    \caption{Posterior summary for ORI mode — EMCCD-only test case.}

    \label{img:ORI_mode_best_EMCCD}
\end{figure}

\begin{table}[h!]
    \centering
    \caption{Posterior summary for ORI mode — EMCCD-only test case.}

    \renewcommand{\arraystretch}{1.2}
    \setlength{\tabcolsep}{4pt}
    \resizebox{\textwidth}{!}{
    \begin{tabular}{lrrrrrrrrr}
    \hline
    Parameter & 2.5CI & True Value & Best Guess & Mode & Mean & Median & 97.5CI & Abs.Error & Rel.Error\% \\
    \hline
    $v_0$ [km/s] & 67.18 & 67.38 & 67.25 & 67.29 & 67.3 & 67.28 & 67.5 & 0.1354 & 0.20 \\
    $m_0$ [kg] & 1.408$\times10^{-6}$ & 1.687$\times10^{-6}$ & 1.514$\times10^{-6}$ & 1.95$\times10^{-6}$ & 1.63$\times10^{-6}$ & 1.591$\times10^{-6}$ & 2.025$\times10^{-6}$ & 1.737$\times10^{-7}$ & 10.29 \\
    $\rho$ [kg/m$^3$] & 54.06 & 134.6 & 519.5 & 125 & 262.4 & 265 & 1143 & 384.9 & 285.9 \\
    $\sigma$ [kg/MJ] & 0.01486 & 0.02167 & 0.02269 & 0.02788 & 0.022 & 0.02176 & 0.03054 & 0.001027 & 4.73 \\
    $h_e$ [km] & 115 & 115.7 & 116.1 & 115.1 & 116 & 116.1 & 116.9 & 0.3767 & 0.32 \\
    $\eta$ [kg/MJ] & 0.08734 & 0.1725 & 0.4986 & 0.1679 & 0.2966 & 0.3038 & 0.8481 & 0.3261 & 189 \\
    $s$ & 2.031 & 2.143 & 2.119 & 2.095 & 2.163 & 2.156 & 2.328 & 0.02345 & 1.09 \\
    $m_{l}$ [kg] & 5.601$\times10^{-12}$ & 1.325$\times10^{-11}$ & 1.774$\times10^{-11}$ & 2.208$\times10^{-11}$ & 1.484$\times10^{-11}$ & 1.438$\times10^{-11}$ & 4.809$\times10^{-11}$ & 4.487$\times10^{-12}$ & 33.86 \\
    $m_{u}$ [kg] & 2.285$\times10^{-9}$ & 5.083$\times10^{-9}$ & 5.669$\times10^{-9}$ & 1.309$\times10^{-8}$ & 6.786$\times10^{-9}$ & 6.028$\times10^{-9}$ & 4.564$\times10^{-8}$ & 5.853$\times10^{-10}$ & 11.51 \\
    $\sigma_{lag}$ [m] & 15.48 & 17.85 & 17.85 & 20.97 & 20.32 & 20.02 & 27.04 & 0 & 0 \\
    $\sigma_{lum}$ [W] & 2.03 & 2.323 & 2.323 & 2.93 & 2.794 & 2.742 & 3.899 & 0 & 0 \\
    \hline
    \end{tabular}}
    \label{tab:posterior_summary_EMCCD_ORI_mode}
\end{table}

\newpage

\subsection{ORI mean EMCCD only test case}

\begin{figure}[ht]
    \centering
    \includegraphics[width=\linewidth]{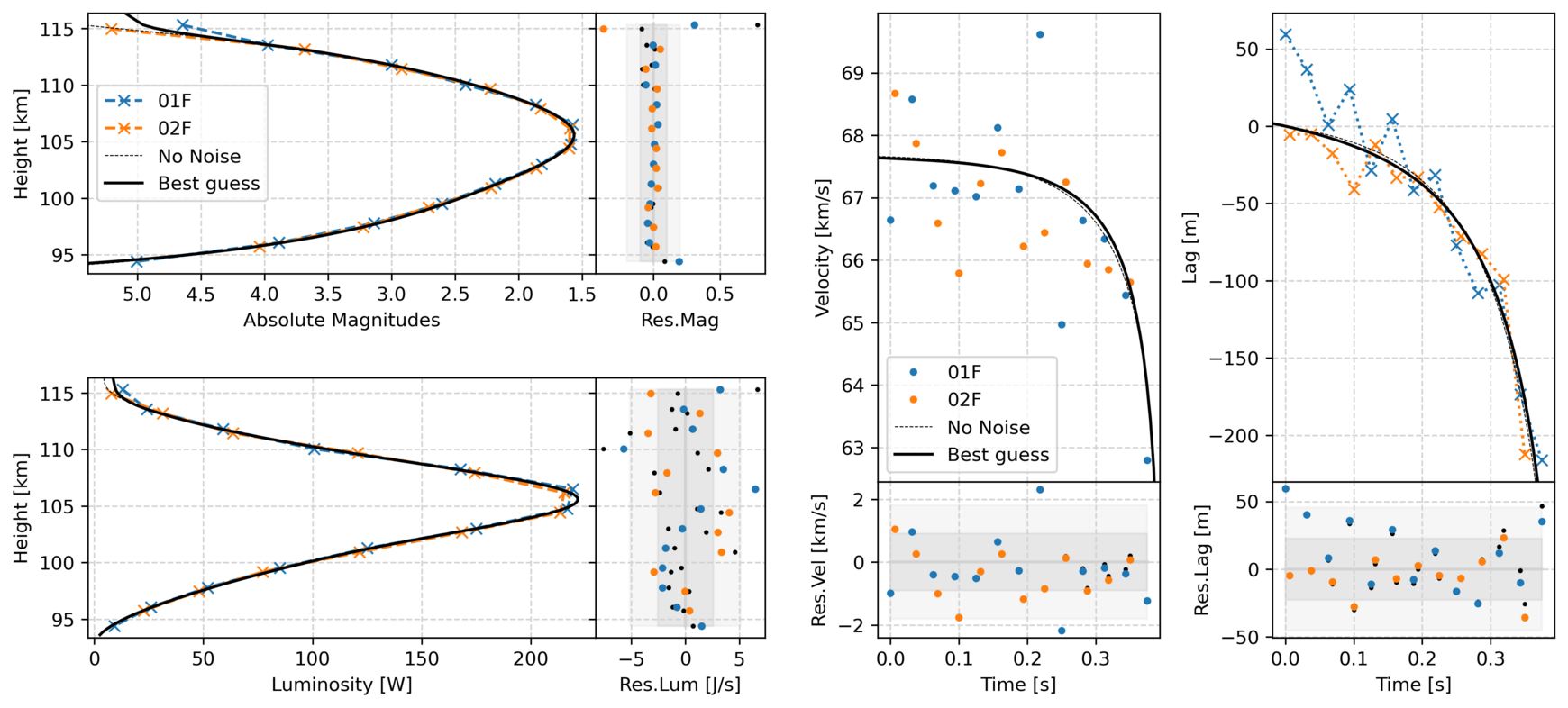}
\caption{Orionid mean test case — EMCCD-only synthetic data.}

    \label{img:ORI_mean_best_EMCCD}
\end{figure}

\begin{table}[h!]
    \centering
\caption{Posterior summary for ORI mean — EMCCD-only test case.}

    \renewcommand{\arraystretch}{1.2}
    \setlength{\tabcolsep}{4pt}
    \resizebox{\textwidth}{!}{
    \begin{tabular}{lrrrrrrrrr}
    \hline
    Parameter & 2.5CI & True Value & Best Guess & Mode & Mean & Median & 97.5CI & Abs.Error & Rel.Error\% \\
    \hline
    $v_0$ [km/s] & 67.61 & 67.73 & 67.73 & 67.66 & 67.69 & 67.7 & 67.77 & 0.004158 & 0.01 \\
    $m_0$ [kg] & 3.401$\times10^{-6}$ & 3.778$\times10^{-6}$ & 4.41$\times10^{-6}$ & 3.753$\times10^{-6}$ & 3.974$\times10^{-6}$ & 3.947$\times10^{-6}$ & 4.61$\times10^{-6}$ & 6.318$\times10^{-7}$ & 16.72 \\
    $\rho$ [kg/m$^3$] & 84.71 & 195.2 & 110.1 & 279.8 & 260.6 & 226.6 & 1311 & 85.12 & 43.6 \\
    $\sigma$ [kg/MJ] & 0.01871 & 0.02033 & 0.02619 & 0.02231 & 0.02532 & 0.02548 & 0.03135 & 0.005855 & 28.79 \\
    $h_e$ [km] & 115.4 & 116.2 & 115.8 & 117 & 116.2 & 116.2 & 117.1 & 0.4254 & 0.36 \\
    $\eta$ [kg/MJ] & 0.1204 & 0.2281 & 0.1399 & 0.2398 & 0.2691 & 0.2452 & 0.8115 & 0.0882 & 38.67 \\
    $s$ & 2.035 & 2.131 & 2.129 & 2.083 & 2.124 & 2.125 & 2.209 & 0.002382 & 0.11 \\
    $m_{l}$ [kg] & 8.071$\times10^{-12}$ & 1.39$\times10^{-11}$ & 2.194$\times10^{-11}$ & 9.314$\times10^{-12}$ & 2.27$\times10^{-11}$ & 2.393$\times10^{-11}$ & 4.837$\times10^{-11}$ & 8.033$\times10^{-12}$ & 57.77 \\
    $m_{u}$ [kg] & 1.027$\times10^{-8}$ & 1.538$\times10^{-8}$ & 5.4$\times10^{-8}$ & 2.406$\times10^{-8}$ & 3.803$\times10^{-8}$ & 3.961$\times10^{-8}$ & 9.587$\times10^{-8}$ & 3.862$\times10^{-8}$ & 251.1 \\
    $\sigma_{lag}$ [m] & 18.58 & 22.73 & 22.73 & 27.7 & 23.87 & 23.57 & 30.85 & 0 & 0 \\
    $\sigma_{lum}$ [W] & 2.437 & 2.559 & 2.559 & 4.423 & 3.324 & 3.271 & 4.579 & 0 & 0 \\
    \hline
    \end{tabular}}
    \label{tab:posterior_summary_EMCCD_ORI_mean}
\end{table}

\newpage

\subsection{CAP mode EMCCD only test case}

\begin{figure}[ht]
    \centering
    \includegraphics[width=\linewidth]{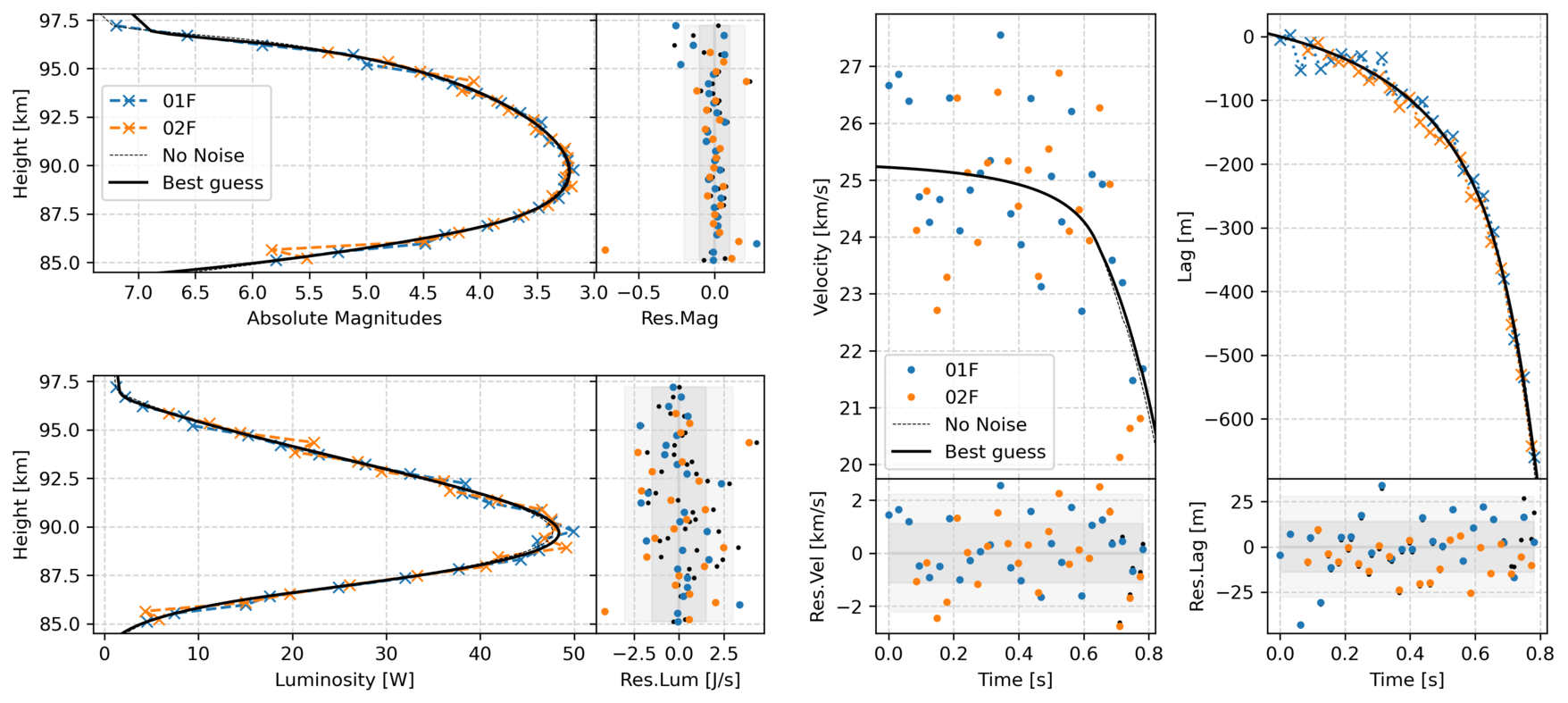}
\caption{Capricornid mode test case — EMCCD-only synthetic data.}

    \label{img:CAP_mode_best_EMCCD}
\end{figure}

\begin{table}[h!]
    \centering
\caption{Posterior summary for CAP mode — EMCCD-only test case.}

    \renewcommand{\arraystretch}{1.2}
    \setlength{\tabcolsep}{4pt}
    \resizebox{\textwidth}{!}{
    \begin{tabular}{lrrrrrrrrr}
    \hline
    Parameter & 2.5CI & True Value & Best Guess & Mode & Mean & Median & 97.5CI & Abs.Error & Rel.Error\% \\
    \hline
    $v_0$ [km/s] & 25.28 & 25.36 & 25.34 & 25.34 & 25.34 & 25.34 & 25.39 & 0.01175 & 0.04 \\
    $m_0$ [kg] & 3.596$\times10^{-6}$ & 3.667$\times10^{-6}$ & 3.925$\times10^{-6}$ & 3.851$\times10^{-6}$ & 3.928$\times10^{-6}$ & 3.897$\times10^{-6}$ & 4.435$\times10^{-6}$ & 2.581$\times10^{-7}$ & 7.03 \\
    $\rho$ [kg/m$^3$] & 841.9 & 1125 & 1154 & 1228 & 1176 & 1157 & 1785 & 28.85 & 2.56 \\
    $\sigma$ [kg/MJ] & 0.0178 & 0.0215 & 0.02866 & 0.02696 & 0.02799 & 0.02737 & 0.04208 & 0.007157 & 33.29 \\
    $h_e$ [km] & 96.98 & 97.53 & 97.24 & 97.75 & 97.44 & 97.43 & 97.99 & 0.289 & 0.29 \\
    $\eta$ [kg/MJ] & 0.08448 & 0.135 & 0.1384 & 0.1433 & 0.137 & 0.1353 & 0.2337 & 0.003382 & 2.50 \\
    $s$ & 1.333 & 2.05 & 1.715 & 2.171 & 1.984 & 1.941 & 2.809 & 0.3351 & 16.34 \\
    $m_{l}$ [kg] & 5.914$\times10^{-12}$ & 3.042$\times10^{-11}$ & 9.604$\times10^{-12}$ & 1.426$\times10^{-10}$ & 4.116$\times10^{-11}$ & 4.576$\times10^{-11}$ & 2.388$\times10^{-10}$ & 2.081$\times10^{-11}$ & 68.42 \\
    $m_{u}$ [kg] & 6.073$\times10^{-10}$ & 5.9$\times10^{-9}$ & 8.352$\times10^{-9}$ & 1.445$\times10^{-8}$ & 1.132$\times10^{-8}$ & 1.244$\times10^{-8}$ & 8.025$\times10^{-8}$ & 2.452$\times10^{-9}$ & 41.55 \\
    $\sigma_{lag}$ [m] & 13.52 & 13.94 & 13.94 & 13.12 & 16.54 & 16.4 & 20.25 & 0 & 0 \\
    $\sigma_{lum}$ [W] & 1.315 & 1.507 & 1.507 & 1.74 & 1.607 & 1.594 & 1.979 & 0 & 0 \\
    \hline
    \end{tabular}}
    \label{tab:posterior_summary_EMCCD_CAP_mode}
\end{table}

\newpage

\subsection{CAP mean EMCCD only test case}

\begin{figure}[ht]
    \centering
    \includegraphics[width=\linewidth]{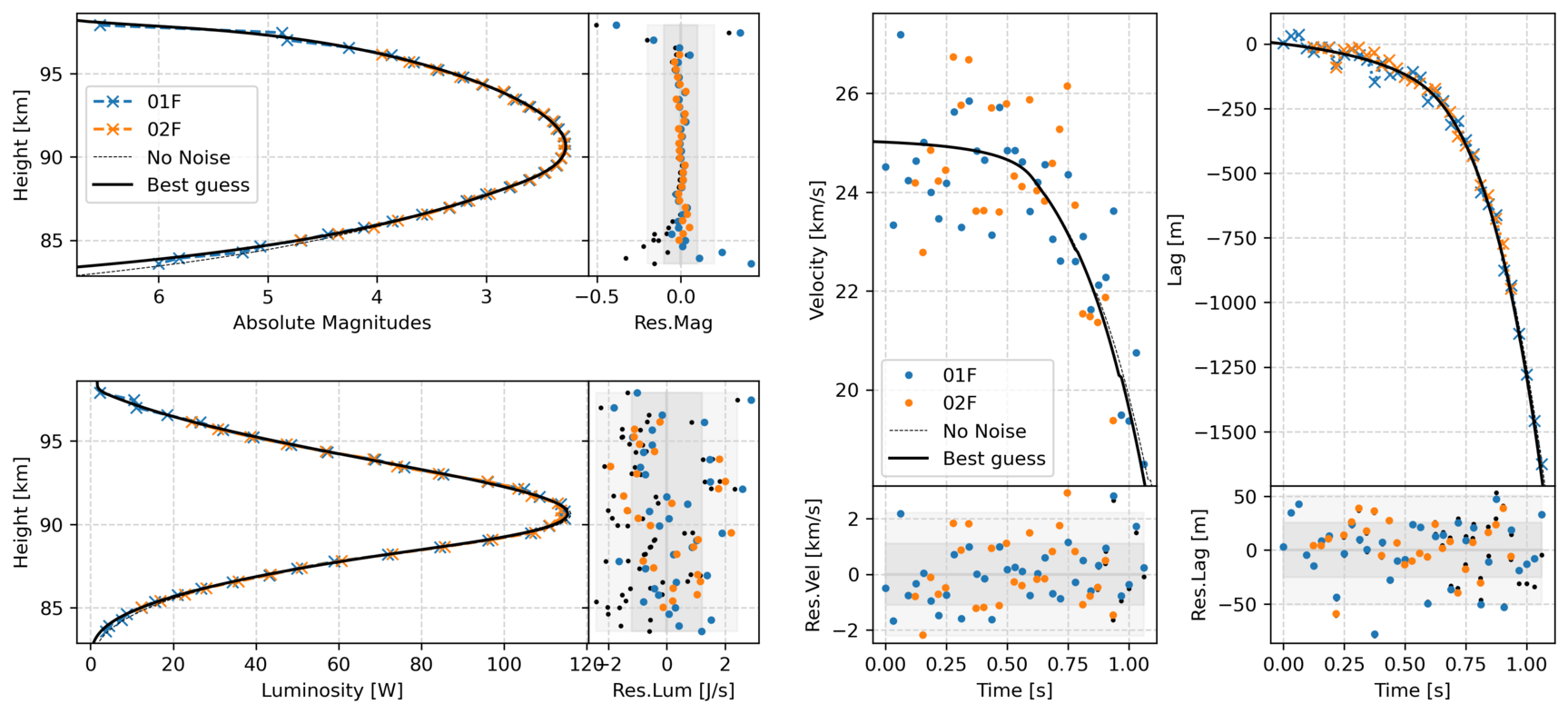}
    \caption{Capricornid mean test case — EMCCD-only synthetic data.}

    \label{img:CAP_mean_best_EMCCD}
\end{figure}

\begin{table}[h!]
    \centering
    \caption{Posterior summary for CAP mean — EMCCD-only test case.}
    \renewcommand{\arraystretch}{1.2}
    \setlength{\tabcolsep}{4pt}
    \resizebox{\textwidth}{!}{
    \begin{tabular}{lrrrrrrrrr}
    \hline
    Parameter & 2.5CI & True Value & Best Guess & Mode & Mean & Median & 97.5CI & Abs.Error & Rel.Error\% \\
    \hline
    $v_0$ [km/s] & 25.06 & 25.13 & 25.1 & 25.1 & 25.13 & 25.13 & 25.19 & 0.02315 & 0.09 \\
    $m_0$ [kg] & 9.938$\times10^{-6}$ & 1.025$\times10^{-5}$ & 1.021$\times10^{-5}$ & 1.007$\times10^{-5}$ & 1.032$\times10^{-5}$ & 1.031$\times10^{-5}$ & 1.08$\times10^{-5}$ & 3.672$\times10^{-8}$ & 0.35 \\
    $\rho$ [kg/m$^3$] & 624.1 & 875 & 1019 & 1211 & 890.4 & 860.2 & 1398 & 144.3 & 16.49 \\
    $\sigma$ [kg/MJ] & 0.01825 & 0.01844 & 0.02222 & 0.02143 & 0.0214 & 0.0214 & 0.02449 & 0.003781 & 20.51 \\
    $h_e$ [km] & 98.24 & 98.5 & 98.55 & 98.57 & 98.45 & 98.43 & 98.72 & 0.04682 & 0.04 \\
    $\eta$ [kg/MJ] & 0.1879 & 0.2444 & 0.2793 & 0.3324 & 0.2505 & 0.245 & 0.3751 & 0.03488 & 14.27 \\
    $s$ & 1.743 & 2.098 & 1.843 & 1.839 & 1.943 & 1.907 & 2.336 & 0.2551 & 12.16 \\
    $m_{l}$ [kg] & 3.409$\times10^{-11}$ & 7.438$\times10^{-11}$ & 8.332$\times10^{-11}$ & 1.19$\times10^{-10}$ & 7.862$\times10^{-11}$ & 7.104$\times10^{-11}$ & 2.833$\times10^{-10}$ & 8.95$\times10^{-12}$ & 12.03 \\
    $m_{u}$ [kg] & 1.19$\times10^{-8}$ & 1.825$\times10^{-8}$ & 1.414$\times10^{-8}$ & 1.142$\times10^{-8}$ & 1.635$\times10^{-8}$ & 1.566$\times10^{-8}$ & 3.241$\times10^{-8}$ & 4.107$\times10^{-9}$ & 22.5 \\
    $\sigma_{lag}$ [m] & 23.4 & 25.32 & 25.32 & 30.43 & 27.58 & 27.44 & 32.95 & 0 & 0 \\
    $\sigma_{lum}$ [W] & 1.107 & 1.21 & 1.21 & 1.413 & 1.325 & 1.319 & 1.599 & 0 & 0 \\
    \hline
    \end{tabular}}
    \label{tab:posterior_summary_EMCCD_CAP_mean}
\end{table}

\newpage

\section{posterior-optimal solutions}\label{sec:Apx wake}

This appendix presents the posterior-optimal solutions, referred to as the \textit{Best Guess}, derived from Dynamic Nested Sampling for each analyzed meteor. The Best Guess corresponds to the simulation sample with the highest posterior probability—representing the most plausible set of physical parameters given the observational data.

The Dynamic Nested Sampling runs were conducted using a combination of lag and luminosity data obtained from both EMCCD and CAMO systems. The results shown here reflect not only the best-fitting synthetic profiles for the observed data, but also a corresponding match to the meteor wake morphology observed by CAMO, providing an additional qualitative validation of the inferred solution.

\newpage

\subsection{Orionids (ORI)}

\subsubsection{20191023\_084916}
\begin{figure}[h!]
\centering
\begin{minipage}{0.85\linewidth}
    \centering
    \includegraphics[width=\linewidth]{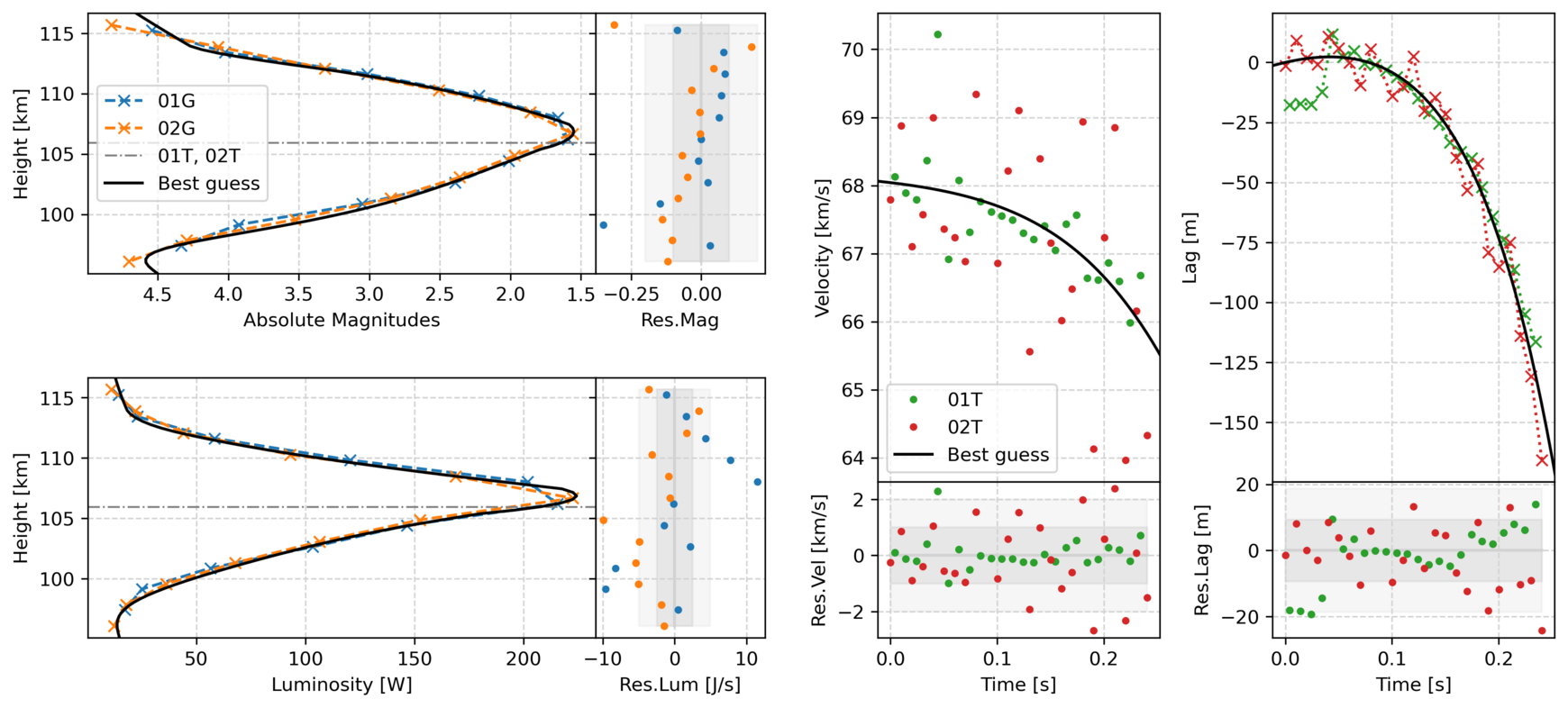}
\end{minipage}

\begin{minipage}{0.7\linewidth}
    \centering
    \includegraphics[width=\linewidth]{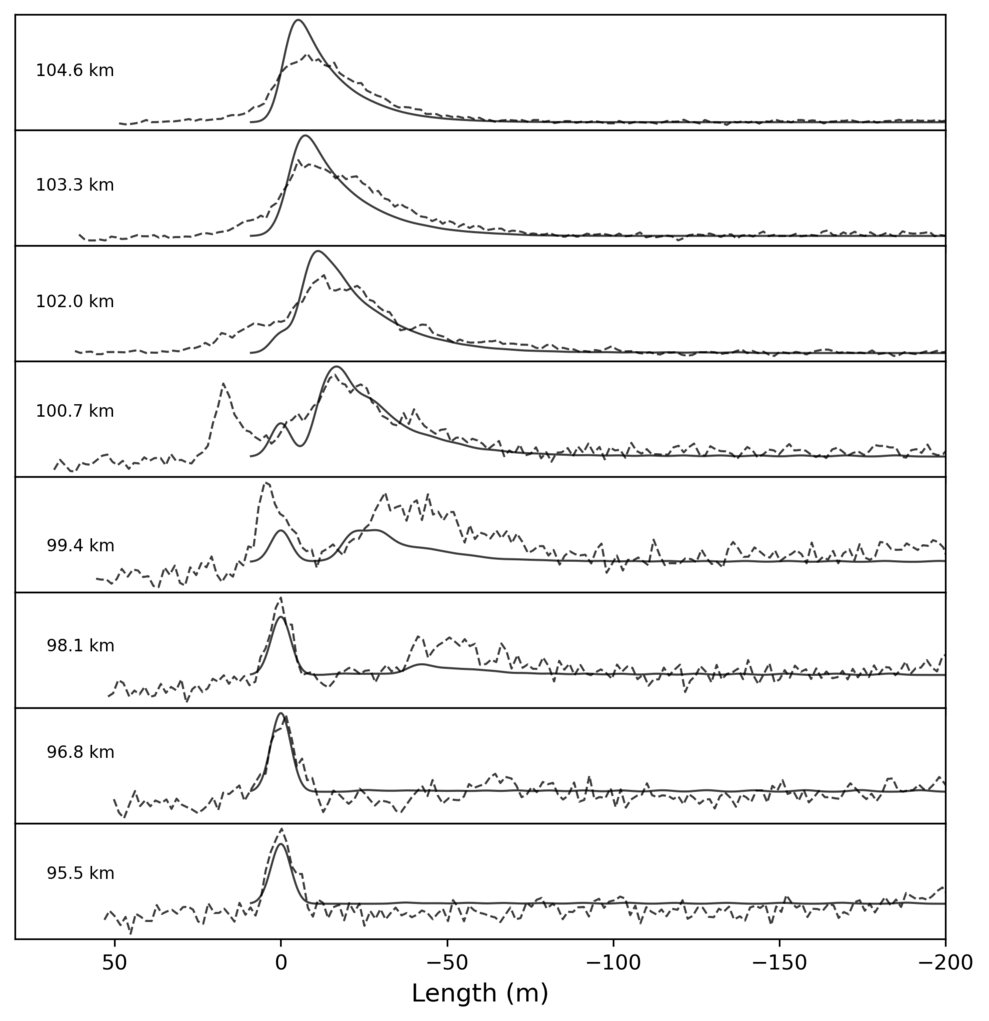}
\end{minipage}

\caption{Top: Best-fit simulation results for the 20191023\_084916 meteor, showing the synthetic luminosity and lag curves matched to EMCCD and CAMO observations. Bottom: Wake comparison using CAMO data for the same simulation.}
\label{img:ORI_20191023_084916}
\end{figure}

\newpage

\subsubsection{20191023\_091225}
\begin{figure}[h!]
\centering
\begin{minipage}{0.85\linewidth}
    \centering
    \includegraphics[width=\linewidth]{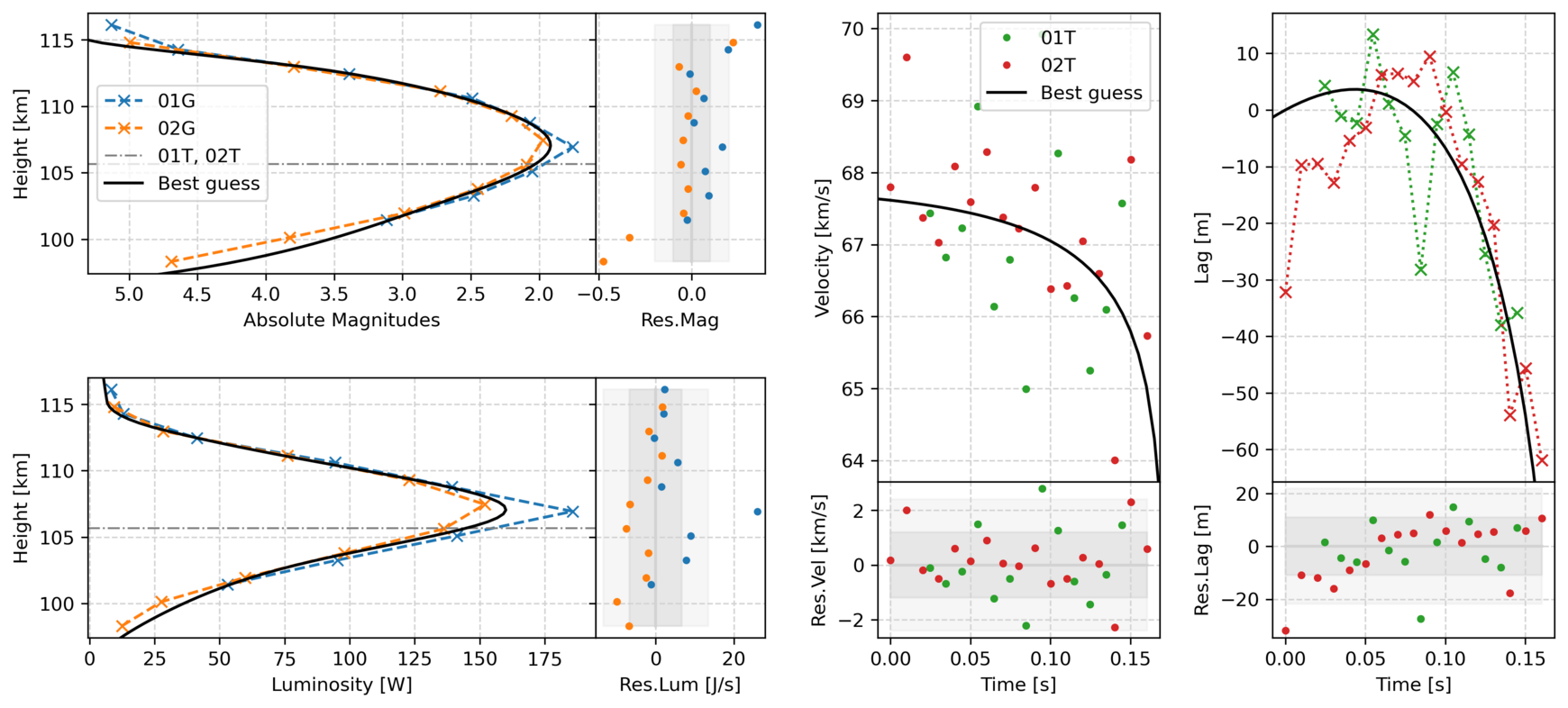}
\end{minipage}

\begin{minipage}{0.7\linewidth}
    \centering
    \includegraphics[width=\linewidth]{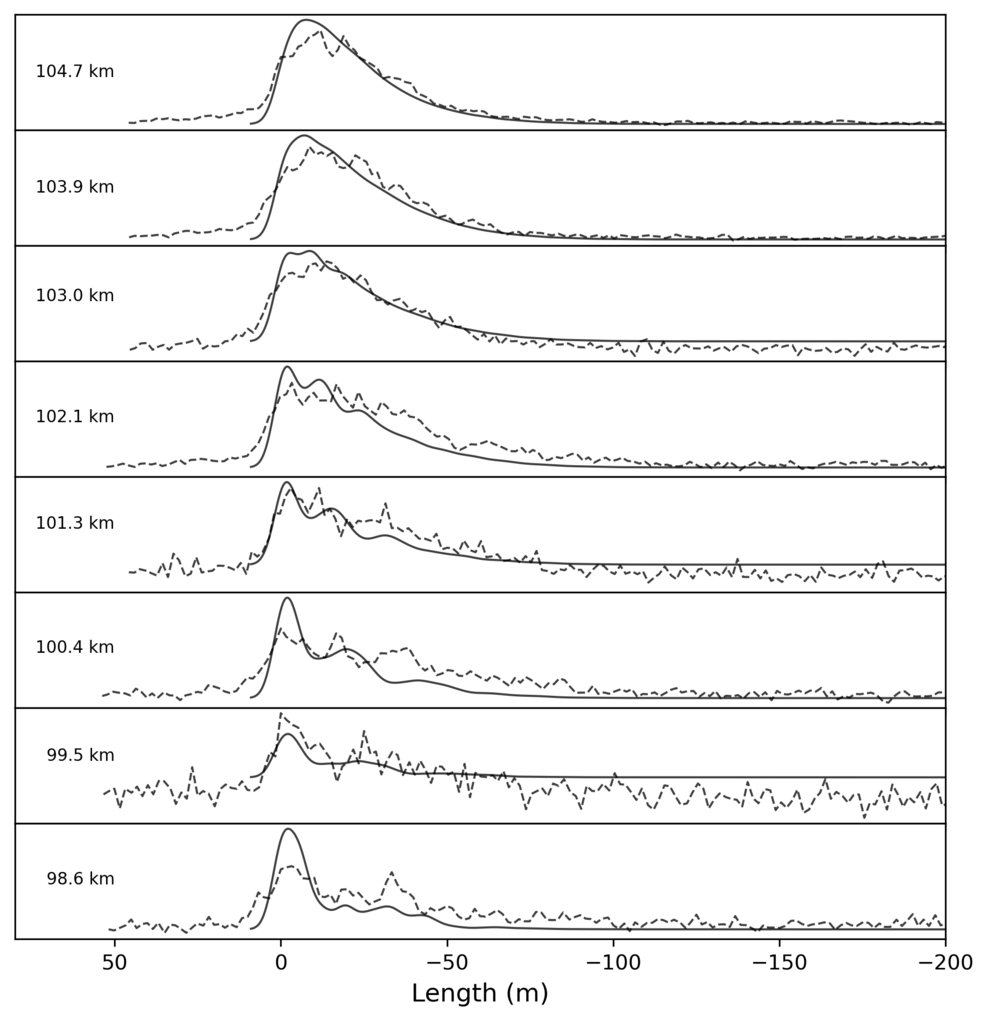}
\end{minipage}

\caption{Top: Best-fit simulation results for the 20191023\_091225 meteor, showing the synthetic luminosity and lag curves matched to EMCCD and CAMO observations. Bottom: Wake comparison using CAMO data for the same simulation.}
\label{img:ORI_20191023_091225}
\end{figure}

\newpage

\subsubsection{20191023\_091310}
\begin{figure}[h!]
\centering
\begin{minipage}{0.85\linewidth}
    \centering
    \includegraphics[width=\linewidth]{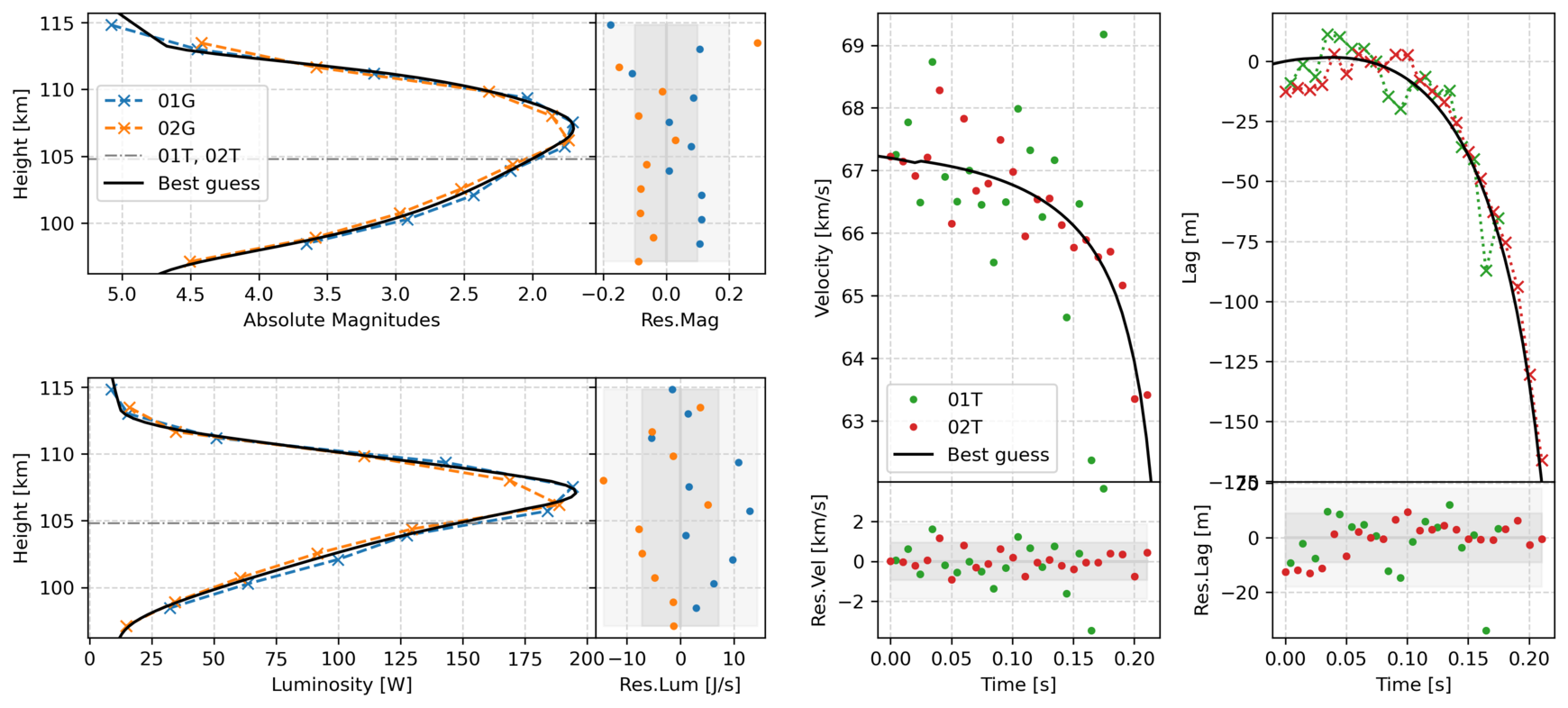}
\end{minipage}

\begin{minipage}{0.7\linewidth}
    \centering
    \includegraphics[width=\linewidth]{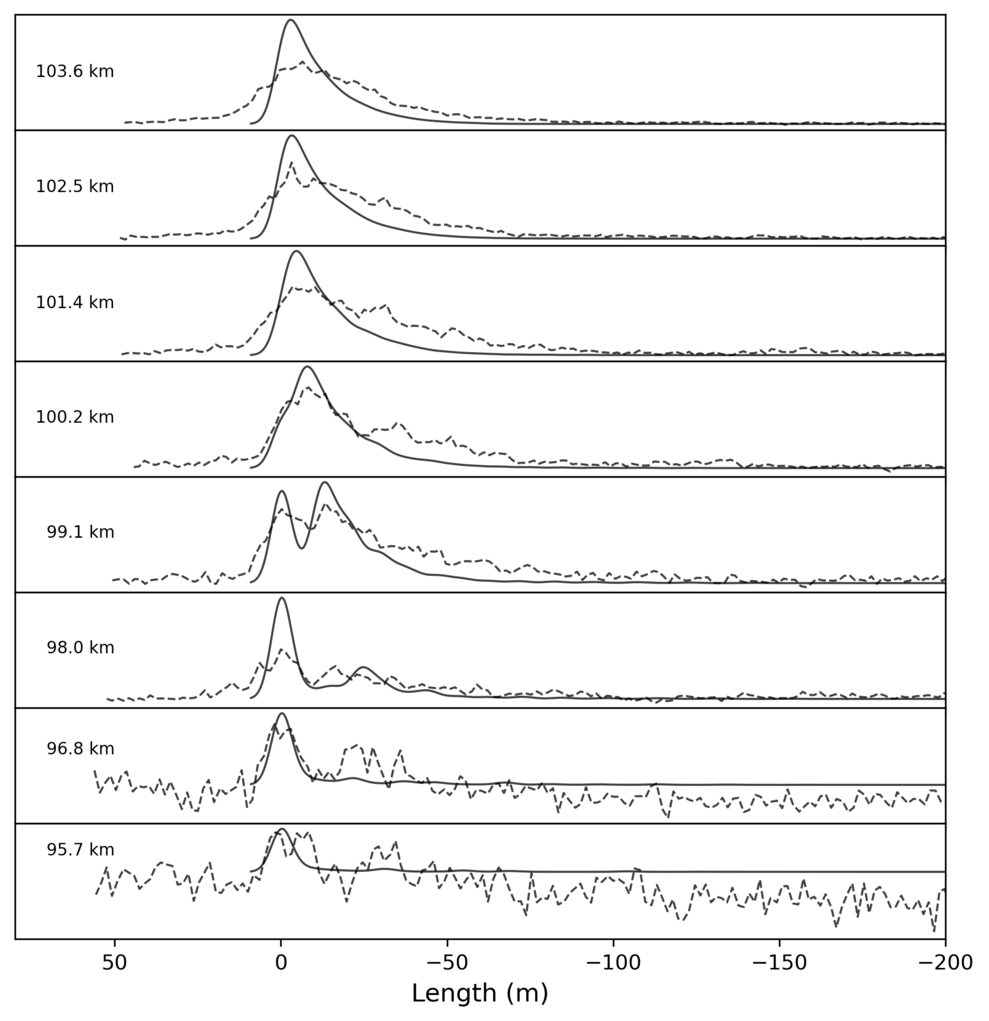}
\end{minipage}

\caption{Top: Best-fit simulation results for the 20191023\_091310 meteor, showing the synthetic luminosity and lag curves matched to EMCCD and CAMO observations. Bottom: Wake comparison using CAMO data for the same simulation.}
\label{img:ORI_20191023_091310}
\end{figure}

\newpage

\subsubsection{20191026\_065838}
\begin{figure}[h!]
\centering
\begin{minipage}{0.85\linewidth}
    \centering
    \includegraphics[width=\linewidth]{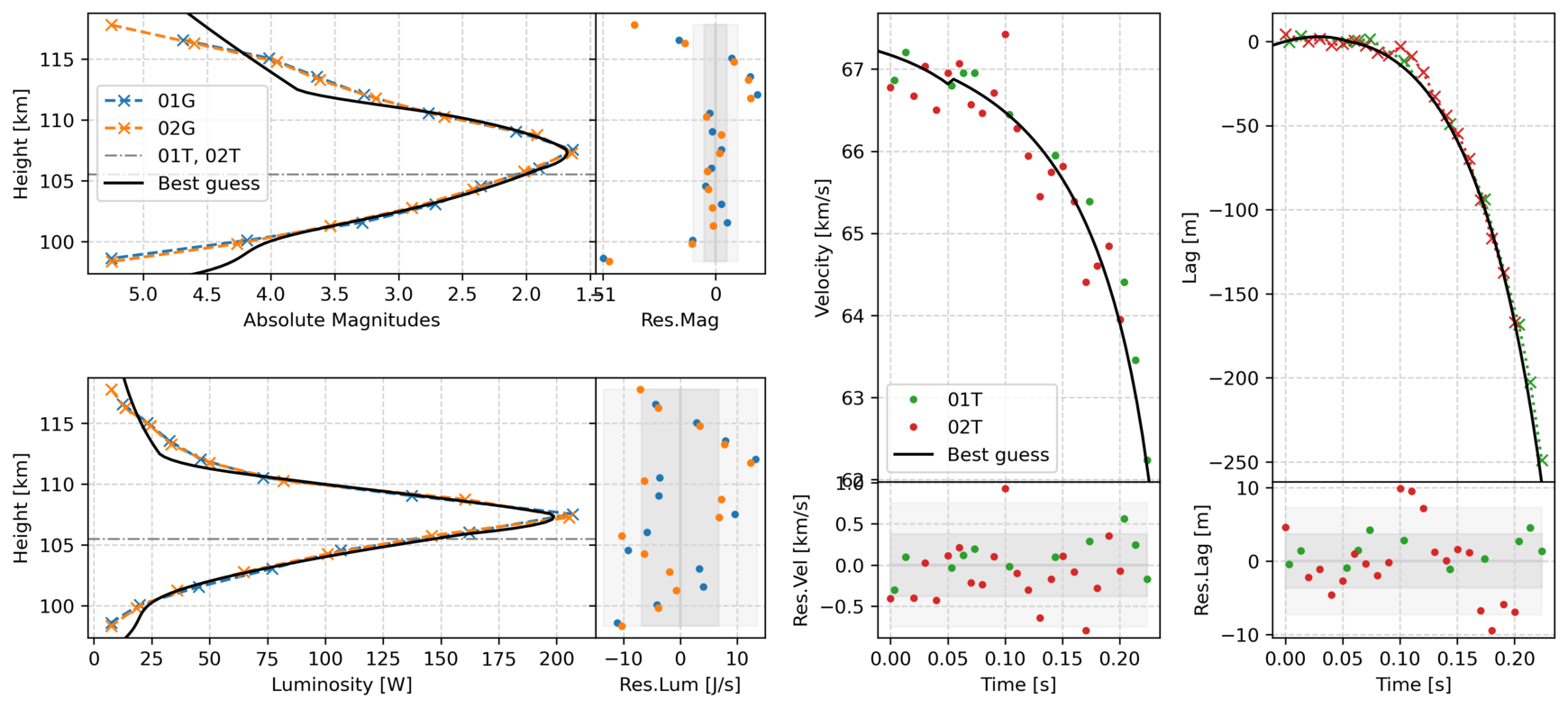}
\end{minipage}

\begin{minipage}{0.7\linewidth}
    \centering
    \includegraphics[width=\linewidth]{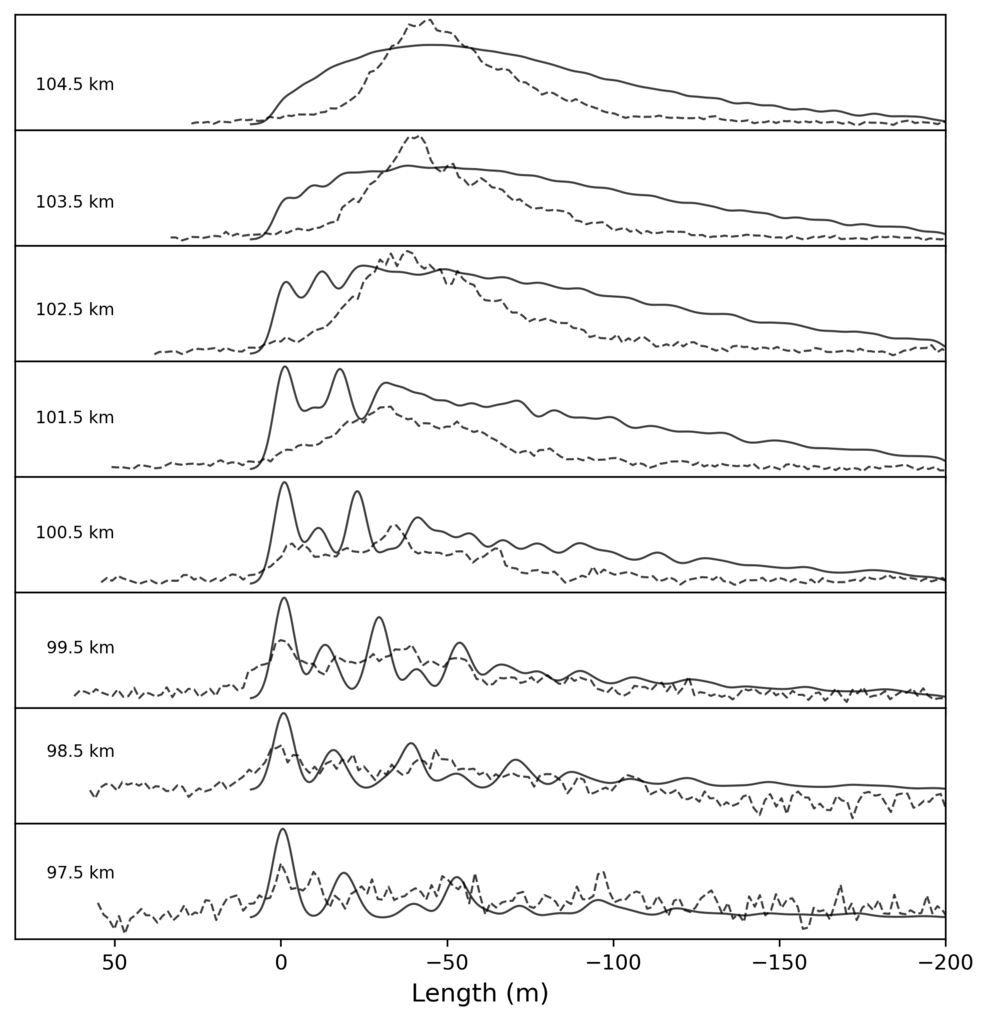}
\end{minipage}

\caption{Top: Best-fit simulation results for the 20191026\_065838 meteor, showing the synthetic luminosity and lag curves matched to EMCCD and CAMO observations. Bottom: Wake comparison using CAMO data for the same simulation.}
\label{img:ORI_20191026_065838}
\end{figure}

\newpage

\subsubsection{20191028\_050616}
\begin{figure}[h!]
\centering
\begin{minipage}{0.85\linewidth}
    \centering
    \includegraphics[width=\linewidth]{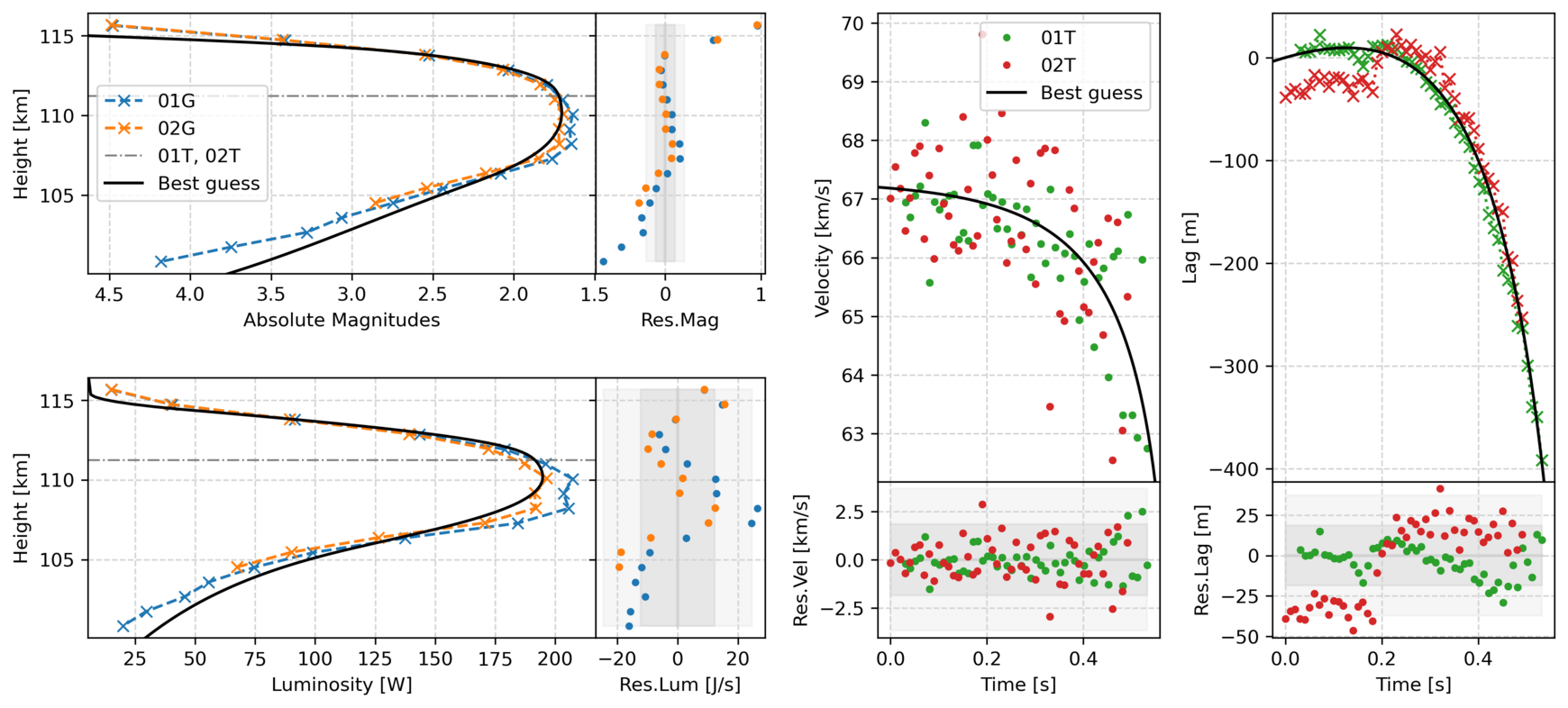}
\end{minipage}

\begin{minipage}{0.7\linewidth}
    \centering
    \includegraphics[width=\linewidth]{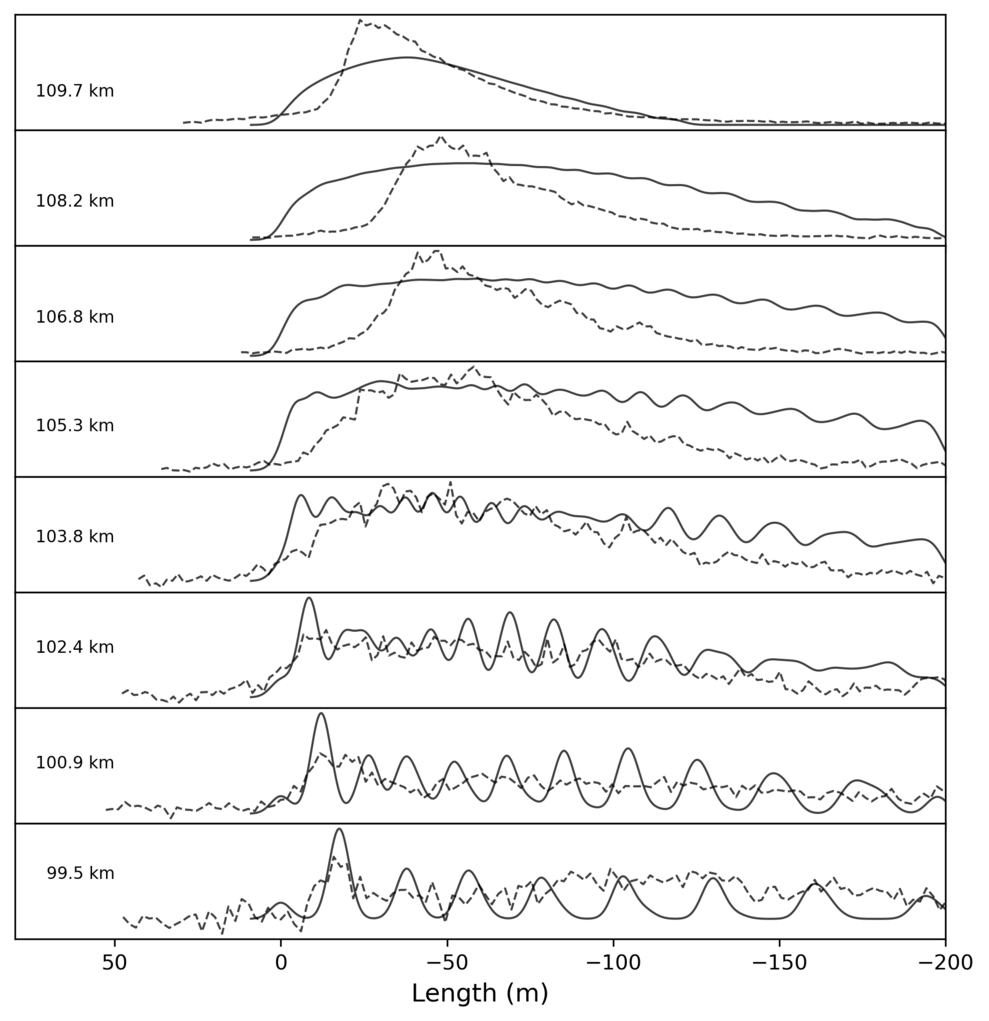}
\end{minipage}

\caption{Top: Best-fit simulation results for the 20191028\_050616 meteor, showing the synthetic luminosity and lag curves matched to EMCCD and CAMO observations. Bottom: Wake comparison using CAMO data for the same simulation.}
\label{img:ORI_20191028_050616}
\end{figure}

\newpage

\subsubsection{20201012\_081716}
\begin{figure}[h!]
\centering
\begin{minipage}{0.85\linewidth}
    \centering
    \includegraphics[width=\linewidth]{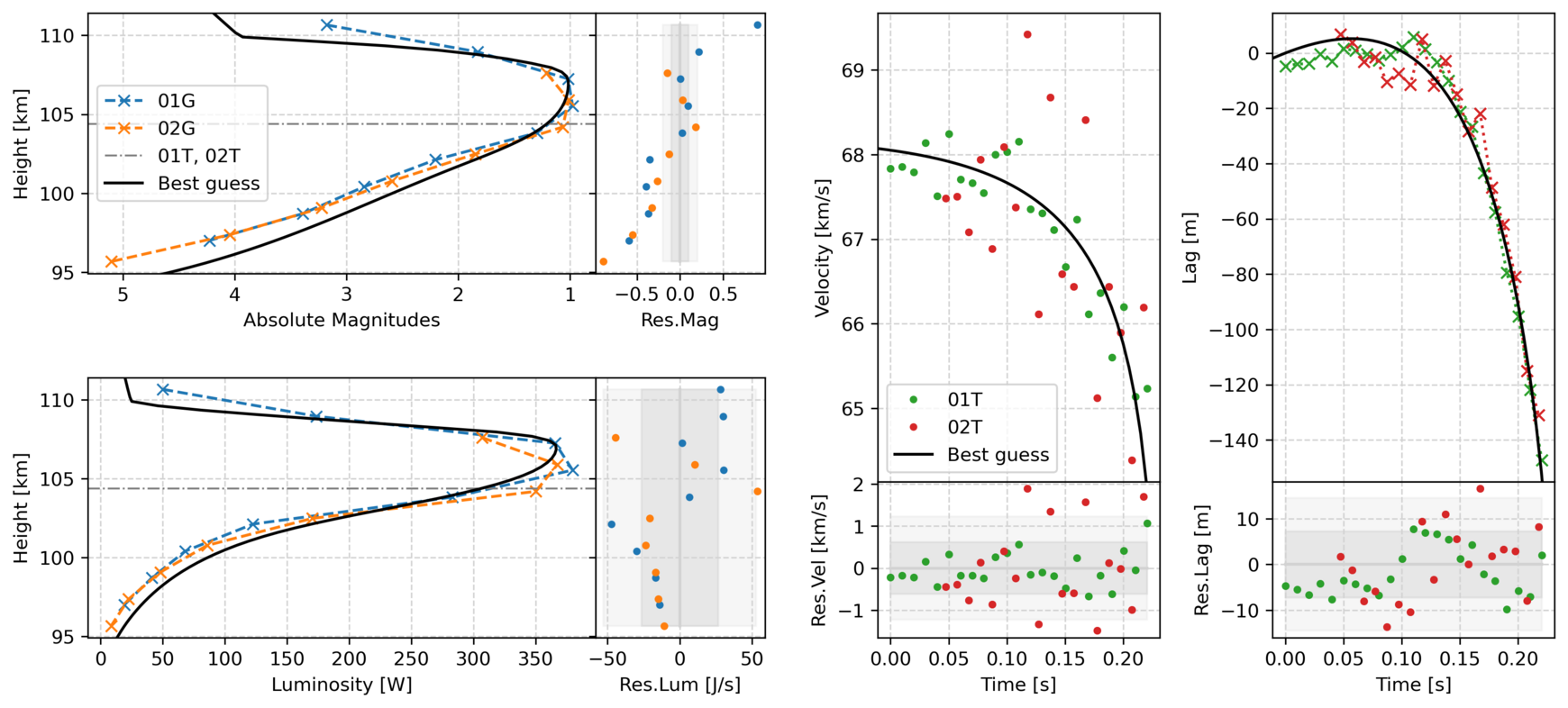}
\end{minipage}

\begin{minipage}{0.7\linewidth}
    \centering
    \includegraphics[width=\linewidth]{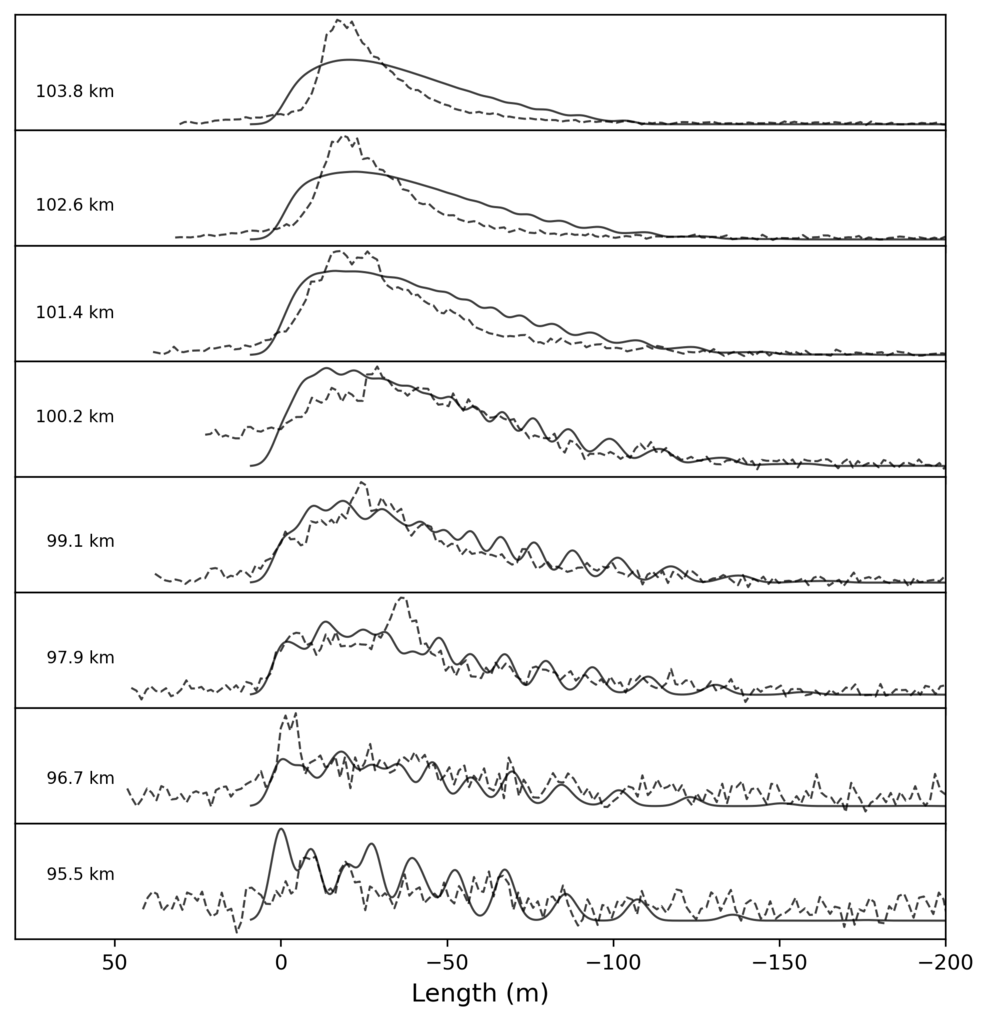}
\end{minipage}

\caption{Top: Best-fit simulation results for the 20201012\_081716 meteor, showing the synthetic luminosity and lag curves matched to EMCCD and CAMO observations. Bottom: Wake comparison using CAMO data for the same simulation.}
\label{img:ORI_20201012_081716}
\end{figure}




\newpage

\subsubsection{20201017\_074007}
\begin{figure}[h!]
\centering
\begin{minipage}{0.85\linewidth}
    \centering
    \includegraphics[width=\linewidth]{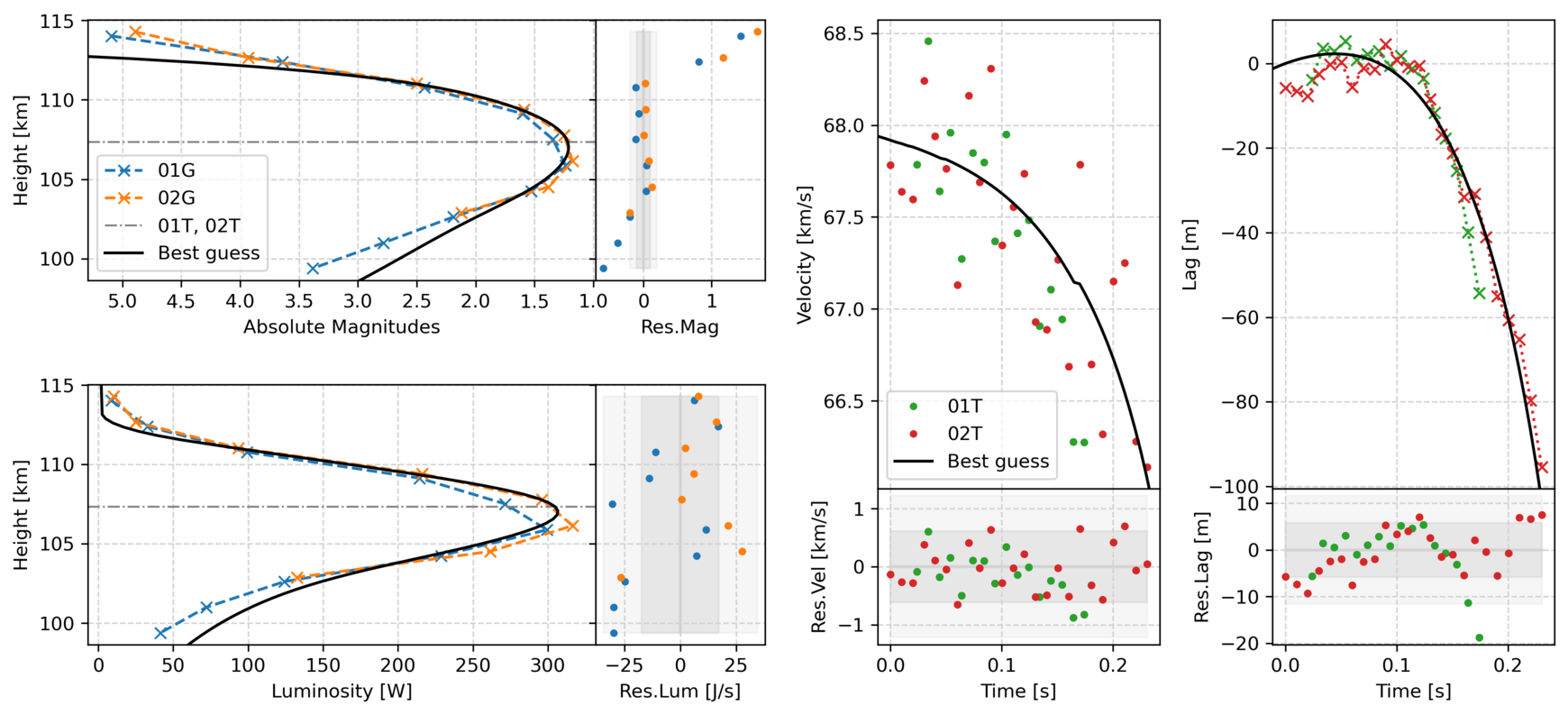}
\end{minipage}

\begin{minipage}{0.7\linewidth}
    \centering
    \includegraphics[width=\linewidth]{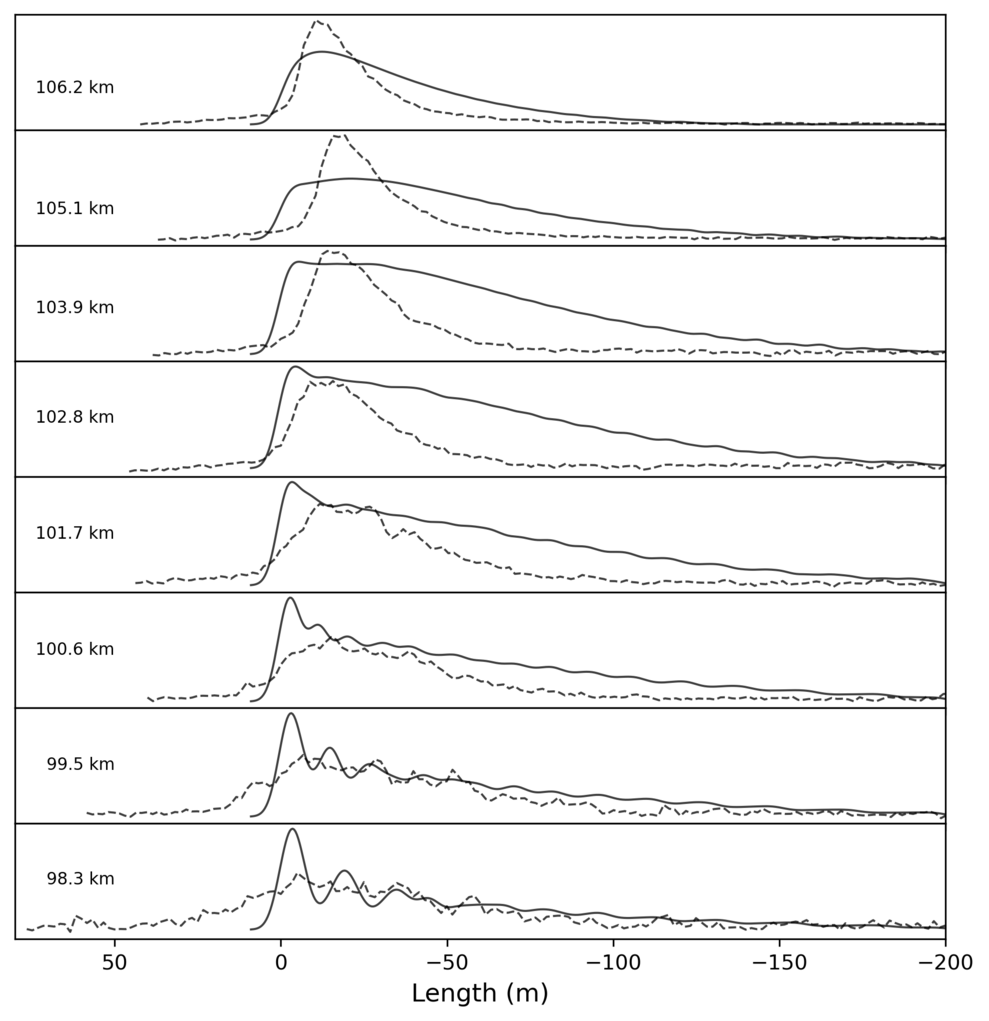}
\end{minipage}

\caption{Top: Best-fit simulation results for the 20201017\_074007 meteor, showing the synthetic luminosity and lag curves matched to EMCCD and CAMO observations. Bottom: Wake comparison using CAMO data for the same simulation.}
\label{img:ORI_20201017_074007}
\end{figure}




\newpage

\subsubsection{20221022\_075829}
\begin{figure}[h!]
\centering
\begin{minipage}{0.85\linewidth}
    \centering
    \includegraphics[width=\linewidth]{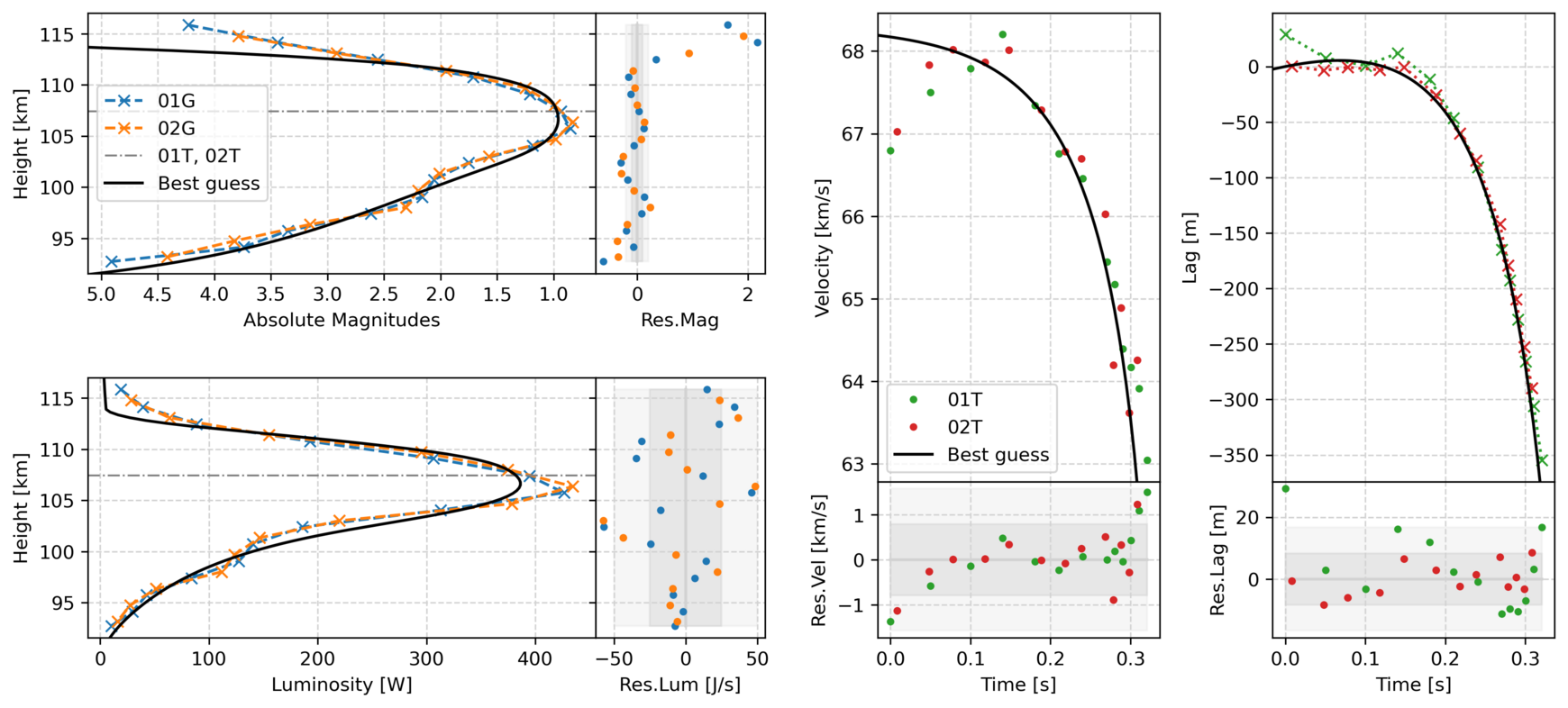}
\end{minipage}

\begin{minipage}{0.7\linewidth}
    \centering
    \includegraphics[width=\linewidth]{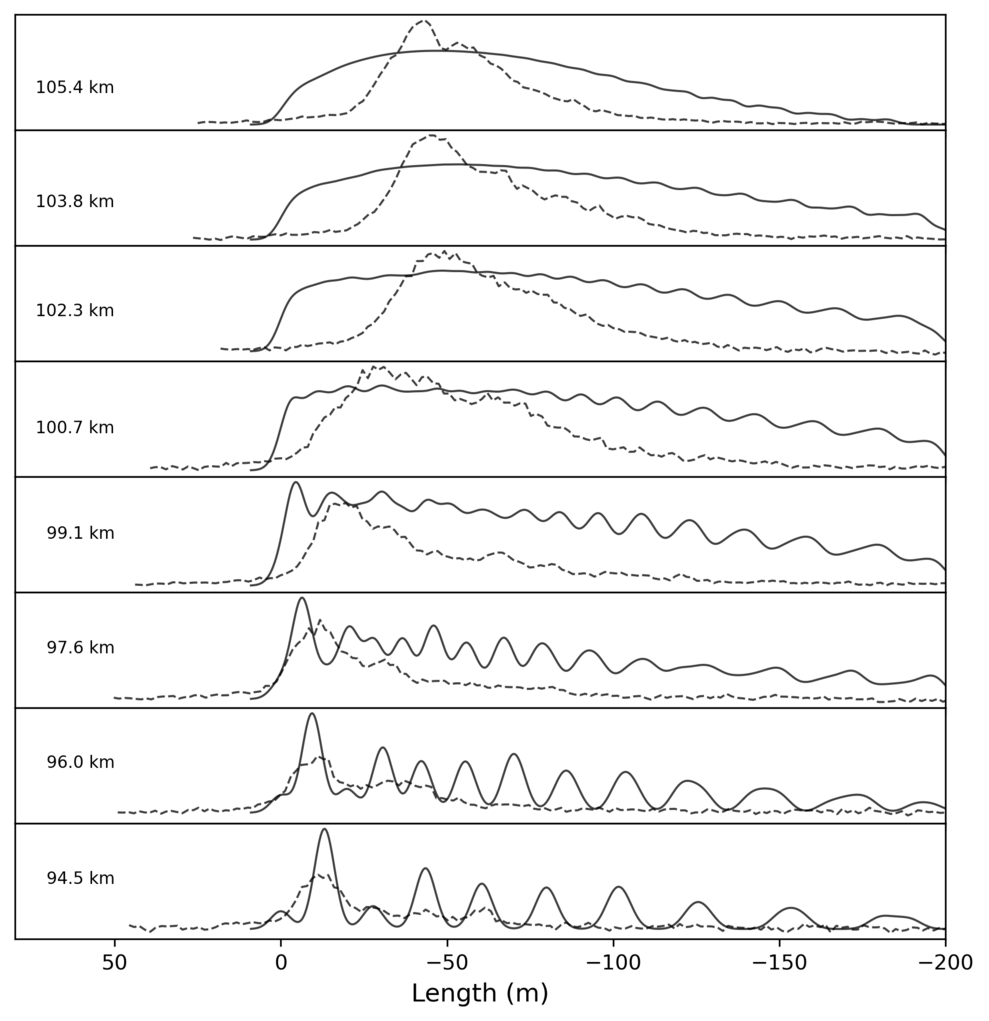}
\end{minipage}

\caption{Top: Best-fit simulation results for the 20221022\_075829 meteor, showing the synthetic luminosity and lag curves matched to EMCCD and CAMO observations. Bottom: Wake comparison using CAMO data for the same simulation.}
\label{img:ORI_20221022_075829}
\end{figure}

\newpage

\subsubsection{20221022\_081606}
\begin{figure}[h!]
\centering
\begin{minipage}{0.85\linewidth}
    \centering
    \includegraphics[width=\linewidth]{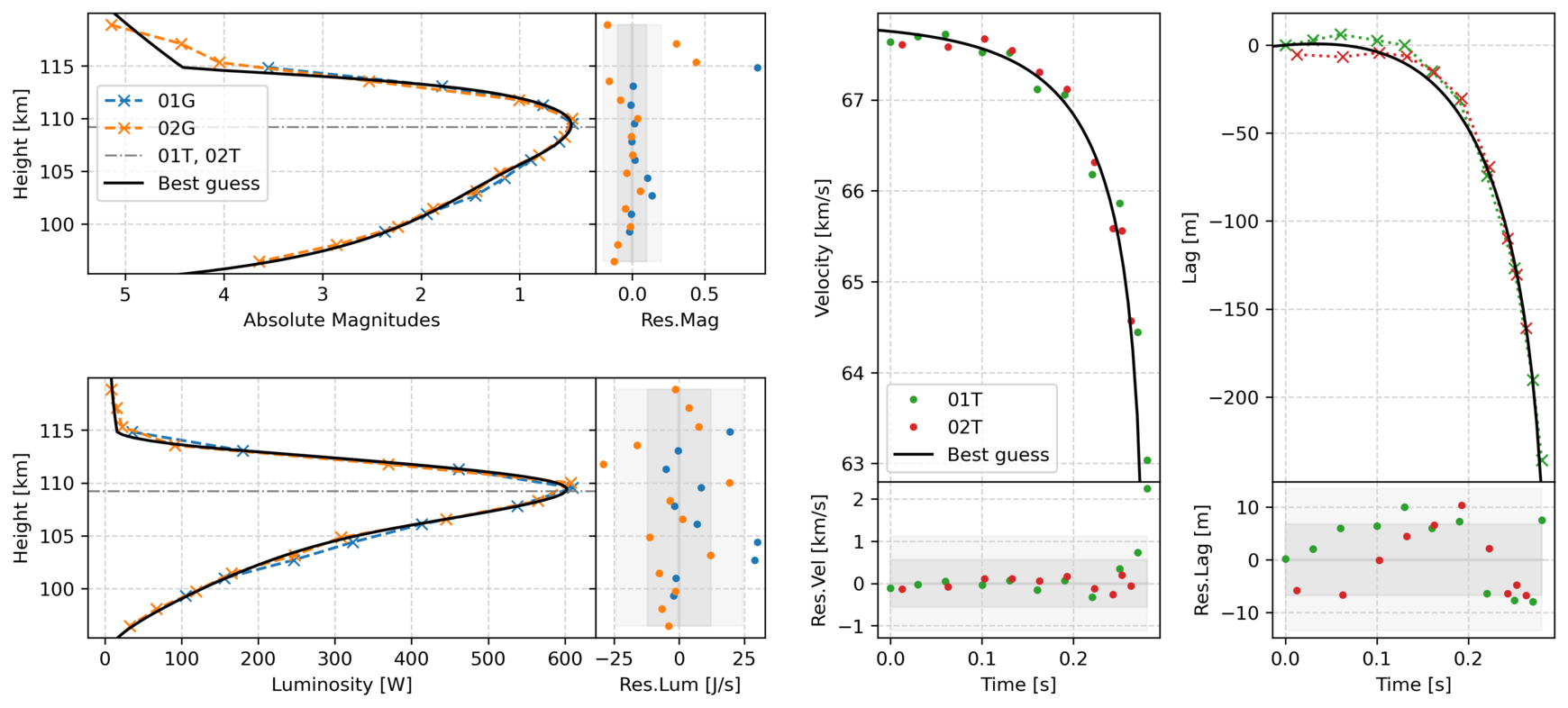}
\end{minipage}

\begin{minipage}{0.7\linewidth}
    \centering
    \includegraphics[width=\linewidth]{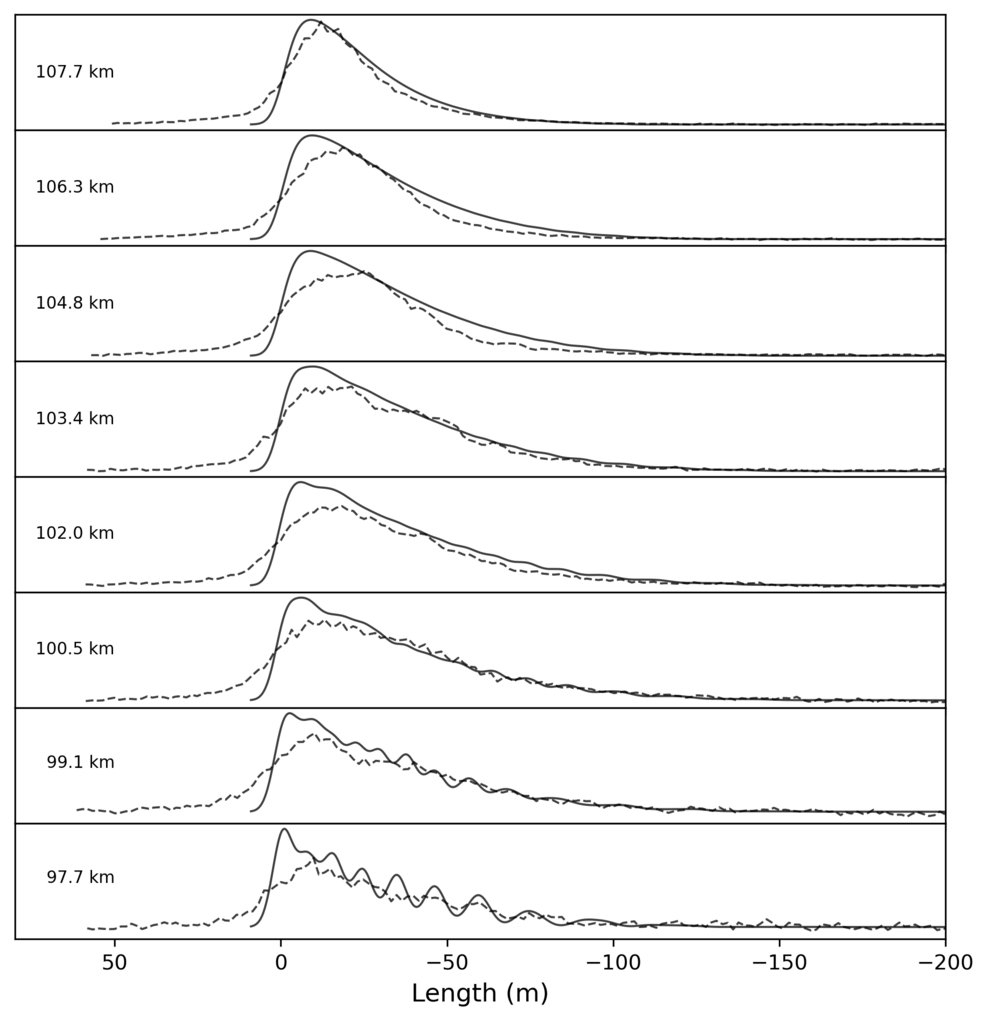}
\end{minipage}

\caption{Top: Best-fit simulation results for the 20221022\_081606 meteor, showing the synthetic luminosity and lag curves matched to EMCCD and CAMO observations. Bottom: Wake comparison using CAMO data for the same simulation.}
\label{img:ORI_20221022_081606}
\end{figure}

\newpage

\subsection{Capricornids (CAP)}

\subsubsection{20190726\_024150}
\begin{figure}[h!]
\centering
\begin{minipage}{0.85\linewidth}
    \centering
    \includegraphics[width=\linewidth]{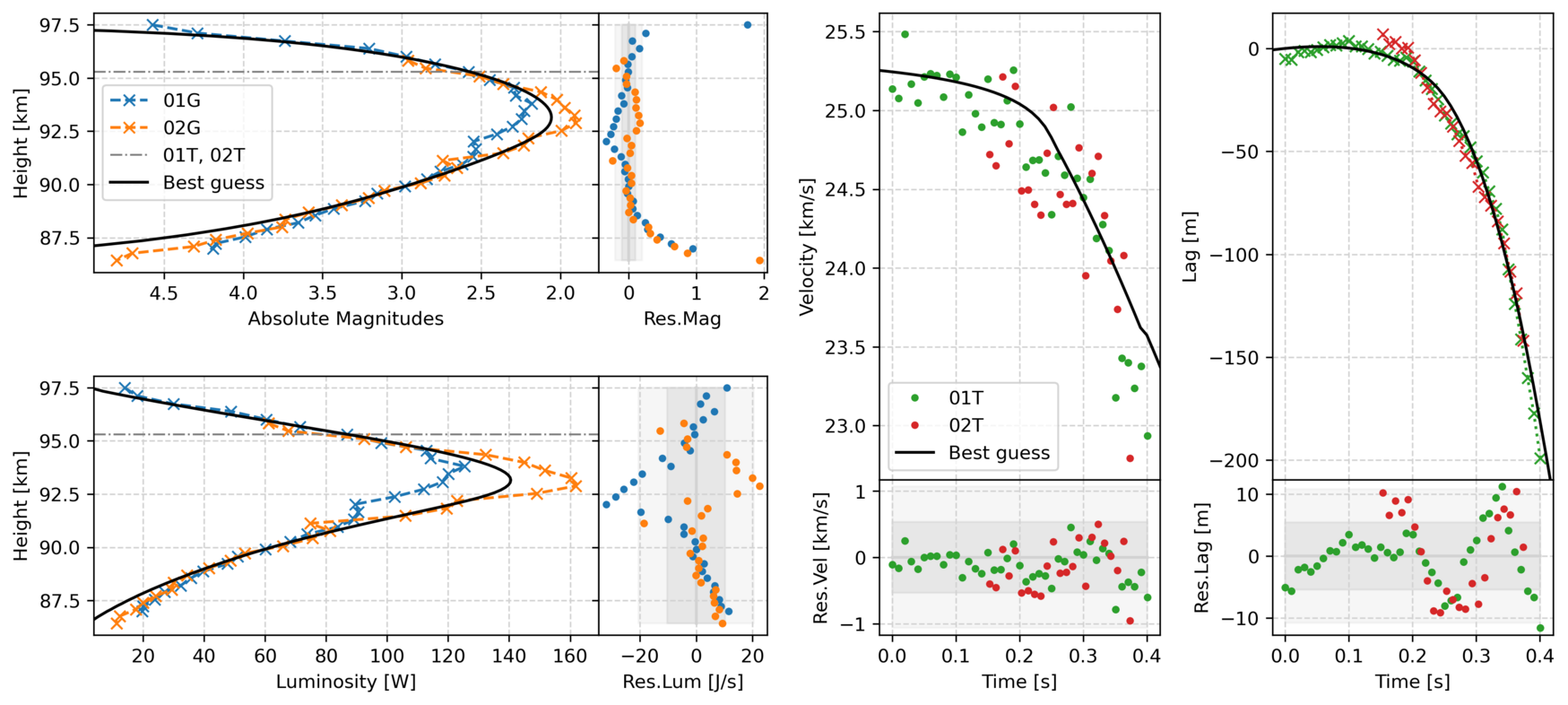}
\end{minipage}

\begin{minipage}{0.7\linewidth}
    \centering
    \includegraphics[width=\linewidth]{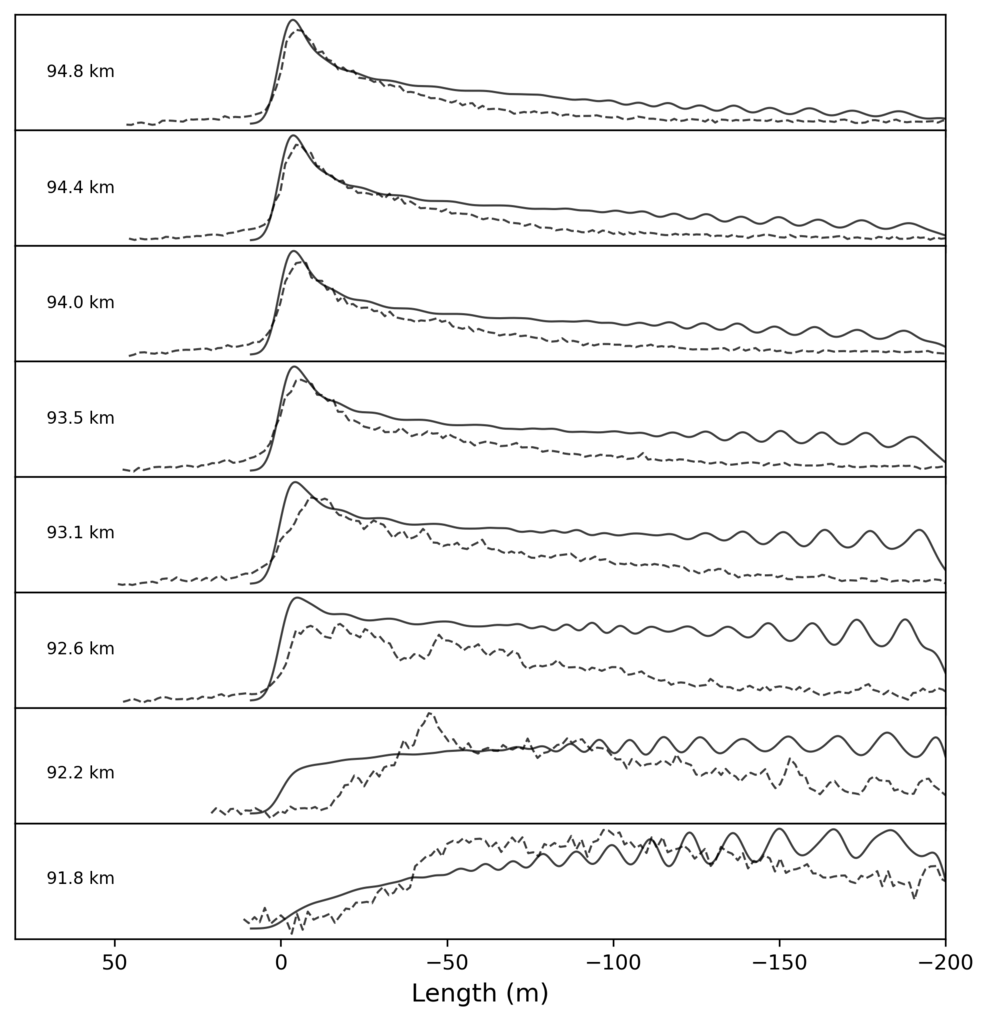}
\end{minipage}

\caption{Top: Best-fit simulation results for the 20190726\_024150 meteor, showing the synthetic luminosity and lag curves matched to EMCCD and CAMO observations. Bottom: Wake comparison using CAMO data for the same simulation.}
\label{img:CAP_20190726_024150}
\end{figure}

\newpage

\subsubsection{20190801\_024424}
\begin{figure}[h!]
\centering
\begin{minipage}{0.85\linewidth}
    \centering
    \includegraphics[width=\linewidth]{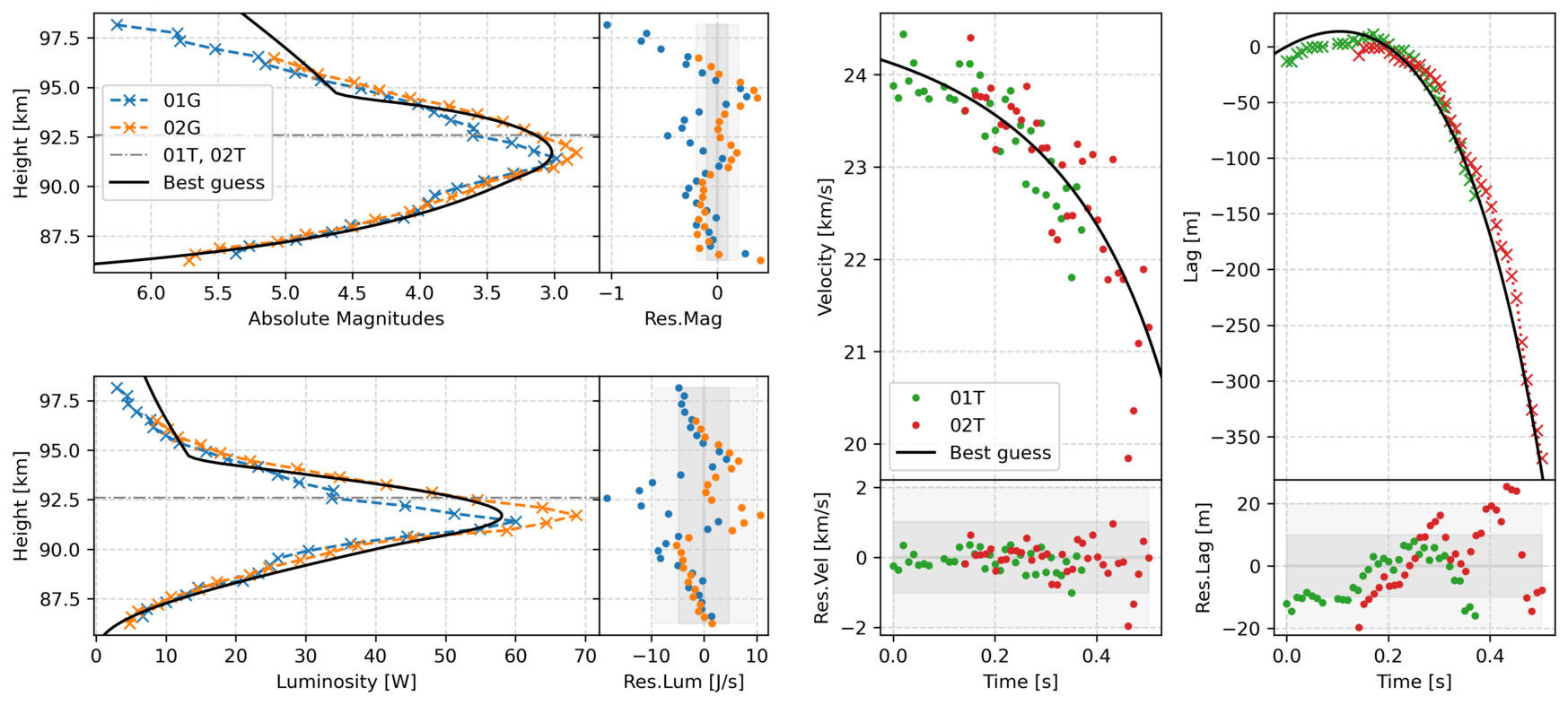}
\end{minipage}

\begin{minipage}{0.7\linewidth}
    \centering
    \includegraphics[width=\linewidth]{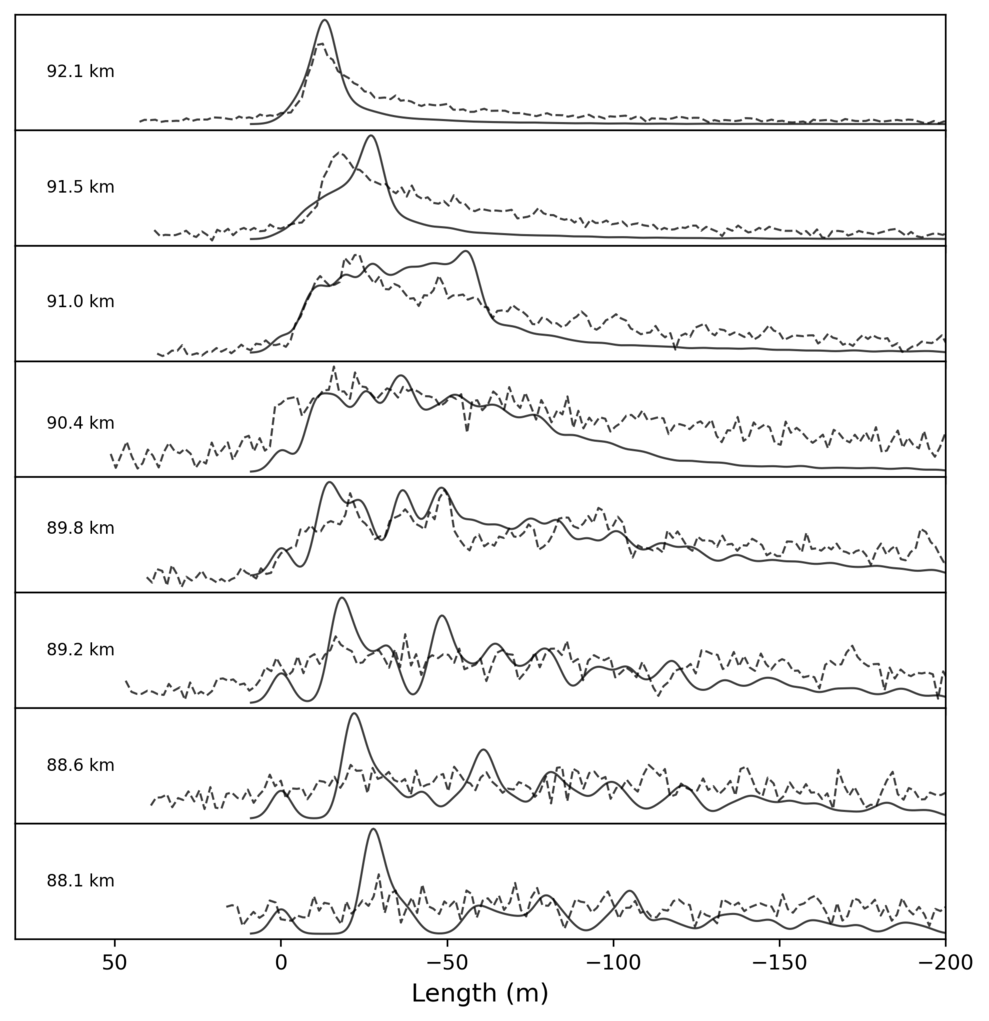}
\end{minipage}

\caption{Top: Best-fit simulation results for the 20190801\_024424 meteor, showing the synthetic luminosity and lag curves matched to EMCCD and CAMO observations. Bottom: Wake comparison using CAMO data for the same simulation.}
\label{img:CAP_20190801_024424}
\end{figure}







\newpage

\subsubsection{20200726\_032722}
\begin{figure}[h!]
\centering
\begin{minipage}{0.85\linewidth}
    \centering
    \includegraphics[width=\linewidth]{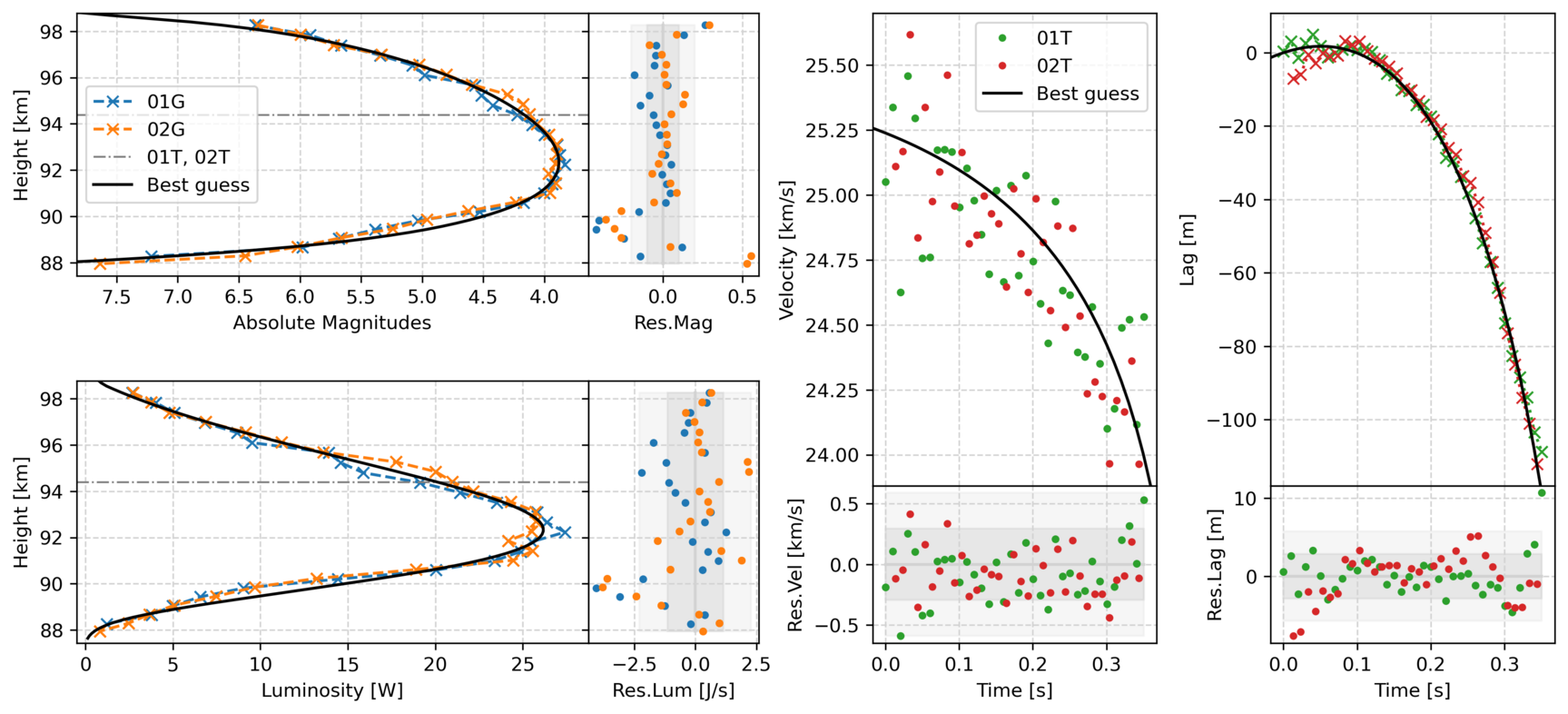}
\end{minipage}

\begin{minipage}{0.7\linewidth}
    \centering
    \includegraphics[width=\linewidth]{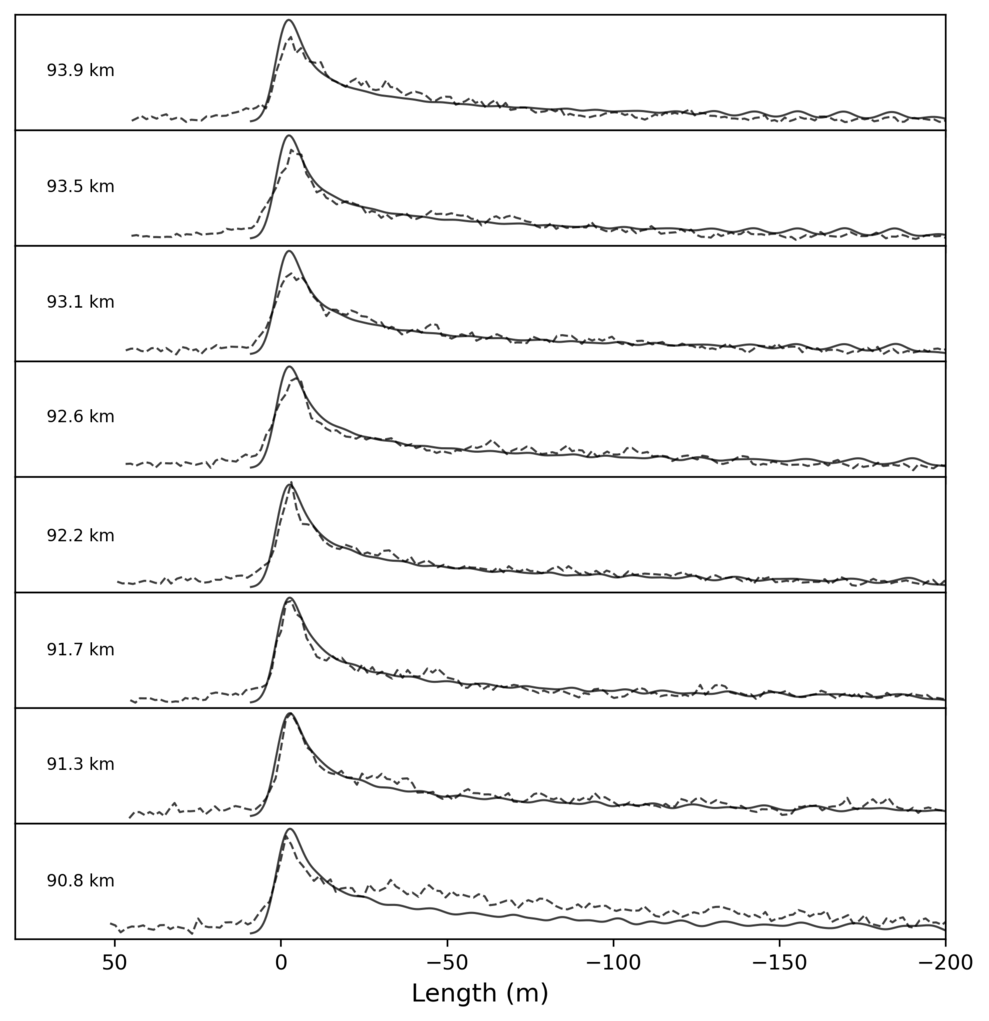}
\end{minipage}

\caption{Top: Best-fit simulation results for the 20200726\_032722 meteor, showing the synthetic luminosity and lag curves matched to EMCCD and CAMO observations. Bottom: Wake comparison using CAMO data for the same simulation.}
\label{img:CAP_20200726_032722}
\end{figure}

\newpage  

\subsubsection{20200726\_060419}
\begin{figure}[h!]
\centering
\begin{minipage}{0.85\linewidth}
    \centering
    \includegraphics[width=\linewidth]{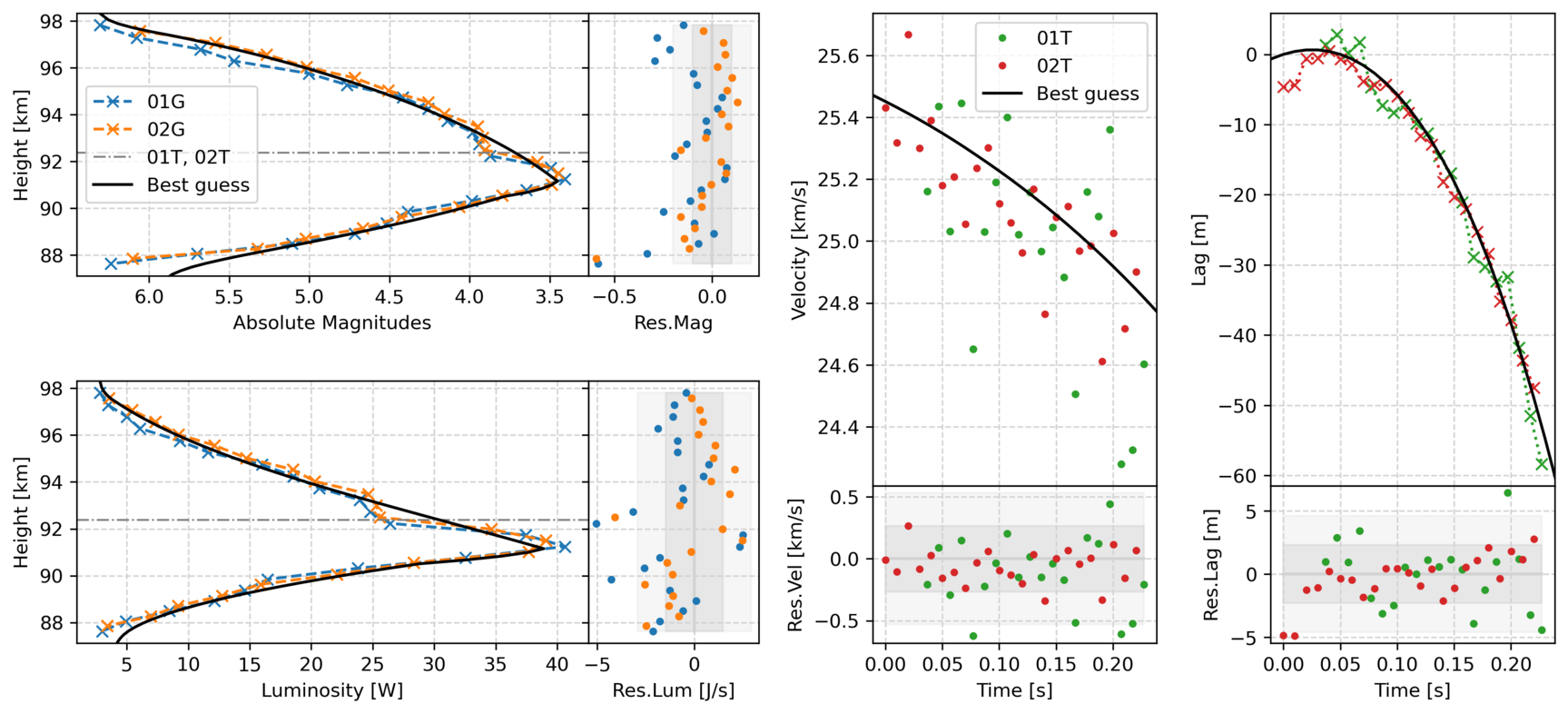}
\end{minipage}

\begin{minipage}{0.7\linewidth}
    \centering
    \includegraphics[width=\linewidth]{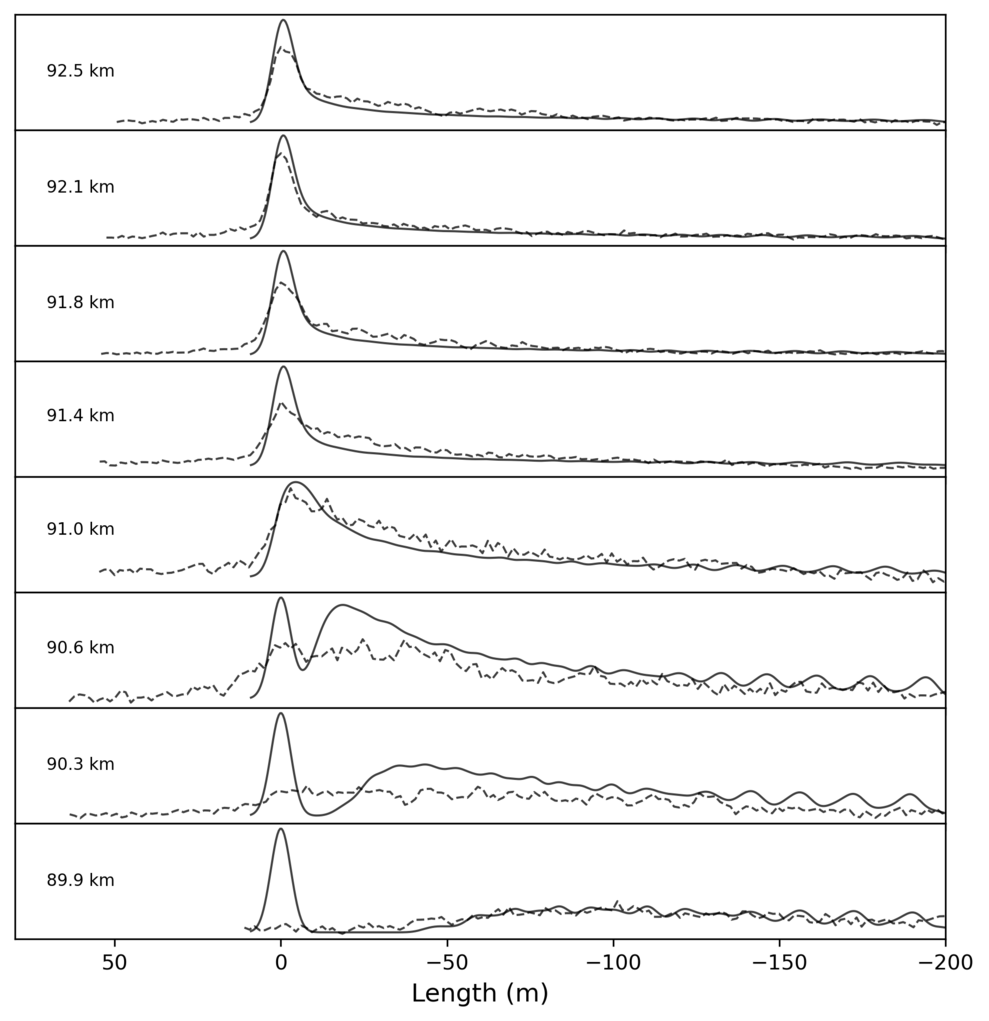}
\end{minipage}

\caption{Top: Best-fit simulation results for the 20200726\_060419 meteor, showing the synthetic luminosity and lag curves matched to EMCCD and CAMO observations. Bottom: Wake comparison using CAMO data for the same simulation.}
\label{img:CAP_20200726_060419}
\end{figure}

\newpage

\subsubsection{20220726\_070831}
\begin{figure}[h!]
\centering
\begin{minipage}{0.85\linewidth}
    \centering
    \includegraphics[width=\linewidth]{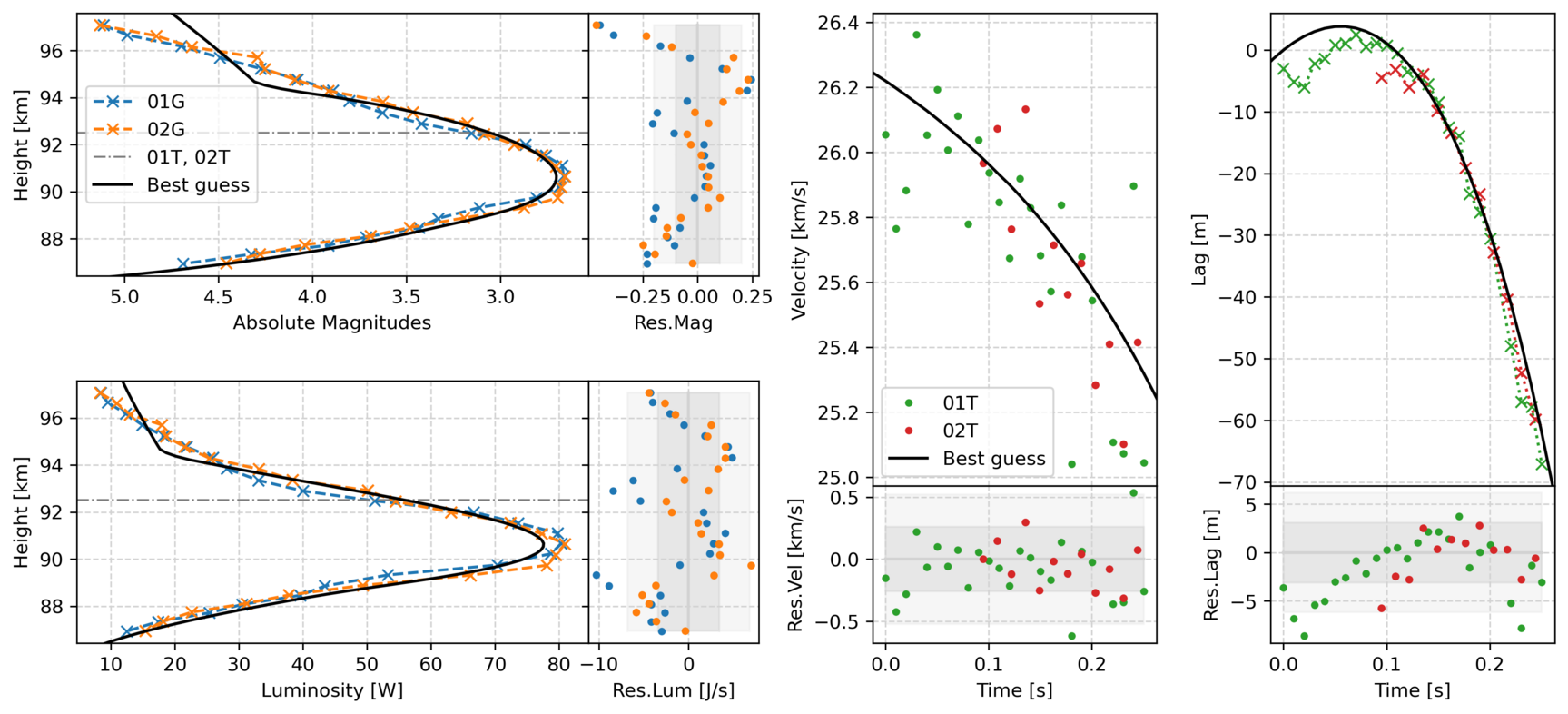}
\end{minipage}

\begin{minipage}{0.7\linewidth}
    \centering
    \includegraphics[width=\linewidth]{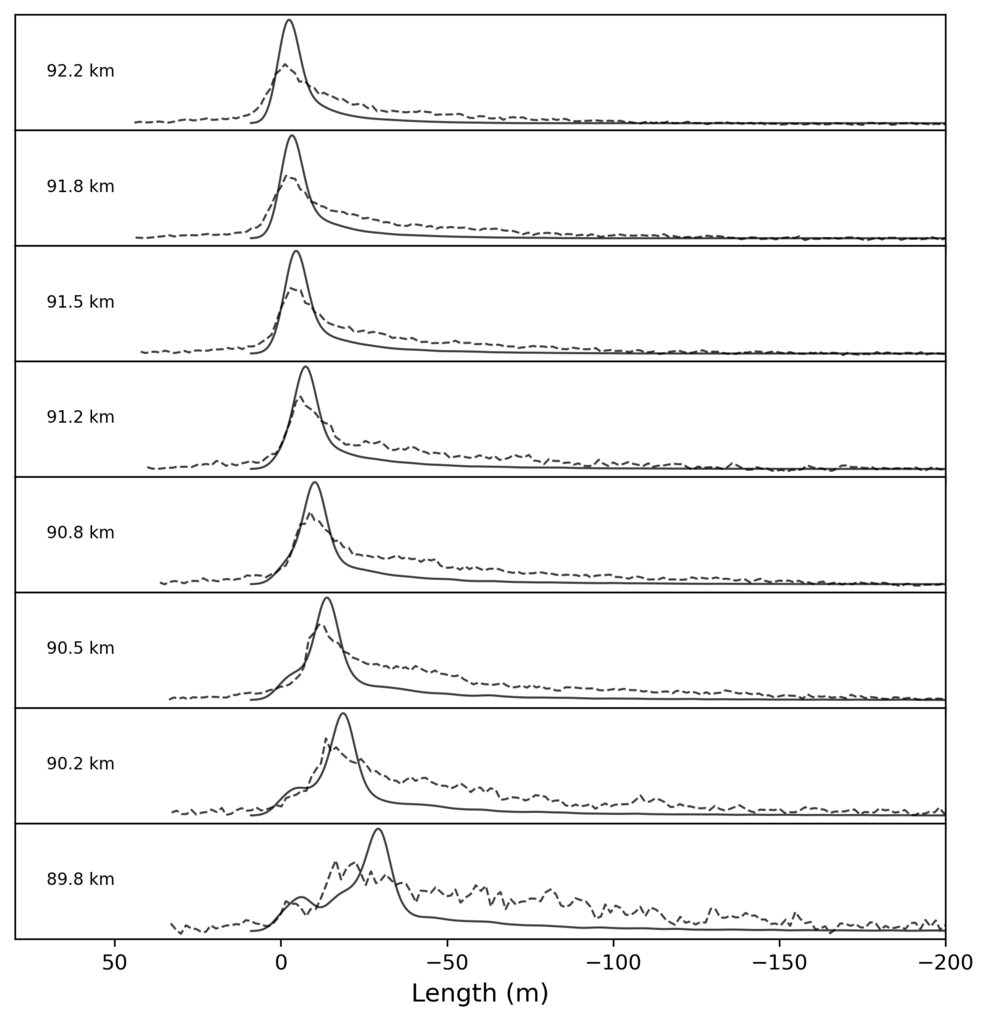}
\end{minipage}

\caption{Top: Best-fit simulation results for the 20220726\_070831 meteor, showing the synthetic luminosity and lag curves matched to EMCCD and CAMO observations. Bottom: Wake comparison using CAMO data for the same simulation.}
\label{img:CAP_20220726_070831}
\end{figure}

\newpage

\subsubsection{20220729\_044924}
\begin{figure}[h!]
\centering
\begin{minipage}{0.85\linewidth}
    \centering
    \includegraphics[width=\linewidth]{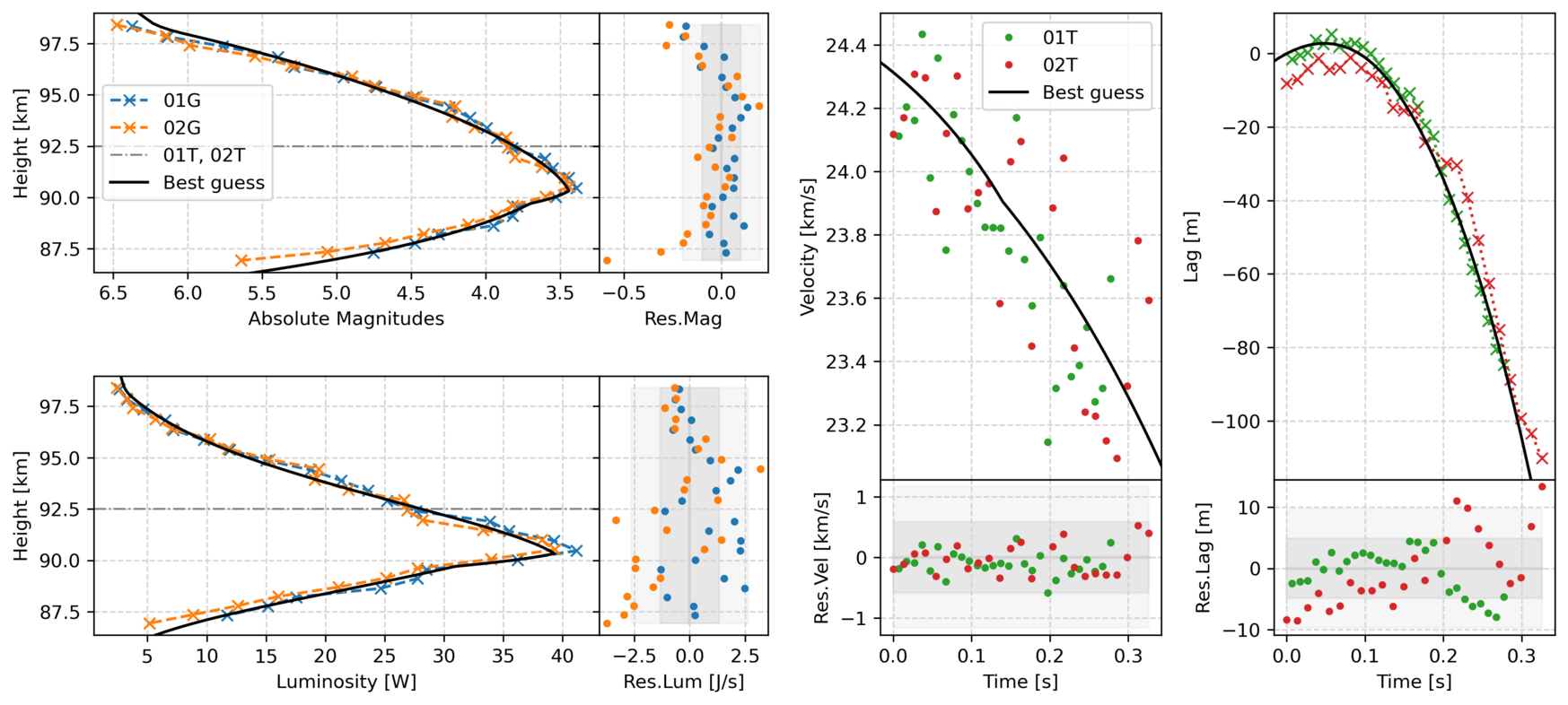}
\end{minipage}

\begin{minipage}{0.7\linewidth}
    \centering
    \includegraphics[width=\linewidth]{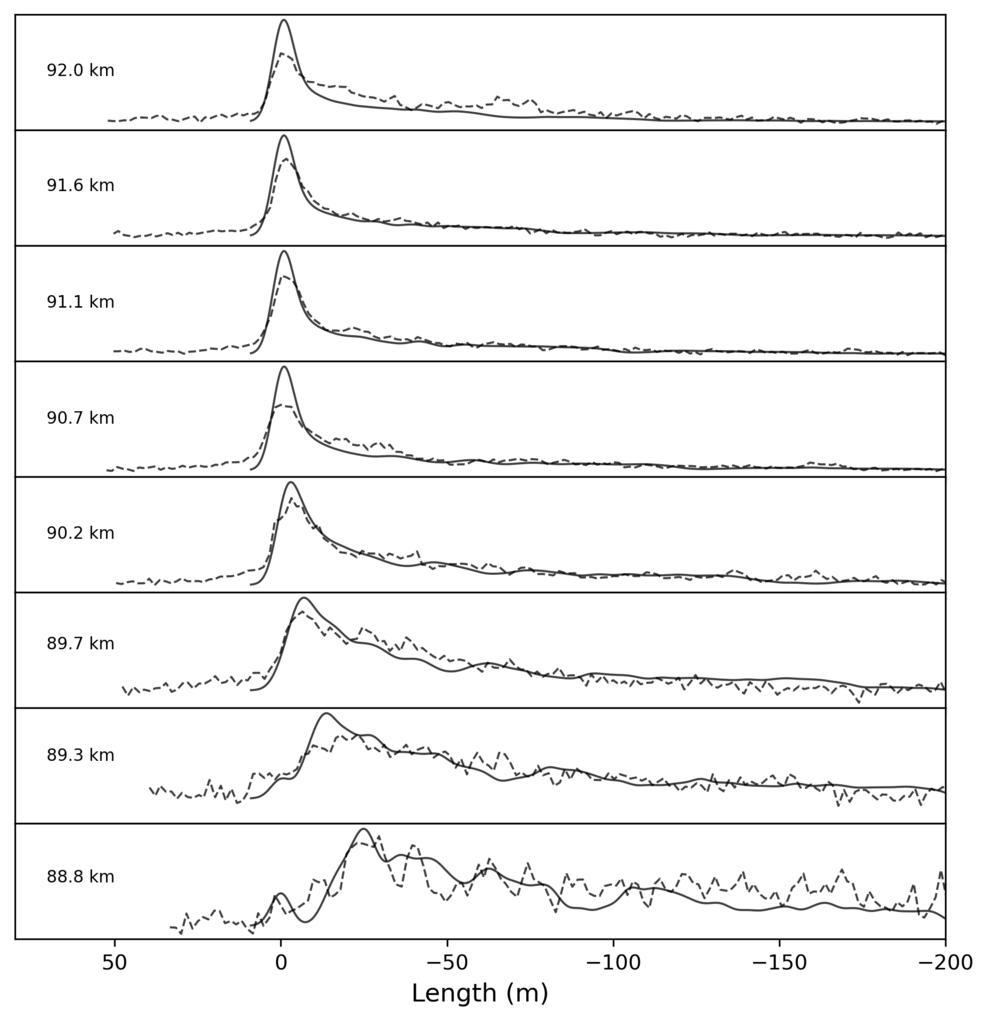}
\end{minipage}

\caption{Top: Best-fit simulation results for the 20220729\_044924 meteor, showing the synthetic luminosity and lag curves matched to EMCCD and CAMO observations. Bottom: Wake comparison using CAMO data for the same simulation.}
\label{img:CAP_20220729_044924}
\end{figure}

\newpage

\section{Sensitivity analysis for grain density and different camera system}\label{sec:Apx sensitivity plots}

\subsection{Grain density change}\label{sec:Apx sensitivity rhoGr plots}

Changing the assumed grain bulk density from \(3000\) to \(3500~\mathrm{kg\,m^{-3}}\) does not significantly alter the inferred physical characteristics of either Orionid or Capricornid meteors. All retrieved parameters remain well within the credible intervals defined by the dynamic nested sampling posteriors, indicating that the model is relatively insensitive to moderate variations in the adopted grain density. This suggests that the baseline value of \(3000~\mathrm{kg\,m^{-3}}\) provides a robust representation of the grain composition for both meteor showers.

\begin{figure}[h!]
    \centering
    \includegraphics[width=0.9\linewidth]{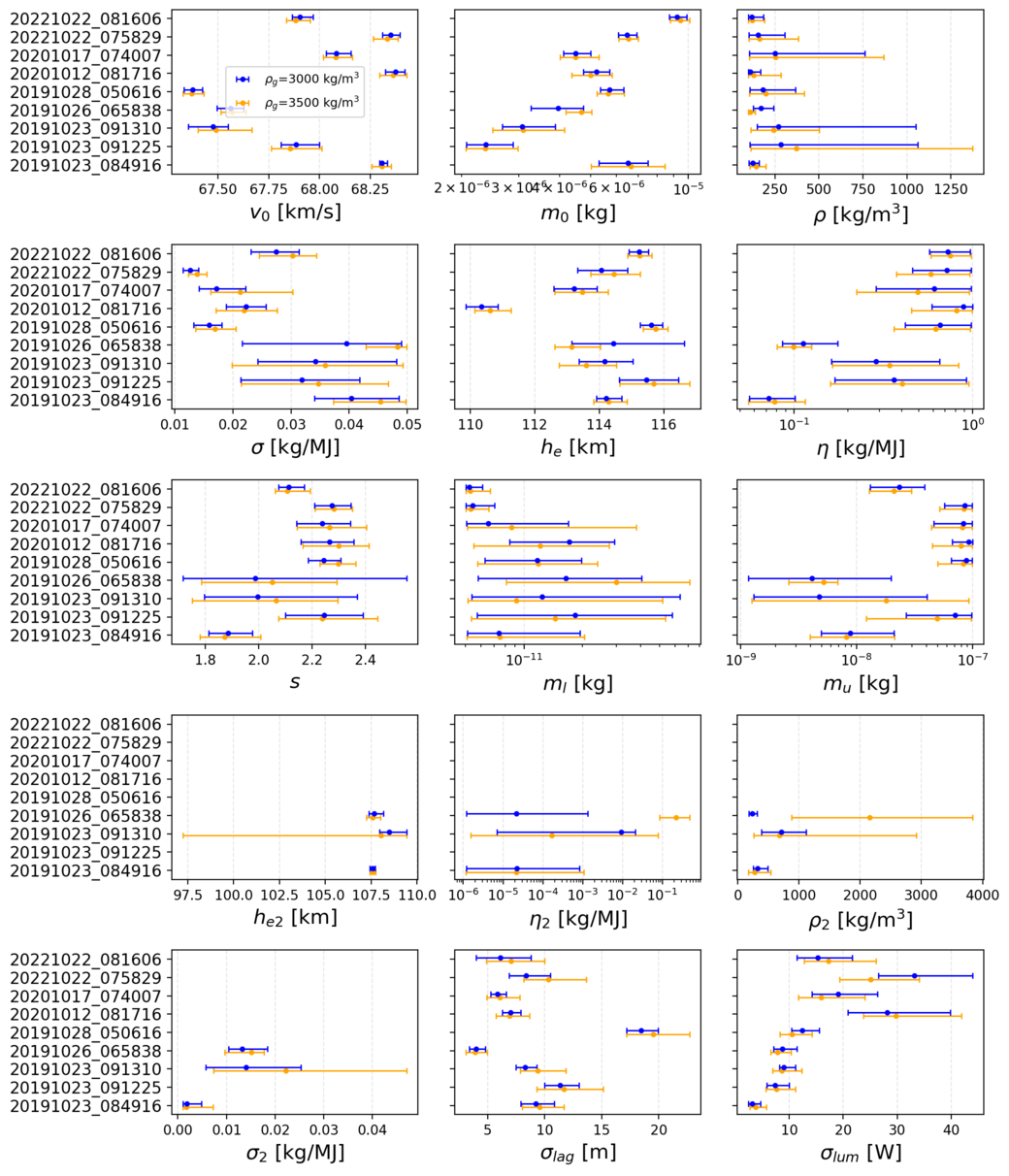}
    \caption{ORI sensitivity of inferred parameters to grain density \(\rho_g\). Points mark posterior medians; bars denote 95\% credible intervals. Blue: \(\rho_g = 3000~\mathrm{kg\,m^{-3}}\); orange: \(3500~\mathrm{kg\,m^{-3}}\).}
    \label{img:ORI_rhoGrCheck}
\end{figure}

\begin{figure}[h!]
    \centering
    \includegraphics[width=0.9\linewidth]{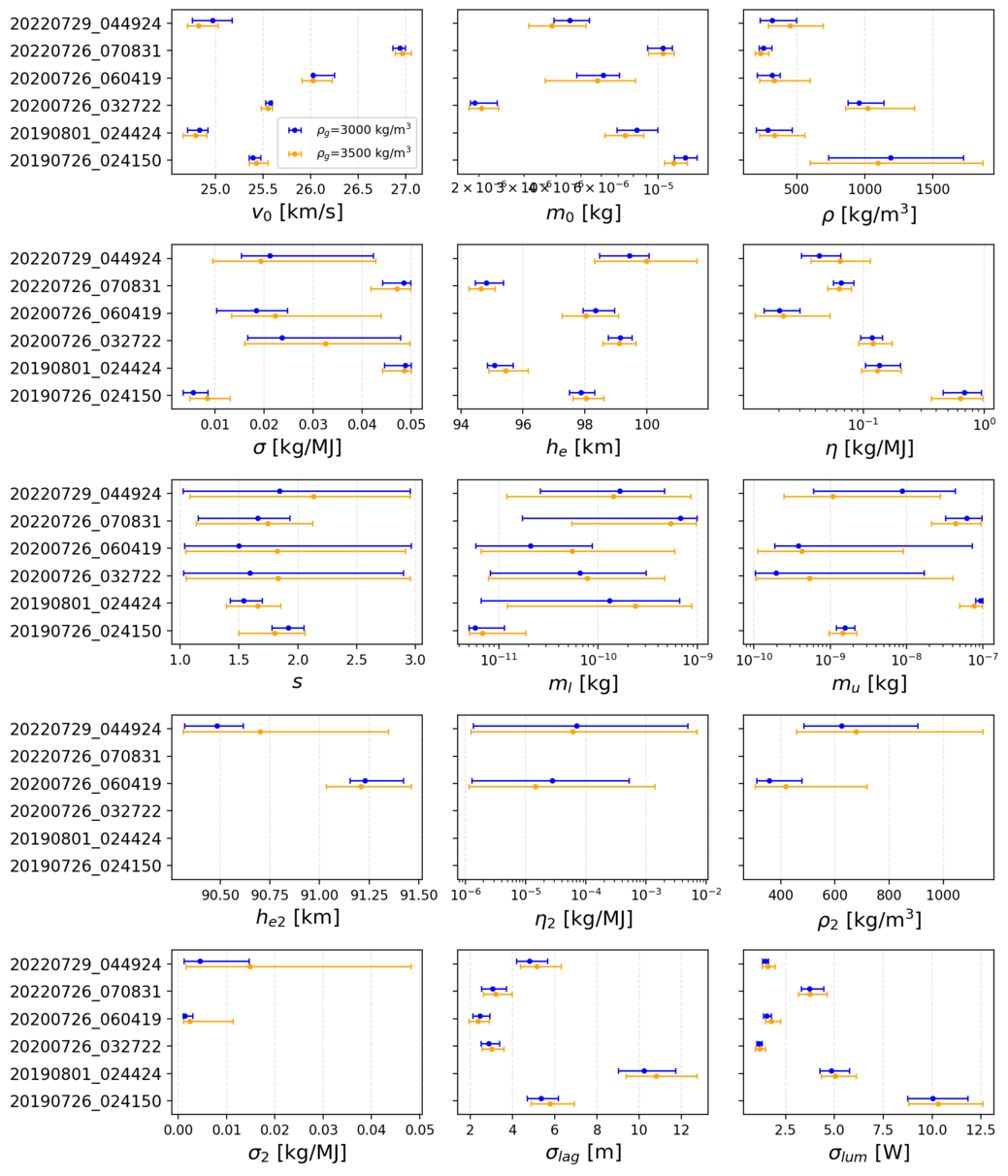}
    \caption{CAP sensitivity of inferred parameters to grain density \(\rho_g\).}
    \label{img:CAP_rhoGrCheck}
\end{figure}

\newpage

\clearpage

\subsection{EMCCD only}\label{sec:Apx sensitivity cam plots}

When EMCCD lag measurements align well with CAMO, the inferred physical parameters remain largely consistent, although the uncertainties increase due to the higher noise levels inherent in EMCCD lag data—consistent with the patterns observed in our dedicated EMCCD validation tests (Section~\ref{sec:EMCCD_validation}). However, when the EMCCD and CAMO lag profiles diverge, the physical solutions begin to shift, particularly for parameters tied to the second fragmentation stage, such as $\rho_2$, $\sigma_2$, and $\eta_2$.

\begin{figure}[h!]
    \centering
    \includegraphics[width=0.9\linewidth]{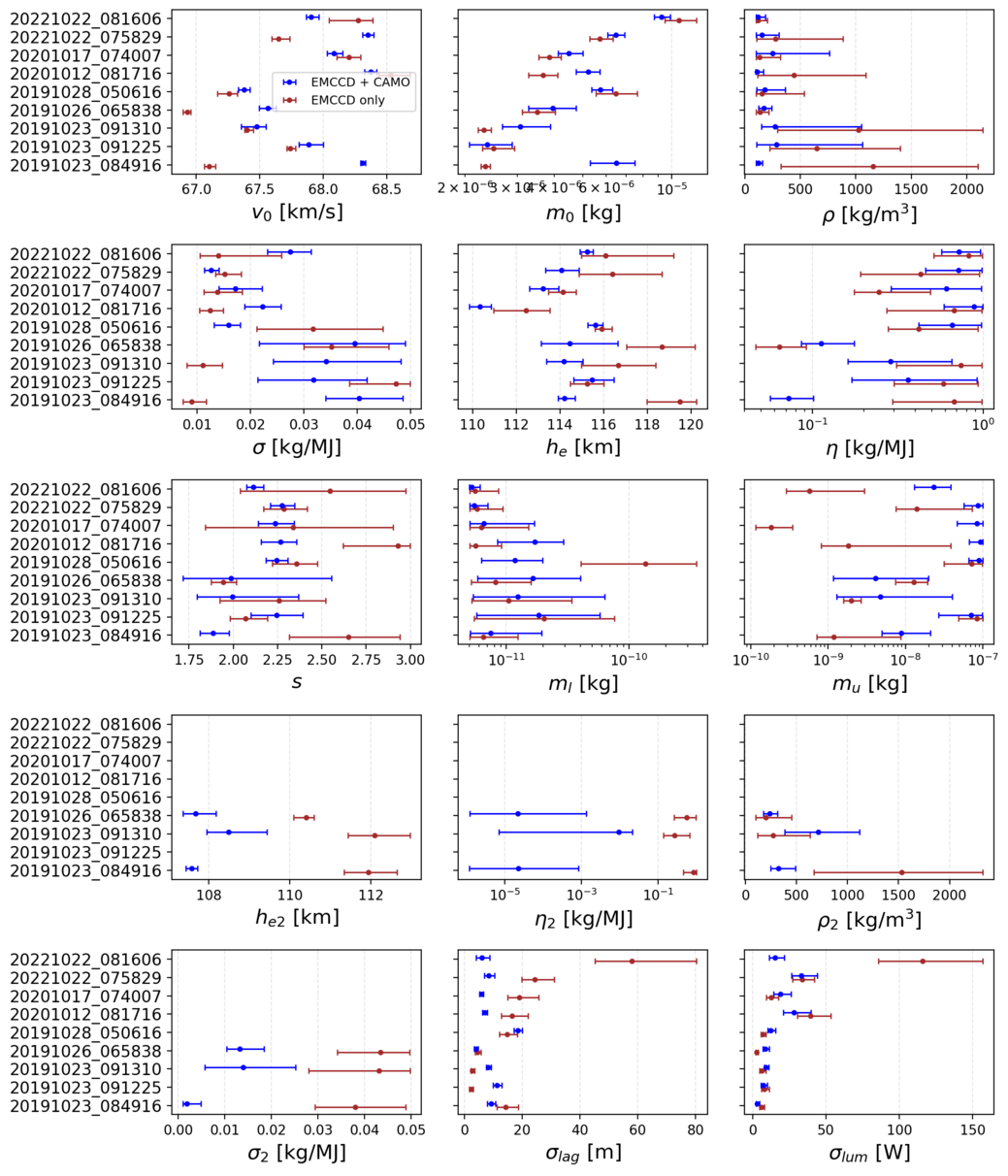}
    \caption{ORI sensitivity of inferred parameters to EMCCD only compare to this study where EMCCD light-curve is used and while CAMO narrow-field is used for the lag. Points mark posterior medians; bars denote 95\% credible intervals. Blue: this study; brown: only EMCCD data.}
    \label{img:ORI_camCheck}
\end{figure}

\begin{figure}[b!]
    \centering
    \includegraphics[width=0.9\linewidth]{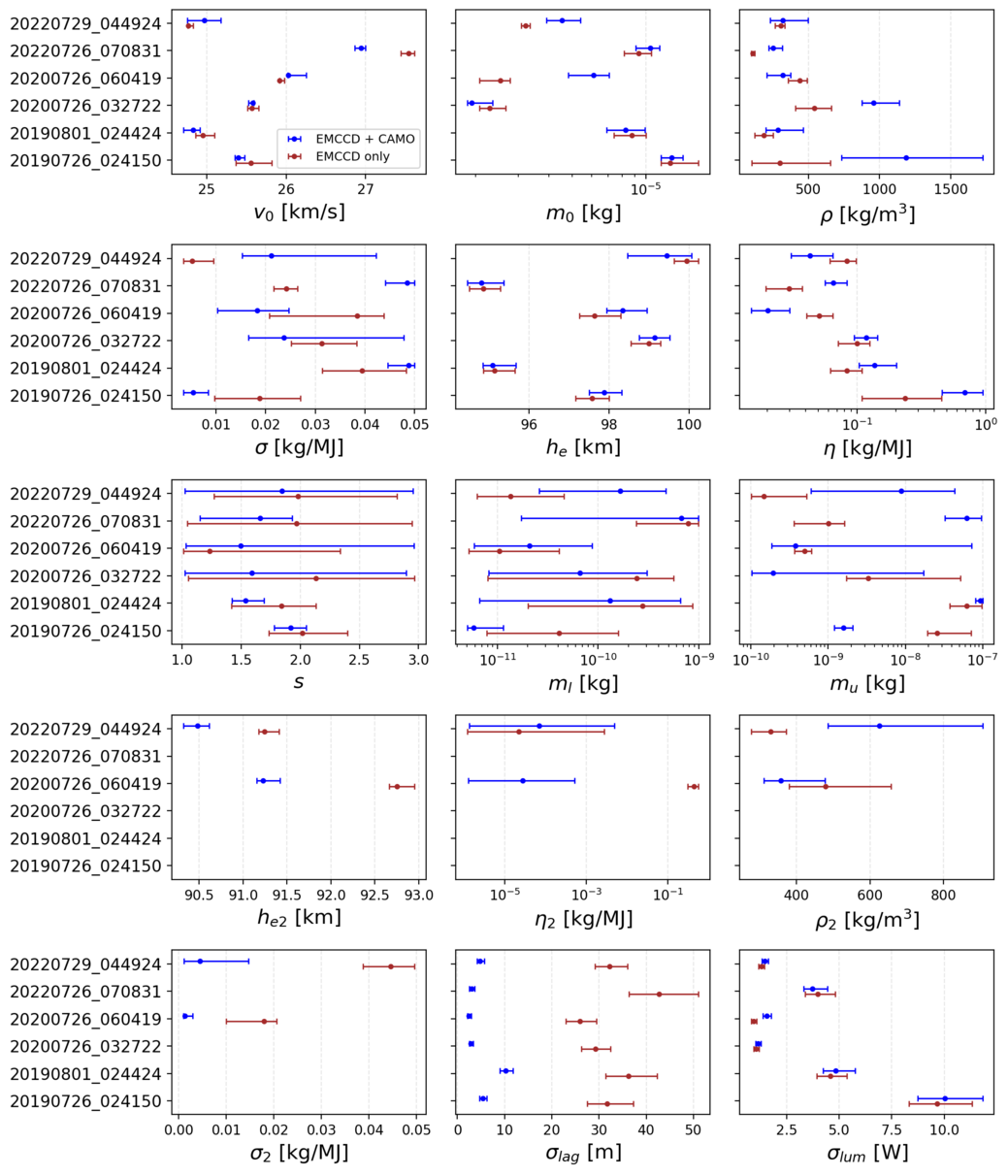}
    \caption{CAP sensitivity of inferred parameters to EMCCD only compare to this study where EMCCD light-curve is used and while CAMO narrow-field is used for the lag.}
    \label{img:CAP_camCheck}
\end{figure}

This is likely because CAMO narrow-field data are better equipped to capture faint secondary fragments that are often difficult to resolve in EMCCD measurements alone. As a result, EMCCD-only solutions tend to indicate ongoing erosion after the second fragmentation, pushing the inferred second-fragment densities and ablation coefficients toward higher values. By contrast, parameters more directly constrained by the initial light-curve shape—such as the initial mass $m_0$ and erosion start height $h_e$—remain relatively stable, and the initial velocity $v_0$ also shows only minimal variation. The ablation coefficient $\sigma$ and erosion coefficient $\eta$ maintain broadly consistent distributions across both datasets; however, we observe that the upper ($m_u$) and lower ($m_l$) erosion mass limits, particularly for the CAP events, tend to be less skewed toward their extreme values when using EMCCD-only data. The erosion index $s$, while noisier, retains its characteristic peak in both ORI and CAP showers, although it typically shifts by about 0.2 toward higher values. Notably, the bulk density $\rho$ exhibits a tendency to broaden under EMCCD-only analyses, reflecting the increased uncertainty introduced by the noisier lag measurements. While the lag error $\sigma_{lag}$ increases by an order of magnitude almost in all cases especially for the CAP as they have more data points compared to ORI meteors.

\newpage

\clearpage

\bibliographystyle{Alphab-elsarticle-num-names} 
\bibliography{Bibliog.bib}

\end{document}